\documentclass[12pt,english]{article}
\pdfoutput=1

\usepackage[authordate,
backend=biber,
doi=only,
isbn=false,
sorting=nyt,
maxcitenames=3,
minbibnames=7,
maxbibnames=7,
uniquename=false,
sortcites=true]{biblatex-chicago}

\bibliography{bib/references.bib}

\AtEveryBibitem{\clearlist{note}\clearlist{language}\clearlist{issn}} 
\AtEveryBibitem{%
	\ifentrytype{online}{%
		\clearfield{urlyear}
		\clearfield{urlmonth}
		\clearfield{urlday}
		\clearfield{note}
		\clearlist{language}
	}{%
		\clearfield{eprint}%
		\clearfield{urlyear}
		\clearfield{urlmonth}
		\clearfield{urlday}
		\clearfield{note}
		\clearlist{language}
	}
}%

\usepackage{mathptmx}
\usepackage{microtype}
\everypar{\looseness=-1}
\usepackage{amsmath}
\usepackage{bbm}
\usepackage{bm}

\usepackage[utf8]{inputenc}
\usepackage[T1]{fontenc}
\usepackage{babel}

\usepackage{setspace}

\usepackage{breakurl}
\usepackage{graphicx}
\usepackage{booktabs,dcolumn}
\usepackage{array}
\usepackage{multirow}
\usepackage[font=singlespacing, skip=3pt]{caption}
\usepackage[usenames,dvipsnames,svgnames,table]{xcolor}
\usepackage{float}
\usepackage[export]{adjustbox}[2011/08/13]
\usepackage{enumitem}
\usepackage{tikz}
\usepackage{float}

\usepackage[hang, flushmargin, bottom, symbol]{footmisc}
\usepackage[colorlinks=true, linkcolor=blue, citecolor=blue, plainpages=false, pdfpagelabels=true, urlcolor=blue]{hyperref}
\usepackage[text={16cm,24cm}]{geometry}
\usepackage{ragged2e}

\usepackage{xspace}

\usepackage[nolists, nomarkers]{endfloat}

\usepackage{subcaption}
\usepackage{caption}
\makeatletter
\date{September 8, 2021}

\newcolumntype{L}[1]{>{\raggedright\let\newline\\\arraybackslash\hspace{0pt}}m{#1}}
\newcolumntype{C}[1]{>{\centering\let\newline\\\arraybackslash\hspace{0pt}}m{#1}}
\newcolumntype{R}[1]{>{\raggedleft\let\newline\\\arraybackslash\hspace{0pt}}m{#1}}

\newcommand{\sym}[1]{\ifmmode^{#1}\else\(^{#1}\)\fi}

\usepackage{calc}
\usepackage{footnotebackref}
\newcommand{\PAPERKEYWORDS}{\textbf{Keywords}: China, School Closure, Education Infrastructure, Rural Education}
\newcommand{\PAPERJEL}{\textbf{JEL}: H40, I21, O15}

\newcommand{\PAPERTITLE}{Estimating the Effects of Educational System Consolidation: The Case of China's Rural School Closure Initiative}

\newcommand{\AUTHORHANNUM}{Emily Hannum}
\newcommand{\AUTHORHANNUMURL}{https://orcid.org/0000-0003-2011-9984}
\newcommand{\AUTHORHANNUMINFO}{\href{\AUTHORHANNUMURL}{\AUTHORHANNUM}: Department of Sociology and Population Studies Center, University of Pennsylvania, 3718 Locust Walk, Philadelphia, PA 19104 (email:hannumem@sas.upenn.edu)}

\newcommand{\AUTHORLIU}{Xiaoying Liu}
\newcommand{\AUTHORLIUURL}{https://orcid.org/0000-0001-9897-1291}
\newcommand{\AUTHORLIUINFO}{\href{\AUTHORLIUURL}{\AUTHORLIU}: Population Studies Center, University of Pennsylvania, 3718 Locust Walk, Philadelphia, PA 19104 (email:xiaoyliu@sas.upenn.edu)}

\newcommand{\AUTHORWANG}{Fan Wang}
\newcommand{\AUTHORWANGURL}{https://orcid.org/0000-0003-2640-5420}
\newcommand{\AUTHORWANGINFO}{\href{\AUTHORWANGURL}{\AUTHORWANG} (\emph{corresponding}): Department of Economics, University of Houston, 3623 Cullen Boulevard, Houston, TX 77204 (email: fwang26@uh.edu)}

\newcommand{\ACKNOWLEDGMENTS}{
We gratefully acknowledge support from Penn's University Research Foundation and School of Arts and Sciences Research Opportunity Grant Programs, support from Grand Challenges Canada (PI: Jere Behrman), and support from the Chiang Ching-Kuo Foundation (Scholar Grant GS040-A-18) for coverage of Wang’s time. We thank Jere Behrman, Aimee Chin, Anthony Howell, Elaine Liu, Stephanie Psaki, Ma Xiang, Yu Xie, Hongliang Zhang, and Shuang Zhang for suggestions and comments.}

\newcommand{\PAPERABSTRACT}{
Global trends of fertility decline, population aging, and rural outmigration are creating pressures to consolidate school systems, with the rationale that economies of scale will enable higher quality education to be delivered in an efficient manner, despite longer travel distances for students. Yet, few studies have considered the implications of system consolidation for educational access and inequality, outside of the context of developed countries. We estimate the impact of educational infrastructure consolidation on educational attainment using the case of China's rural primary school closure policies in the early 2000s. We use data from a large household survey covering 728 villages in 7 provinces, and exploit variation in villages' year of school closure and children's ages at closure to identify the causal impact of school closure. For girls exposed to closure during their primary school ages, we find an average decrease of 0.60 years of schooling by 2011, when children's mean age was 17 years old. Negative effects strengthen with time since closure. For boys, there is no corresponding significant effect. Different effects by gender may be related to greater sensitivity of girls' enrollment to distance and greater responsiveness of boys' enrollment to quality.\\
\PAPERJEL}

\newcommand{\PAPERDOIURL}{https://doi.org/10.1086/711654}
\newcommand{\PAPERINFO}{
This paper is published as: Hannum, Emily, Xiaoying Liu, and Fan Wang. ``Estimating the Effects of Educational System Consolidation: The Case of China’s Rural School Closure Initiative.'' Economic Development and Cultural Change 70, no. 1 (October 1, 2021): 485–528. \url{\PAPERDOIURL}.
}

\newcommand{\regone}{1}
\newcommand{\regtwogirl}{1}
\newcommand{\regtwoboy}{1}
\newcommand{\reginteract}{0.91}

\newcommand{\aptxspc}{1.8}

\newcommand{\baselinegroup}{Baseline group: 14--21 years old at village primary school closure year}
\newcommand{\newtablepanfemale}{Female only Regressions}
\newcommand{\newtablepanmale}{Male only Regressions}
\newcommand{\closeinteragestart}{\textit{Closure} $\times$ age at closure}
\newcommand{\closeinteragestardura}{\textit{Closure} $\times$ child age at village primary school closure year was}
\newcommand{\enrolldistbase}{Categorical distance (compare to 0 km)}
\newcommand{\enrollqualbase}{Categorical quality (compare to 0-3)}

\newcolumntype{L}[1]{p{#1}}
\newcolumntype{C}[1]{>{\centering\let\newline\\\arraybackslash\hspace{0pt}}m{#1}}

\newcommand{\footnotegap}{\addlinespace[-1.3em]}
\newcommand{\innerheadwidth}{9cm}
\newcommand{\subheadwidth}{15cm}
\newcommand{\lablcolwidth}{7.1cm}
\newcommand{\footwidth}{18.5cm}
\newcommand{\innerheadwidthenroll}{11.1cm}
\newcommand{\subheadwidthenroll}{12.95cm}
\newcommand{\lablcolwidthenroll}{6.75cm}
\newcommand{\footwidthenroll}{20cm}
\newcommand{\innerheadwidthsummeight}{11.1cm}
\newcommand{\subheadwidthsummeight}{23cm}
\newcommand{\lablcolwidthsummeight}{7.5cm}
\newcommand{\footwidthsummeight}{22.5cm}

\newcommand{\formatpanelheadsumm}[3]{\multicolumn{#1}{#2}{\textbf{{#3}}}\\[0.50ex]}
\newcommand{\formatpanelheadsupregs}[3]{\multicolumn{#1}{#2}{\textbf{\textit{#3}}}\\[0.25ex]}
\newcommand{\formatpanelheadregs}[3]{\multicolumn{#1}{#2}{\textbf{{#3}}}\\[1.00ex]}
\newcommand{\formatpanelheadsubregs}[3]{\multicolumn{#1}{#2}{\textit{#3}}\\[0.75ex]}
\newcommand{\exclcontrol}{\multicolumn{7}{L{\subheadwidth}}{\textbf{\textit{\normalsize Exclusions and controls:}}} \\[0.25ex]}
\newcommand{\exclcontrolcont}{Village and province-age FE and controls\dag & Yes &     Yes &     Yes &     Yes &     Yes &     Yes    \\
Exclude villages that never had schools\ddag&   &     Yes &   &     Yes &   &     Yes    \\}

\newcommand{\footsummstatsmain}{Village level summary statistics. {\large\dag} Column 4 shows the p-values from tests of the difference of the variable means between closure and non-closure villages conditional on provincial fixed effects. {\large\ddag} Column 5 shows the p-values of a linear trend test across the year of closure for each variable among villages experiencing school closure.}

\newcommand{\footsummstatsabc}{\footsummstatsmain\xspace All variables are from the village-head survey except for fraction of woman and fraction of individuals below age 30 which are calculated by authors.}

\newcommand{\footeduattain}{Table shows means of variables. Individuals in Group A are those that are  fully exposed to consolidated primary schools. Group B are individuals that were 6 to 9 at the year of closure and were exposed to consolidated primary school for more than half of their primary school years. Group C consists of individuals who were 10 to 13 at year of closure and transitioned from village schools to consolidated primary schools during the final years of primary school. Group D and E consist of individuals who are from villages that experienced closure, but were beyond primary school age at the year of closure. Group F consists of individuals from villages without school closure.}

\newcommand{\footgraphmain}{Each dot represents the impact of school closure on grades completed by 2011 for each age group ($a$--$b$) defined at the time of school closure. These results, estimated for finer age-at-closure groups, correspond to the results as shown in Column 1 of Table \ref{regone} which had 5 age-at-closure groups (see Section \ref{sec:ageeffects}). All coefficients are from estimating Equation \eqref{eq:startsOnly}.}
\newcommand{\footgraphfiner}{\footgraphmain}

\newcommand{\footattaincore}{\dag\thinspace Controls include ethnicity, household size and relative household wealth.\thinspace
\ddag\thinspace Odd columns check robustness by excluding category 2 villages (22 villages with closure between 1999 and 2010, but also have a school in 2011) and category 4 villages (48 villages that never had a school and 35 villages that only had a village primary school before 1999) discussed in Section \ref{sec:data}.\thinspace \\
Statistical significance:\thinspace* 0.10 ** 0.05 *** 0.01. Robust standard errors clustered at village level. Each column/panel is a separate regression. Estimates compare children with different starting age of exposure (and length of exposure) to children in baseline group who should not be impacted by school closure. Columns 3 to 6 restrict the sample to smaller 2011 age ranges (see Table \ref{summthree}).}

\newcommand{\footattain}{\footattaincore\xspace Sample individuals are all below 45 years of age in year 2011, and below 29 at the year-of-closure for those who experienced school closure.}

\newcommand{\footenrollcore}{Statistical significance:\thinspace* 0.10 ** 0.05 *** 0.01. Standard errors clustered at village level. Each column is a separate regression. Distance to closest primary school and school facility information are reported by village head. School facilities include pipe water, library, computers, non-dilapidated buildings and others shown in Table \ref{summtwo}. All regressions include county fixed effects, province-specific age fixed effects, controls for village per capita income, village per capita land area, village population size, household relative wealth, the number of household members and household ethnicity.}
\newcommand{\footenroll}{\footenrollcore}

\newcommand{\footattainmoreage}{\footattaincore\xspace Individuals included in regressions from all columns are below 35, 30 and 25 years of age in 2011 in each of the three Panels, and below 21 at the year-of-closure for those that experienced closure between 1999 and 2010.}

\newcommand{\footattaininter}{\footattain}

\newcommand{\footenrollinter}{\footenrollcore}

\def\tikzageedusmall{
\begin{tikzpicture}[overlay]
	\node[draw=none,text width=11cm, line width=0mm]
	    at (14.00, 3.55)
	        {{ Tables \ref{regone}, \ref{regtwogirl} and \ref{regtwoboy} columns 3 and 4 regression data}};
	\draw[black,dashed, thick, opacity=.65, line width=0.75mm]
	    (8.5, -5.5) rectangle (17.90, 3.3);
\end{tikzpicture}
}

\begin{document}

\title{
\vspace*{-2em}
\onehalfspacing
\PAPERTITLE
\thanks{\PAPERINFO}}

\author{
\AUTHORHANNUM, \AUTHORLIU, and \AUTHORWANG\thanks{
\AUTHORHANNUMINFO;
\AUTHORLIUINFO;
\AUTHORWANGINFO.
\ACKNOWLEDGMENTS}}

\date{
October 1, 2021}
\maketitle

\vspace*{-1em}
\begin{abstract}
\singlespacing
\PAPERABSTRACT
\end{abstract}
\vfil
\hfil \small\PAPERKEYWORDS \hfil
\vfil
\thispagestyle{empty}
\clearpage

\pagenumbering{arabic}
\setcounter{page}{1}
\renewcommand*{\thefootnote}{\arabic{footnote}}
\section{Introduction}
\label{subsec:intro}

Educational infrastructure consolidation has been a long-standing policy response to declining student populations in high-income countries, with the rationale that economies of scale will enable higher quality education to be delivered in an efficient manner. However, the pressure to consolidate is expanding beyond high-income country settings, as more middle-income countries experience demographic trends of declining fertility, population aging, and rural outmigration.\footnote{Press reports suggest that school consolidation is also emerging as an important policy response to changing demographics in middle income countries with large rural populations such as Thailand \Autocite{saengpassa_ministrys_2017} and India \Autocite{chowdhury_cramped_2017}, among other countries. See news reports and press releases about current or planned consolidation efforts in \textcite{setiawati_schools_2010, harun_schools_2017, tawie_sarawak_2017}.} In low- and middle-income countries, past research has addressed the impact of school expansion \Autocite{duflo_2001_school, ANDRABI20131, BurdeLinden_2013, Kazianga_2013} and programs to improve access to school \Autocite{Muralidharan_2017} on school enrollment and attainment. However, to date, relatively few studies have considered the impact of school system contraction on student outcomes, outside of high-income countries.\footnote{One important exception is a study of high school closures at the end of the Cultural Revolution in China, which found large declines in high school completion rates and significant negative long term labor market outcomes among individual who were exposed to high school closures \autocite{Zhang2018}.} This paper estimates the impact of school consolidation on educational attainment\footnote{By educational attainment, we mean the highest grade level (in years) that an individual has completed. Each year of additional grade completed adds to the current educational attainment of a child. We consider both educational attainment for children who are still going to school and for children who have completed schooling.} in a developing country. In the context of rural villages in developing countries, school consolidation could significantly increase the cost of school enrollment by increasing the travel cost of attendance, but might also increase the perceived return to education due to improvements in school quality.

China is at the vanguard of the consolidation trend. Nationally, China faced dramatic declines in school-aged cohorts in the early 2000s due to fertility reduction.\footnote{For example, China's 0 to 14 population dropped from 321,937,264 in 2000, to 266,616,527 in 2005, to 240,183,007 in 2010, to 233,556,402 in 2015 \autocite{UScensus}.} While fertility is generally somewhat higher in rural areas than urban areas, depopulation through unprecedented rural-urban migration has occurred in these settings.\footnote{The urban population in China increased from 26.41 percent in 1990 to 49.95 percent in 2010 \autocite{textor_china_2021}} The State Council in 2001 initiated a massive national push to consolidate educational infrastructure. The school consolidation initiative intended to address sparse demand, inefficiencies in provision, and perceived quality problems in rural education \Autocite{mei_school_2015}. Consolidation has happened extremely rapidly in China. Official data suggests that the total number of primary schools decreased by about 53 percent, from 491,273 to 228,585, between 2001 and 2012 \autocite{ceic_cn_2021}.

We analyze the impact of these changes on educational attainment using the 2011 China Household Ethnic Survey (CHES 2011), which is a household and village survey implemented in 728 villages in 7 provinces and autonomous regions with substantial minority populations. Driven by the national directive to consolidate schools, 215 of the villages in the sample experienced village primary school closure between 1999 and 2010.\footnote{We use 1999 as the empirical cut-off year because school closures were sporadic before this date in the data. Even though the central directive for school consolidation was officially issued in 2001, from our data, it seems that the policy was in place in some counties before the nation-wide policy announcement (also see \textcite{dai_cost_2017}).} Typically, closure decisions were made by county administrators who eliminated village schools and required students to attend schools that were farther away--generally located in township centers--but better appointed. In the CHES survey, compared to villages that had not experienced a school closure, villages that had experienced school closures are on average 3.8 km further away from the closest primary school. Compared to schools in villages without a closure, schools serving villages that had experienced closure have better buildings and technological equipment. We find that villages with and without school closure in our survey have similar income levels, a similar fraction of agricultural and migrant workers, and similar gender compositions. Villages with closure have, on average, 13 percent fewer households (415 vs 469 households per village), indicating that county administrators tended to close primary schools in smaller villages.

Our estimation strategy is close to that employed by \textcite{duflo_2001_school}, which explores variation in policy exposures across age cohorts to estimate the labor market impact of school expansion program in Indonesia. Using cross-sectional data, we exploit the different calendar years in which village-level closure policies were rolled out. We compare the educational attainment of individuals exposed to the effects of the policy against that of counterparts too old to be exposed to the policy (i.e., individuals who were already beyond primary school age at year of school closure). To identify the policy effect, we then compare changes in educational attainment across these cohorts to corresponding cross-cohort changes in villages unaffected by closures. We also decompose the policy effects into age-at-exposure effects and duration-since-exposure effects (hereafter age-at-closure effects and duration-since-closure effects, respectively)\footnote{The age-at-closure effect is determined by the age of exposed children at year of closure. For example, given a primary school cycle of 6 years, an 8 year old child who was in 3rd grade in the year of closure is potentially exposed to the consolidated primary school for 3 years. The duration-since-closure effect is the time since policy initiation and is determined by the calendar year of school closure. For example, children in villages that experienced school closure in the year 2000 have been exposed to closure for 11 years by the year 2011, when the CHES survey took place.} in the same vein as \textcite{KingBehrman2009WBRO,BehrmanParkerTodd2011}, which emphasize that the timing and duration of exposure are both important dimensions in evaluating social programs.\label{rr:aetwoidentifyintro} In the current context, age-at-closure effects capture the heterogeneous effects on children depending on their age when they experienced school closure, while years-since-closure effects capture the dynamic effects of school closure on children's education as they progress through school, which might strengthen or fade over time.

Specifically, to identify the effects of the school consolidation policy, within each province, we first compare the difference in educational attainment (number of grades completed by 2011) between those from closure villages who were exposed to closure to individuals of the same age cohorts from non--closure villages. We also compare the difference in educational attainment between unexposed individuals from closure villages and individuals of the same age cohorts from non-closure villages. We interpret the difference in these differences as the impact of the policy. We estimate the effects of the policy first over subgroups based on age at year of closure, and then subgroups based both on age at year of closure and the number of years since closure. We interpret results for years since closure as short-, medium- and long-run impacts of the policy on educational progression conditional on age at year of closure.

We find that the school closure policy had a significant negative impact on educational attainment for children exposed to closure. For example, for children who were between age 10 and 13 in the year of closure, we find that school closure reduced grades completed by 0.42 years by 2011, when children are on average 17 years of age. Analyzing girls and boys separately for this subset of children, we find that the reduction in attainment for girls is much greater at 0.60 years, while the reduction for boys is insignificant at 0.24 years. Dividing individuals into subgroups based on age at year of closure as well as the number of years since closure, we also find that the negative effects strengthen with time since closure. For example, for girls who were 6 to 9 years of age at year of closure, there is no significant impact of the policy on their grades completed in the 3 years after closure. However, 4 to 6 years after closure, grades completed is reduced by 0.56 years, and 7 to 12 years after closure, grades completed is lowered by 0.77 years.

To understand the mechanisms that drive these results within the constraints of our data, we considered the relationship between enrollment and school distance and facility quality. Focusing on children who are between 5 and 12 years of age in 2011, we find that each additional kilometer to school is associated with 1.1 percentage point lower school enrollment for girls. Boys' enrollment is also negatively associated with distance to school, but not significantly so. Additionally, we find that boys' enrollment is higher when the closest primary school to the village has better school facilities, but girls' enrollment does not respond to differences in school facility features. By extension, it is possible that increased distance associated with closure tended to impede the education of girls, while improved quality of facilities tended to encourage the education of boys.

The remainder of this paper is organized in the following sections. Section II provides background on school consolidation policies, in comparative perspective and in China. In Section III, we describe the data. Section IV presents our estimation strategy and estimation equations. The first part of Section V presents results from a regression model that differentiates the impact of the policy on children who were in different age groups in the year of closure. The second part of Section V shows differential short-, medium- and long-term impacts of the policy for children in different age groups in the year of closure. Section VI provides a discussion of mechanisms, with a focus on the potential impact of school quality and distance on school enrollment. Section VII concludes.

\section{Literature and Background\label{literature-and-background}}

\subsection{Global Background and Significance}
\label{background-and-significance}

In the United States, school closures have been a common policy response to declining student populations in sparsely populated rural communities, with the rationale that economies of scale would enable higher quality education to be delivered in an efficient manner \autocites[106-107] {post_district_1999}{howley_consolidation_2011}. For example, a trend of consolidating small schools during much of the 20\textsuperscript{th} century reduced the total number of public schools: in 1929 to 30, there were 248,117 public schools, compared to 98,271 in 2013 to 14 \Autocite[][Table 214.10]{united_states_department_of_education_chapter_2016}.\footnote{While in recent years, the number of public schools has held relatively stable, with closures balanced by openings, consolidation remained a non-trivial phenomenon: in 2017 to 2018, for example, there were 1,310 school closures, affecting an estimated 266,777 students who had been enrolled in the prior school year (2016 to 2017) \autocite{united_states_department_of_education_closed_2021}.} While rural population decline has created formidable challenges to maintaining rural schools \Autocite{blauwkamp_school_2011}, closure policies have also emerged in urban areas in recent decades \autocites[e.g.,][]{lee_impact_2017}. For example, in 2013, the third largest school district in the United States, Chicago Public Schools, announced a plan to close 54 primary schools with the expectation of saving 43 million USD annually \autocite{lee_impact_2017}.\footnote{In the United States, this pressure was tied in part to policy pressures to turn around or close ``failing schools'' \autocites{deeds_organizational_2015}{kirshner_tracing_2010}.} The closure decision may be based on a combination of declining enrollments and low achievement, with the idea that economies of scale would enable a higher quality educational experience for those who experience a closure \autocites[for example, see][]{engberg_closing_2012}.

Beyond the United States, school district mergers and school closures have occurred in many countries \autocites[for example, see][]{kearns_status_2009, bartl_economisation_2013, slee_school_2015}. In Chile, between 2002 and 2011, the educational system changed significantly: 1,282 schools closed, constituting about one-tenth of the contemporary stock, and 2,350 new schools were established--most of which were private-voucher schools \autocite{grau_school_2018}. In the Netherlands, consolidation reforms implemented in the 1990s reduced the number of primary schools by about 15 percent in just a few years \autocite[817]{de_haan_school_2016}. In Hong Kong, student enrollment per school started falling around the turn of the millennium, and over 36 percent of primary schools were closed in the decade that followed a closure policy established in 2003 \autocite[3]{chiu_effects_20161}. Press reports suggest that school consolidation is emerging as an important policy response to changing demographics in middle income countries with large rural populations.\footnote{For example, Thailand's Ministry of Education recently announced a plan to merge thousands of small schools with fewer than 120 students each with other schools within a six-kilometer radius \autocite{saengpassa_ministrys_2017}. In Rajasthan, India in 2014, the government merged 17,000 of the over 80,000 government schools in the state with other schools, with more mergers planned \autocite[3]{chowdhury_cramped_2017}.} In the case of Brazil, official education statistics show that the number of rural primary schools dropped 31 percent between 2007 and 2017, from 88,386 rural primary schools to 60,694 \autocite{brazil_ministry_of_education_sinopses_2020}.

\subsection{Impact of Closures}
\label{evidence-about-impact}

Literature on the impact of school closures on affected students shows inconsistent results. Some studies suggest negative effects on performance and outcomes. \textcite{grau_school_2018} estimate that school closure increases the probability of high-school dropout between 49 and 68 percent (1.8 and 2.5 percentage points). The authors also identify large causal effects of school closure on grade repetition in primary school. In the United States, one study of the experiences of Latino and African American students in an urban high school in the year following the closure of their school showed declines in the transition cohort's academic performance after transferring to new schools \autocite{kirshner_tracing_2010}. A study in another urban school district showed that students displaced by school closures can experience adverse effects on test scores and attendance, but these effects can be minimized when students move to higher quality schools \autocite{engberg_closing_2012}. The same study showed that a negative effect on attendance for students displaced by school closures disappears after the first year in the new school. A study of school consolidations in Denmark from 2010 to 2011 showed that school consolidation had adverse effects on achievement in the short run, but effects appeared to weaken over time, suggesting that part of the effect was due to disruption \autocite{beuchert_short_term_2016}.

Other studies do not show negative effects. A study of closing poor performing primary schools in Amsterdam showed no negative impacts on student performance \autocite{de_witte_influence_2014}. Another study in the Netherlands indicated that consolidation reforms led to increased student achievement on a nationwide exit examination \autocite[818]{de_haan_school_2016}. One study of over 200 school closings in Michigan found, on average, no persistent detrimental effect on the achievement of displaced students, and that students displaced from relatively low-performing schools experienced achievement gains \autocite[108]{brummet_effect_2014}. An analysis of closure of charter schools in Ohio indicated that closing low-performing charter schools led to longer-term achievement gains of around 0.2 to 0.3 standard deviations in reading and math for students attending these schools at the time they were identified for closure \autocite[31]{carlson_charter_2016}.

\subsection{Hypothesized Mechanisms of Impact}
\label{hypothesized-mechanisms-of-impact}

The inconsistencies in observed impact described in the preceding section could stem from differences in context, or from which among the disparate mechanisms of impact of school closings on student outcomes dominates. By design, school closure typically implies three changes for students in affected communities: they experience disruption, they must attend schools farther away from home, and they attend schools that are larger and better-resourced than the schools that were shuttered. A literature on the disruptive effects of moving schools suggests negative effects from switching schools, but a literature on school and teacher quality suggests the possibility of improvement associated with moving from lower-performing to higher-performing schools \autocite[110]{sacerdote_when_2012}.

Under conditions of mobility associated with closure, students' emotional reactions to the change---anger or disenchantment at school closing, and experiences of stress in a new school and peer context---may impede student achievement and persistence \autocite[for example, see][]{kirshner_tracing_2010}. One study of student mobility in Texas unrelated to school closing indicated that while cross-district moves tended to be associated with improvements in school quality, within district moves did not, and were associated with short-run achievement costs \autocite{hanushek_disruption_2004}.  \textcite{sacerdote_when_2012} investigated the impact of displacement due to Hurricanes Katrina and Rita on long-term academic performance and college going for students in New Orleans. Analyses showed a short-term decline in academic performance, but long-term improvement, with gains concentrated among students initially in the lowest quintiles of the test score distribution. However, evacuees did not show gains in college-going relative to earlier cohorts from the same pre-hurricane high schools.

Research in a variety of contexts has indicated that the likelihood of attending a school declines as distance to the school increases, a finding possibly due to greater costs such as those involving transportation \autocite[28]{schwartz_small_2013}. Press reports have raised concerns about distance and student safety in consolidating Chicago Public Schools for students who will need to traverse city neighborhoods \autocite[for example, see][]{chicago_tribune_editorial_board_case_2017}. In developing countries, distance could also be an important determinant of school participation, particularly if long distances are involved or there are safety concerns \autocite{kremer_challenge_2013}. A multi-level analysis of survey data from 220,000 children in 340 districts of 30 developing countries estimated that parental decisions regarding children's enrollment were associated with distance from school, net of a host of other school, family and community characteristics \autocite{huisman_effects_2009}. One study in Afghanistan implemented a randomized trial to estimate the effects of establishing village-based schools on enrollment and test scores for a sample of 1,479 boys and girls aged six to eleven in 31 villages in Afghanistan \autocite{BurdeLinden_2013}. Results one year out showed significant enrollment effects, even more for girls than for boys, despite the non-significant observed correlation between distance to school and enrollment of children in the control group. Results also showed a sizeable achievement effect. These findings illustrate a potential gender difference in the implications of distance for enrollment opportunities. However, this difference in distance effect for girls and boys is not consistently found: one multi-national study found similar magnitudes of effect of distance on enrollment at ages 8 to 11 \autocite{huisman_effects_2009}.

The quality of the new school environment may be an important factor conditioning the impact of closure on student outcomes. In the United States, a study in one urban school district showed that adverse effects of moving schools on test scores and attendance were minimized when students moved to higher quality schools \autocite{engberg_closing_2012}. Emerging literature in developing country contexts suggests that children who attend better quality schools are more likely to remain enrolled \autocite{hanushek_students_2008}. One multinational study found that parental decisions regarding children's education were associated with quality-related characteristics of the available educational facilities such as number of teachers \autocite{huisman_effects_2009}.

One caveat is important to mention. While the larger schools students transfer into may provide more resources, research in the United States about the impact of attending larger, presumably better-resourced schools, is inconsistent \autocite{gershenson_effect_2015, schwartz_small_2013}. One study using the 1980 census to estimate the effects of changes in school size indicated that students born in states where average school size increased obtained lower returns to education and completed fewer years of schooling (relative to the national population) than did earlier cohorts born in the same state \autocite{berry_growing_2010}.

\subsection{School Closure in China}
\label{closure-china}

China's Compulsory Educational Law, promulgated in 1986, provided the legal foundation for nine years of compulsory education and established the principle that primary schools should be located in close proximity to rural children \autocite{ministry_of_education_compulsory_1986, dai_cost_2017}. A legacy of this principle was a widely distributed network of schools across the country \autocite{YangWang_2013}. Schools included both complete and ``incomplete'' (early grades) primary schools. However, demographic changes were already exerting pressures on provision of education at the village-level in the 1990s \autocite[124]{cai_has_2017}. \textcite{dai_cost_2017} report that consolidation experiments were piloted in some provinces in 1993.

National school consolidation policies commenced in 2001 \autocite{dai_cost_2017}. On May 29\textsuperscript{th}, 2001, the State Council issued a document entitled ``Decision on Basic Education Reform and Development'' \autocite{state_council_state_2001}. This document required local governments to make reasonable adjustments to schools' geographic distribution to improve efficiency.\footnote{Concurrently, two other national policies--national tax reform (which terminated an agricultural surtax) and compulsory school education policy adjustment reform (which emphasized the financial responsibility of county level government in providing compulsory education)--were issued that gave county level officials greater autonomy and also imposed more budgetary pressures on them \autocite{ding2015dismantling}.} As seen elsewhere, the case for school closures is made in terms of quality and efficiency considerations \autocites[see, for example,][]{fan_reasons_2013,liu_closures_2013, xie_consolidating_2013}.

The number of rural schools decreased from 512,993 in 1997 to 210,894 in 2010, while teaching points (incomplete primary schools) dropped from 186,962 in 1997 to 65,447 in 2010 \autocite{ChinaEduYearBook}. The number of students also decreased, from 95.6 million enrolled students in 1997 to 53.5 million in 2010 \autocite{ChinaEduYearBook}. However, the pace of school closures generally outstripped the pace of decline of students. \textcite{YangWang_2013} calculated an ``average closure intensity parameter'' as a ratio of the percent decline in number of schools and the percent decline in number of students during the same period to denote intensity of school consolidation in each province from 2000 to 2010. By this measure, 22 out of 27 provinces had average closure intensities greater than 1, and the highest reached 13.

Studies of county and provincial government policy documents have indicated that considerations about efficiency and economies of scale dominated decisions about school closures. As school consolidations rolled out across the nation, scholars and journalists raised concerns about the degree to which consolidation policies might be employed to avoid costs associated with compulsory school provision. \textcite{ding2015dismantling} analyze aggregate provincial educational expenditure data from 1996 and 2009 and find that provinces with a greater rate of school consolidation significantly reduced their financial expenditure share on primary education. In 2008, the National Development and Reform Commission issued standards prescribing that at least one primary school should be planned in each town \autocite[cited in][3]{dai_cost_2017}. In 2012, the Ministry of Education and then the General Office of the State Council issued documents calling for an end to consolidation \autocite{state_council_2012}, but persistent population decline in rural China continues to exert immense pressures toward further consolidation, and the number of schools continued to decrease after 2011.\footnote{The number of schools further decreased to 118,381 in 2015, and the number of students enrolled decreased further, to below 30 million in 2015 \autocite{ChinaEduYearBook}. Exceptionally, the number of teaching points bounced back to 81,818 in 2015, possibly reflecting efforts by the central government to counteract the consolidation policy.}

Scott Rozelle, Hongmei Yi, and their co-authors have studied the implications of school consolidation policy for student achievement in three adjacent provinces in the north to northwestern part of China: Shanxi, Sha'anxi and Ningxia \autocite{liu_effect_2010, mo_transfer_2012, chen_poor_2014}. Using data from ten counties in Sha'anxi Province and four in neighboring Ningxia Hui Autonomous Region, \textcite{liu_effect_2010} find that primary school closures between 2002 to 2006 did not negatively impact the academic performance of students in 2006, but the timing of mergers in students' lives mattered: higher-grade students' grades rose after merging, while grades of younger students fell. In three counties in Sha'anxi Province and one county in neighboring Shanxi Province, \textcite{mo_transfer_2012, chen_poor_2014} find that elementary school students' academic performance improved when they transferred from less centralized schools to more-centralized schools.\label{rr:citechen} However, as discussed in \textcite{chen_poor_2014}, the need to board at school at early ages may jeopardize the benefits of centralized schools. Parents may not wish to avail themselves of centralized schools if these schools are too far for daily commuting.

By design, achievement studies must focus on students who remain in school to take tests. A limitation of this approach is the lack of attention to dropout, continuation, or attainment. Non-continuation might be expected to be a crucial mechanism of impact of consolidation. To understand the full implications of consolidation, including implications for performance, attention must be paid to short and longer-term implications for educational continuation and attainment. Presumably, quality improvements in primary schools attended could increase the chances of school continuation. At the same time, concerns about safety of children associated with traveling long distances or boarding at schools could detract from continuation, and it is possible that safety concerns might be more pronounced for girls than for boys.

Separate concerns may relate to distance-associated cost burdens on rural families. Using Chinese Household Income Project data, \textcite{cai_has_2017} study the effects of the consolidation policy on 209 households and find that the compulsory school consolidation program increased educational expenditures, including expenditures on transportation and boarding due to greater distance to school. Using data from one county in Guangdong Province, \textcite{zhao_increasingly_2015} find that children from poorer families have difficulties paying for a bus or boarding at school and are more likely to endure longer commutes. It is possible that poor rural families would be more likely to shoulder costs for boys than girls: some research in China suggests that girls' educational attainment has been more susceptible than boys' to poverty \autocite{liu2017early}.

In summary, school consolidation has been a major policy initiative in China, but the implications are not yet well understood. In particular, existing studies of impact on students have focused on important questions of impact on short-term school performance, but have not considered the impact on school continuation or attainment. In addition, existing studies of impact on students have had limited geographic coverage, collectively and individually, and have not distinguished short- and long-term consequences. The current study begins to address these limitations by applying a difference in differences design to investigate short and long-term implications of school consolidation for educational attainment using data from 728 villages across seven provinces.

\section{Data\label{sec:data}}

This paper utilizes data from the rural sample of the China Household Ethnic Survey (CHES 2011), which covers households and villages from 728 villages in 81 counties of 7 provinces with substantial minority populations in China.\footnote{Appendix Section \ref{sec:aloc} presents a map of CHES Survey prefectures \autocite{Howell_2017} and discusses distribution of survey villages across provinces.} CHES 2011 sought to investigate the economic and social conditions of people in minority areas, and so utilized subsamples of the National Bureau of Statistics' Rural Household Survey (RHS) in seven provinces and autonomous regions with substantial minority populations. Household information by the end of 2011 was collected through diaries and single-round visits in early 2012. Routinely-collected RHS data and purpose-designed questionnaires for the CHES project were included in the data.

Village closure information is taken from a village head survey, which was collected in conjunction with household surveys. Village heads were asked if the village currently had a primary school, and asked about the year of school closure if the village school had been closed. Based on the village head survey, there are four categories of closure status. The first category includes 193 villages that did not have village schools in 2011 and experienced school closure between 1999 and 2010. In the second category, which included 22 villages, a school closure year between 1999 and 2010 was reported, but village heads also reported that the village currently had a school in 2011. In this case, it is plausible that new schools were built in these 22 villages after school closure.\footnote{Generally students went to schools in township centers after village school closure, but in these 22 villages, it is possible that a new consolidated school was built inside these villages.} In the third category, 430 villages had village schools in 2011 and did not experience school closure.\footnote{We do not have survey information on the opening year of the schools. The vast majority of these schools should have been established in the 1980s and early 1990s when the central government aimed to have a school in each village to provide education to rural children.} Finally, the fourth category includes 48 villages that had never had a primary school and 35 that do not currently have a school but had a village primary school at some point between 1954 and 1999. In the following analysis, we designate the first and second categories as school closure. The third and fourth categories are coded as non-closure.\footnote{Our main results tables also show specifications that test the robustness of regression results by dropping the second and fourth categories from the closure and nonclosure groups.}

There is heterogeneity in the timing of school closure. Out of the 193 villages in the first category mentioned above, 14 experienced school closure between 1999 and 2001, 28 between 2002 and 2004, 80 between 2005 and 2007, and 71 between 2008 and 2010. School closure took place in all seven provinces in all the year ranges listed. In addition, the intensity of school closure also varied across provinces. Among the villages in this dataset, approximately 49 percent of the surveyed villages from Hunan Province in south-central China and the Inner Mongolia Autonomous Region in north China reported village school closures between 1999 and 2010. Around 24 percent of surveyed villages from Ningxia Hui Autonomous Region and the Xinjiang Uygur Autonomous Regions in northwestern China, and Guizhou Province in southwestern China experienced village school closures between 1999 and 2010. The Guangxi Zhuang Autonomous Region in south-central China and Qinghai Province in northwest China had the lowest prevalence of closure. In these locations, around 18 percent of villages surveyed in 2011 by CHES reported having experienced closure between 1999 and 2010.

\subsection{Comparing Villages With and Without School Closure}
\label{sec:closecomp}

In Table \ref{summtwo}, we compare village-level statistics between villages with and without closure across several sets of variables. All variables are from the village-head survey component of the CHES data. Summary statistics are organized in Panels A through C. The first column shows the overall averages for all villages. The second and third columns show the mean values for villages with and without closure respectively. Column four presents the p-value from a significance test of whether the means differ between non-closure and closure villages.\footnote{We control for provincial fixed effects in these mean tests, but results are generally the same even if provincial fixed effects are not controlled for.} And column five tests, just for villages with closures, whether a linear trend exists for the variables across the year of closure (1999 to 2010).

\begin{table}[t]
\centering
\def\sym#1{\ifmmode^{#1}\else\(^{#1}\)\fi}
\caption{Summary Statistics for Village Characteristics\label{summtwo}}
\begin{adjustbox}{max width=1\textwidth}
\begin{tabular}{m{8cm} >{\centering\arraybackslash}m{1.5cm} >{\centering\arraybackslash}m{1.5cm} >{\centering\arraybackslash}m{1.5cm} >{\centering\arraybackslash}m{1.5cm} >{\centering\arraybackslash}m{1.5cm}}
\toprule
& \multicolumn{5}{L{8.5cm}}{Villages with and without school closures} \\
\cmidrule(l{5pt}r{5pt}){2-6}
& \multicolumn{1}{C{1.5cm}}{\small \textbf{all}} & \multicolumn{2}{C{3cm}}{\small \textbf{group averages}} & \multicolumn{2}{C{3cm}}{\small \textbf{p-values testing}} \\
\cmidrule(l{5pt}r{5pt}){2-2} \cmidrule(l{5pt}r{5pt}){3-4} \cmidrule(l{5pt}r{5pt}){5-6}
& \multicolumn{1}{C{1.5cm}}{\textit{\small mean}} & \multicolumn{1}{C{1.5cm}}{\textit{\small non-closure}} & \multicolumn{1}{C{1.5cm}}{\textit{\small closure}} & \multicolumn{1}{C{1.5cm}}{\textit{\small closure vs non-closure\dag}} & \multicolumn{1}{C{1.5cm}}{\textit{\small years of closure trend\ddag}} \\
\midrule
\formatpanelheadsumm{6}{L{18cm}}{Panel A: Closest Primary School Physical Facility Measures}
Fraction with non-dilapidated buildings&    0.82&    0.79&    0.90&    0.00&    0.40\\
Fraction with heating&    0.25&    0.18&    0.44&    0.00&    0.38\\
Fraction with tap water&    0.81&    0.80&    0.84&    0.19&    0.22\\
Fraction with kitchen&    0.70&    0.66&    0.81&    0.01&    0.15\\
Fraction with shower&    0.19&    0.15&    0.32&    0.00&    0.34\\
Fraction with sufficient desks&    0.94&    0.93&    0.96&    0.23&    0.27\\
Fraction with library&    0.76&    0.72&    0.85&    0.00&    0.49\\
Fraction with personal computers&    0.53&    0.47&    0.68&    0.00&    0.73\\
Fraction with internet access&    0.46&    0.41&    0.58&    0.00&    0.27\\
\midrule
\formatpanelheadsumm{6}{L{18cm}}{Panel B: Distance to Closest Primary School (km)}
Distance measure from village head survey&    2.87&    1.80&    5.67&    0.00&    0.15\\
\midrule
\formatpanelheadsumm{6}{L{18cm}}{Panel C: Village Size and Demographics}
Per household arable land (mu, 1 acre = 6 mu) & 11.59 & 10.15 & 15.09&    0.00&    0.58\\
Number of households&      453.55&      469.35&      415.04&    0.02&    0.62\\
Fraction of Han in village&    0.40&    0.36&    0.48&    0.02&    0.31\\
Fraction of women (all ages)&    0.48&    0.48&    0.47&    0.08&    0.15\\
Fraction of 30 and younger (2011)&    0.46&    0.48&    0.42&    0.00&    0.23\\
Fraction of women 30 and younger (2011)&    0.46&    0.46&    0.46&    0.37&    0.13\\
\midrule
\bottomrule
\footnotegap
\multicolumn{6}{L{18cm}}{\footnotesize\justify\footsummstatsabc}\\
\end{tabular}\end{adjustbox}
\end{table}

In Panels A and B of Table \ref{summtwo}, we compare distance to school and school facility measures. In both panels, village heads reported information for the complete primary school\footnote{Complete primary school is defined as a school that includes all the grades in primary school, i.e., grade 1 to grade 6.} that was closest to the village in 2011. For villages with school closure, the statistics reported are for the ``replacement'' school that village children attend, away from their own village.

Panel A compares nine kinds of school facilities (non-dilapidated buildings, heating, tap water, kitchen, shower, sufficient desks, library, personal computers, and internet access) for the closest primary school between closure villages and non-closure villages. The current schools for closure villages are more likely to have each of the nine kinds of physical facilities. In terms of technology, in current schools for closure villages, 68 percent of schools have computers and 58 percent have internet access. These fractions are 47 percent and 41 percent for schools for non-closure villages. Regarding other kinds of facilities, in schools serving closure villages, 44 percent have heating, 84 percent have tap water, 81 percent have kitchens, and 32 percent have showers. Corresponding figures are 18 percent, 80 percent, 66 percent, and 15 percent in schools serving non-closure villages.\footnote{We do not have information on school facilities in place before closure in closure villages. If these schools had facility measures similar to or worse than those of village-schools in nonclosure villages, then these data indicate that school closure might have brought about a significant improvement in school facility quality. This change would be consistent with the stated goal of the policy to improve quality through consolidation. We do not have information on teachers’ characteristics in these school.} Column 4 shows that the provision of school facilities tends to differ in closure and non-closure villages: 7 out of 9 facilities measures have p-values close to 0. Moreover, for villages with closure, there is no linear trend in quality with the calendar year of school closure (as indicated by larger p values in column 5), which means that we do not find replacement schools' facilities to be systematically better for villages that closed schools more recently than those with earlier closures.

Panel B shows that the distance to school is significantly greater for villages with school closure than without. The average distance is 5.67 kilometers for villages with school closure,\footnote{The 25th percentile, median and 75th percentile in the distribution of distance to school are, respectively, 2 kilometer, 3.5 kilometer, and 8 kilometer.} compared to 1.80 kilometers for the latter. 65 percent of villages without closure report a 0 kilometer distance to the closest school.\footnote{Distance to school is not 0 for all villages without school closure. Households are located in various parts of a village, and the questionnaire did not specify if distance to school should be from the village center or from a village boundary. Some of the non-zero values possibly reflect the vantage point of the village head. It is also possible that a primary school exists in the village, but it is a teaching point rather than a full primary school with all 6 grades, and village heads reported distance to a full primary school further away.} In short, these findings suggest that schools serving closure villages, compared to non-closure villages, are better-resourced and more distant.

In Panels C of Table \ref{summtwo}, we show that villages without school closure have, on average, 469 households, while villages with school closure have, on average, 415 households. Households in closure villages also have significantly more arable land per person and are more likely to be classified as non-minority (ethnic Han).\footnote{In villages with closure, per household arable land is about 15.09 mu (1 acre=6 mu), in villages without closure, per household arable land size is only 10.15 mu. In non-closure villages, non-Han ethnic groups account for on average 64 percent of the village populations. In closure villages, non-Han ethnic groups account for 52 percent of the village population. These means are significantly different, but there are no mean trends in these variables for closure villages across closure years.}

In Panels A, B, C and D of Table \ref{summtwob} in the Appendix, we test how villages with and without school closure differ along several other dimensions, in terms of political connectedness (Panel A), income and labor market participation (Panel B), village expenditures (Panel C) and participation in other national policy schemes such as the ``Grain for Green'' reforestation initiative, collectively-owned medical station initiatives, and the rural medical insurance scheme (Panel D). Controlling for provincial fixed effects, we generally do not find statistical differences between closure and non-closure villages along these dimensions.

\subsection{Closure Year, Children and Attainment}
\label{sec:attainment}

Given the cross-sectional data structure, there are two different dimensions of time: the year when a child was born and the year of school closure. Since children live in different villages with different dates of school closure, we are able to group children based on these two time dimensions: i.e. their ages in the year of school closure (calculated as time elapsed from the year of birth to the year of closure) and their ages in 2011 (calculated as time elapsed from the year of birth to 2011). To distinguish the different impacts of school closure on children at different ages, we divide the children in villages with school closure into 6 groups according to the ages in year of school closure: Group A---preschool period (age 0-5)\footnote{The 0 age group includes children borne after school closure in villages with closure.}, Group B---lower level of primary school (age 6-9), Group C---higher level of primary school (age 10-13), Group D---middle school or beyond (age 14-21), and Group E---an extra control group for parallel trend tests (age 22-29). We simply group age in 2011 by 5--year interval. Table \ref{summthree} presents the distribution of our data along these two dimensions (with rows showing the age at year of school closure and columns showing the age in 2011). In each cell, the top number shows average educational attainment for each group, and the bottom number shows the sample size for each group. This table shows the complete sample, while the cells in the box circled with dashed lines constitute the partial sample that we use in our regressions for a robustness check.

\begin{table}[t!]
\centering
\def\sym#1{\ifmmode^{#1}\else\(^{#1}\)\fi}
\caption{Exposure Groups and Grades Completed by 2011\label{summthree}}
\begin{adjustbox}{max width=1\textwidth}
\tikzageedusmall
\begin{tabular}{m{4cm} >{\centering\arraybackslash}m{1.5cm} >{\centering\arraybackslash}m{1.5cm} >{\centering\arraybackslash}m{1.5cm} >{\centering\arraybackslash}m{1.5cm} >{\centering\arraybackslash}m{1.5cm} >{\centering\arraybackslash}m{1.5cm} >{\centering\arraybackslash}m{1.5cm} >{\centering\arraybackslash}m{1.5cm}}
\toprule
& \multicolumn{8}{L{\innerheadwidthsummeight}}{Age at village-specific year of closure and 2011 age}\\
\cmidrule(l{5pt}r{5pt}){2-9} & \multicolumn{2}{C{3.0cm}}{\small\textit{Age in 2011}} & \multicolumn{2}{C{3.0cm}}{\small\textit{Age in 2011}} & \multicolumn{2}{C{3.0cm}}{\small \textit{Age in 2011}} & \multicolumn{2}{C{3.0cm}}{\small\textit{Age in 2011}} \\
\cmidrule(l{5pt}r{5pt}){2-3} \cmidrule(l{5pt}r{5pt}){4-5} \cmidrule(l{5pt}r{5pt}){6-7} \cmidrule(l{5pt}r{5pt}){8-9}
& \multicolumn{1}{C{1.5cm}}{\textbf{\small 0-4}} & \multicolumn{1}{C{1.5cm}}{\textbf{\small 5-9}} & \multicolumn{1}{C{1.5cm}}{\textbf{\small 10-14}} & \multicolumn{1}{C{1.5cm}}{\textbf{\small 15-19}} & \multicolumn{1}{C{1.5cm}}{\textbf{\small 20-24}} & \multicolumn{1}{C{1.5cm}}{\textbf{\small 25-29}} & \multicolumn{1}{C{1.5cm}}{\textbf{\small 30-34}} & \multicolumn{1}{C{1.5cm}}{\textbf{\small 35-44}} \\
\midrule
\formatpanelheadsumm{9}{L{17.2cm}}{Group A: Age 1 to 5 at Year of Closure}
Mean grades completed  & 0.08&    1.11&    5.18&    9.69&   &   &   &    \\
Observations   &      303&     333&     126&      16&  &  &   & \\
\midrule
\formatpanelheadsumm{9}{L{17.2cm}}{Group B: Age 6 to 9 at Year of Closure}
Mean grades completed &   &    1.98&    5.51&    9.80&    10.7&   &   &    \\
Observations   &   &      98&     211&      69&      17&  &   & \\
\midrule
\formatpanelheadsumm{9}{L{17.2cm}}{Group C: Age 10 to 13 at Year of Closure}
Mean grades completed &   &   &    5.98&    9.40&    10.4&    8.50&   &    \\
Observations   &   &  &     133&     224&     117 & 2&   & \\
\midrule
\formatpanelheadsumm{9}{L{17.2cm}}{Group D: Age 14 to 21 at Year of Closure}
Mean grades completed &   &&  7&    9.80&    10.2&    9.05&    8.59&
\\
Observations   &   &  &      16&     276&     592&     241&      32&\\
\midrule
\formatpanelheadsumm{9}{L{17.2cm}}{Group E: Age 22 to 29 at Year of Closure}
Mean grades completed &   &   &   &   &    9.98&    8.81&    8.14&    7.99\\
Observations   &   &  &  &  &     101&     460&     322&      98\\
\midrule
\formatpanelheadsumm{9}{L{17.2cm}}{Group F: Non-closure Villages Individuals}
Mean grades completed  & 0.09&    1.37&    5.55&    9.33&    9.71&    8.44&    7.37&    6.81\\
Observations   &      783&    1227&    1521&    1774&    2237&    1713&    1420&    1569\\
\bottomrule
\footnotegap
\multicolumn{9}{L{19.75cm}}{\footnotesize\justify \footeduattain} \\
\end{tabular}
\end{adjustbox}
\end{table}

The extent to which a given child is affected by the school closure policy depends on his or her age at the time of school closure and the duration of exposure (calculated as the difference between the age in 2011 and the age at time of school closure). First, the individuals in row groups A, B, and C of Table \ref{summthree} could have been directly affected by the primary school closure policy. Those in row groups D and E were between age 14 to 21 and 22 to 29 in the year of closure and should not have been directly impacted (as students usually enter secondary school at age 14). Individuals in row group F are not exposed to the school closure policy. Second, for the column groups, individuals in the second to fourth column are between age 5 and 19 in 2011 and are mostly still attending school. Individuals in columns 5 through 8 are between age 20 and 44 in 2011, and their 2011 educational attainment generally reflects their final attainment. Third, the group of individuals in each row and column cell were exposed to school closure at different starting ages, and experienced different durations of impact when we observe their educational attainment (number of grades completed) in 2011. To be specific, row group B shows that among individuals who were between 6 to 9 years of age when the village school was closed, 98 were between 5 to 9 years of age, 211 were between 10 to 14, 69 were between 15 to 19, and 17 were between 20 to 24 in the year 2011.

In Table \ref{summthree}, we also compare educational attainment within each age-in-year-2011 group.\footnote{In Appendix Section \ref{sec:agrpedu}, we also compare educational attainment and the proportion of individuals who complete middle schools by gender within each age-in-year-2011 group.} The educational attainment variable is based on years of schooling completed by each individual.\footnote{The survey asks individuals to report years of schooling completed. For example, 4 is recorded for someone who has completed 4th grade in primary school, 9 is recorded for someone who has completed middle school, and 10 is recorded for someone who has completed one year of high school.} By simple comparisons of the mean, we find among those who are too old to be affected by school closure policy (i.e., groups D and E) from closure villages, their average educational attainments are uniformly higher than their counterparts in non-closure villages (group F), no matter what age group in 2011. However, for those who were at school ages at the time of school closure (groups B and C), their average educational attainment is not consistently higher or lower compared to their counterparts, at different ages in 2011. In the following sections, we proceed to study the causal impact of school consolidation policy on children's educational attainment in a regression framework.

\section{Methods\label{sec:method}}
\subsection{Identification Strategy}
\label{sec:identify}

We follow a difference in differences strategy to identify the effects of the school consolidation policy. Within each province, we first compare the difference in educational attainment (number of grades completed by 2011) between those from closure villages who were exposed to closure to individuals of the same cohorts from non--closure villages. This first difference could be due to the school consolidation policy or could be due to existing differences in educational attainment across villages that would have occurred without the policy. Hence, we also compare the difference in educational attainment between individuals not impacted by the consolidation policy from closure villages and individuals of the same cohorts from non-closure villages. This second difference should not be related to school consolidation but measures only existing differences in educational attainment across villages. We allow this second difference to be village-specific. We interpret the difference in the first and second differences as the impact of the policy. We estimate the effects of the policy first over subgroups based on age at year of closure (age effects), and then subgroups based both on age at year of closure and the number of years since closure. We interpret results for years since closure as short-, medium- and long-run impacts of the policy on educational progression conditional on age at year of closure (duration effects).\footnote{Given the cross-sectional data structure, we cannot identify cohort effects separately from the age effects. Note that we distinguish between cohort (age in 2011), and age at year of closure for children exposed to closure in years prior to 2011.}

Our age effects estimation strategy follows \textcite{duflo_2001_school}. Similar to \textcite{duflo_2001_school}, we exploit the fact that individuals were at different ages in the village-specific years of closure to distinguish between individuals who could be impacted and who should not be impacted by the school closure policy. Specifically, the policy could impact both children who were attending elementary school at the time of school closure (Group B and C in Table \ref{summthree}), and children who were yet to attend elementary school (Group A in Table \ref{summthree}). Those who were above elementary school age (age 13) at the time of the school consolidation program are less likely to be affected by this policy\label{rr:roneidentifyone}.\footnote{\label{rr:roneidentifyonefoot}We provide school enrollment age patterns by different subgroups in Appendix Section \ref{sec:enrollage}. Our age-at-closure groupings are based on these analyses.} Specifically, we classify individuals from villages that have experienced closure into cohorts who are fully exposed (0 to 5 years old at the time of primary school closure) or partially exposed (6 to 13 years old at the time of primary school closure) to school closure and the older cohorts who are not exposed (between 14 and 21 at the time of school closure).\label{rr:roneqtwoidentify} Individuals from villages without closure do not have age at closure (Group F from Table \ref{summthree}). Age in 2011 identifies the cohorts that individuals from all villages belong to.

Besides the age effects, we further estimate the effects of duration of exposure to school closure for each age group. Specifically, the age at which school closure takes place determines the \emph{age effect} of the policy--which corresponds to the potential number of years that a child has attended a consolidated primary school. The duration between the survey year, 2011, and the year in which village specific closures took place determines the \emph{duration effect} of the policy--which corresponds to the number of years since village school closes. Under the assumption that closure effects do not vary by year-at-closure, we are able to estimate the age and duration effects of the policy using a cross-sectional dataset because of the variation in the year that village schools were closed. Closures in earlier years provide us with longer-duration effects of the policy on final attainment, and more recent closures provide us with shorter-duration effects of the policy on the number of grades completed for children who are still attending school. The effect of the policy will be conditional on both the starting age at exposure and the length of exposure for each individual. As discussed previously, Table \ref{summthree} shows the distribution of children with different starting ages of exposure and lengths of exposure (by subtracting age at exposure from age in 2011) in the sample.

\label{rr:aetwoidentify}Our conceptual framework is close to \textcite{KingBehrman2009WBRO, BehrmanParkerTodd2011}, which emphasize the importance of distinguishing the effects of educational policy along two dimensions: exposure differential (our age effects) and time since program initiation (our duration effects).\footnote{They study the effects of the Progresa conditional cash transfer program in Mexico on educational attainment (grades completed). Specifically, for the exposure-differential/age-effects, they study program effects for children who were never exposed to the program and children who were 9--10, 11--12, and 13--15 when the policy started. For the time-since-initiation/duration-effects, they study the effects of the program 1.5 years after policy initiation and then 6 years later when children were 15--16, 17--18, and 19--21. \textcite{BehrmanParkerTodd2011} exploit the experimental variations in Progresa to analyze the 1.5 year time-since-initiation effects, but rely on non-experimental data to evaluate the 6 year time-since-initiation policy effects.} As time since program initiation increases, possible initial short run differences due to exposure differential might be magnified or fade. In this sense, our paper differs from most education policy impact evaluations that focus only on the short-run effects of policy.\footnote{As discussed in \textcite{BehrmanParkerTodd2011}, due to data limitation, many educational policy impact evaluations focus on the effects of policy on educational outcomes within 1 or 2 years after policy initiation and when children are in primary and lower middle schools.} Our methodology also differs from \textcite{duflo_2001_school}, which focuses on final educational attainment using the data collected 20 years after the start of school expansion.

Overall, the primary underlying assumption of our strategy is that in the absence of the school closure policy, the change in educational attainment at a certain age of younger cohorts relative to older cohorts would not have been systematically different in treatment (closure) and control (non-closure) villages within the same province, i.e., educational attainment in both groups of villages follows a provincial common trend. Given that the closure policy took place at the village level, we have the advantage of including village-specific fixed effects to account for unobservable differences in educational attainment across villages that are common across cohorts considered. \label{rr:aethreeintro}Controlling for village fixed effects can account for any time-invariant determinants that contribute to the closure of village school, which may also affect children's educational attainment \autocite{duflo_2001_school}. Additionally, the large number of villages experiencing closure and the large number of control villages means that the difference in differences estimates are averaged over many groups that experience policy changes. This means that our estimates are more likely to be robust to the presence of age- and location- specific random effects \autocite{Conley_Taber_2011}.

Our common trend assumption, however, would be violated if county administrators choose to close the schools in villages where educational attainment is already deviating from the provincial trend. In other words, we may suffer from endogeneity bias if the closed schools were getting worse over time relative to the average time trend. In order to test whether the identification assumption is valid, we compare the differences in educational attainment between the cohort groups of age 14 to 21 at the time of exposure, and those of age 22 to 30. If our ``common trend'' assumption holds, the difference of educational attainment between these two groups should not be significantly different once we control for village and province-cohort (age in 2011) fixed effects.

In addition, our estimation may raise concerns about sample selection bias if school closure had led to increased out migration either by individual laborers or by families \autocite{liu_migrate_2016}. The first of these possibilities---increased labor migration caused by school closure---is less of a concern for our estimates, as the survey gathered data on educational attainment for all members of households including those working as migrant workers. For the second possible migration issue, however, we have not seen evidence supporting the proposition that entire households out-migrate as a result of school closure. An additional concern related to sample selectivity could emerge if a child's birth location is endogenous with respect to school consolidation policy. That is to say, if parents chose where to live before children are born with consideration to the quality of schools, which may be related with the risk of school closure, sample selection would be a concern. However, in the rural China context, mandatory household registration and land allocation policies do not allow people to choose their registered household location freely.

\subsection{Regression Model with Only Age Effects}
\label{sec:identifyageeffect}

In Equation \eqref{eq:startsOnly}, we generalize our estimation strategy first to a regression framework in which we assume that the policy has an immediate and constant impact on educational attainment. In the scenario without policy intervention, educational attainment, or the number of grades completed, \(E\) of a child \(i\) from village \(v\) in province \(p\) and whose cohort (age in 2011) is \(a\) could be decomposed into four parts: a village fixed effect \(\beta_{v}\), a province-specific cohort fixed effect \(\rho_{pa}\), and idiosyncratic terms including one part that can be explained by observed characteristic \(X_{i}\) and another unobserved error term \(\epsilon_{i}\). With school closure, the policy's effect is assumed to be additive and captured by \(\tilde{\lambda_{z}}\) that is constant within age group \(z\) defined by age-at-closure \(t_{i}\):\footnote{\(t_{i}\) is equal to zero for children borne in villages with school closure after closure had taken place.}

\begin{singlespace}\vspace*{-\baselineskip}
\begin{eqnarray}
E_{pvia} & = & \phi+\beta_{v} + \rho_{pa} \nonumber \\
& & + \sum_{z=1}^Z \tilde{\lambda_{z}}\cdot \bm{1}\left\{ l_{z}\leq t_{i}\leq u_{z}\right\} \cdot c_{v}\label{eq:startsOnly}\\
& & +X_i\cdot\gamma\nonumber +\epsilon_{i}\nonumber
\end{eqnarray}
\end{singlespace}\noindent\ignorespaces
In Equation \eqref{eq:startsOnly}, \(\phi\) is a constant, and \(c_{v}\) is a binary variable indicating if individual \(i\) is from a village \(v\) with school consolidation (i.e. treatment village). We group children in villages with school closure into \(Z\) groups based on their age at closure \(t_{i}\), with lower and upper bounds for each group \(l_{z}\) and \(u_{z}\). Therefore \(\tilde{\lambda_{z}}\) captures the average treatment effect for age group \(z\). \label{rr:roneqtwoageeffects}For individuals from villages without closure, \(c_{v}=0\). Including individuals from villages without closure helps us to isolate the policy effects from the provincial specific cohort (age in 2011) patterns captured by \(\rho_{pa}\).

This specification imposes three key assumptions. First, \(\tilde{\lambda_{z}}\) is not specific to 2011 age \(a\), which means \(\tilde{\lambda_{z}}\) captures the average effect across the children who were in the same age group when school was closed but maybe in different cohorts (ages in 2011). We relax this assumption in Equation \eqref{eq:startsLengths} which has both age and duration effects. Second, \(\tilde{\lambda_{z}}\) is not specific to the calendar year in which the closure policy was implemented. \label{rr:rtwoqoneageeffect}Third, the \(l_{z}\) and \(u_{z}\) age cut-offs are common for all \(i\).\footnote{This assumes that the enrollment patterns for key demographic sub-groups are similar. To verify that assumption, we analyze the enrollment patterns for boys, girls, minorities and Han individuals and show that they are broadly similar to warrant the same age-at-closure cut-off rules in Appendix Section \ref{sec:enrollage}.}

\subsection{Regression Model with Age and Duration Effects}
\label{sec:methodageduraeffects}

Besides the age at school closure, the impact of school closure on educational attainment may also differ by the number of years of exposure to the policy: short-run effects of closure on a child's educational attainment progression could be dampened or amplified over the medium and long run.\footnote{After an individual completes schooling, duration effects will become constant. In studies with cross-sectional data taken long after a policy has been implemented, \textcite{duflo_2001_school} for example, the duration effect is irrelevant because all educational attainment data is observed long after sample individuals have completed schooling. In our data, a significant proportion of individuals have not completed schooling, allowing us to have meaningful duration effects.} In order to identify both age and duration effects with our cross-sectional data, we exploit the variation in the year of school closure. Under the assumption that the impact of the policy is not specific to the calendar year of closure as well as assumptions for Equation \eqref{eq:startsOnly} stated previously, we can estimate Equation \eqref{eq:startsLengths} to obtain the impact of the policy as a function of both starting age and the length of exposure.

In Equation \eqref{eq:startsLengths}, we use similar notations as in Equation \eqref{eq:startsOnly}, the difference is that the policy's effects are now captured by \(\hat{\lambda_{zl}}\) that varies by age-at-closure variable \(t_{i}\) and years-of-exposure variable \(\tau_{i}\):

\begin{singlespace}\vspace*{-\baselineskip}
\begin{eqnarray}
E_{pvia} & = & \phi+\beta_{v}+\rho_{pa} \nonumber \\
& & + \sum_{z=1}^Z \sum_{l=1}^L \left(
        \hat{\lambda_{zl}} \cdot \bm{1}\left\{ (l_{l}\leq\tau_{i}\leq u_{l} ) \cap  ( l_{z}\leq t_{i}\leq u_{z} ) \right\}
        \right)
        \cdot c_{v} \label{eq:startsLengths}\\
& & +X_i\cdot\gamma\nonumber +\epsilon_{i}\nonumber
\end{eqnarray}
\end{singlespace}\noindent\ignorespaces
In Equation \eqref{eq:startsLengths}, as before, \(c_{v}\) is a binary variable indicating if individual \(i\) is from a village \(v\) with school consolidation (i.e. treatment village). As in Equation \eqref{eq:startsOnly}, we group children in villages with school closure into \(Z\) groups based on their age at closure, with lower and upper bounds for each group, \(l_{z}\) and \(u_{z}\). To capture duration effects, we further divide each of the Z groups of children into \(L\) groups based on the length of exposure \(\tau_i\), defined as the gap between individual \(i\)'s age in 2011 and \(i\)'s age at year of school closure, \(t_{i}\).\footnote{\(\tau_i = \min(a_i, a_i - t_i)\): \(\tau_i\) is the gap between age in 2011 and \(t_{i}\) if individual \(i\) was borne before the year of closure, and it is the age of the child in 2011 if the child was borne after school closure.} Each \(l\) length of exposure group includes those with \(\tau_i\) falling within lower and upper bounds, \(l_{l}\) and \(u_{l}\). The exposure groups allow us to separately estimate the short, medium and long run effects of the consolidation policy on educational attainment. There are \(Z\cdot L\) groups of interest for this regression.\footnote{Ideally, we would estimate the policy effects for each \(t_i\) and \(\tau_i\) combination separately, but we have constructed the \(Z\) and \(L\) groups due to limited sample size.}

\section{Results}
\subsection{Age Effects Only Results}
\label{sec:ageeffects}

Table \ref{regone} presents estimates of \(\lambda_{z}\) in Equation \eqref{eq:startsOnly}. The first panel presents overall results, while the sex-specific results---based on regressions including only individuals from one gender---are shown in Panels B and C. In each panel, we compare three subsets of children below age 14 against baseline group---those between 14 and 21 at the time of school closure. Columns 1 and 2 include all individuals between 1 and 44 years of age in 2011, columns 3 and 4 restrict to individuals between 10 and 34, and columns 5 and 6 restrict further to individuals between 15 and 34 years of age. The even-numbered columns drop villages that never had a school from the villages without closure group (category 4 as defined in data section). All regressions include several individual and household controls.\footnote{All regressions include controls for households size, a dummy for if the individual is Han and a categorical variable for the relative wealth. The relative wealth variable is based on the survey question that asked households if they are better or worse off than village average. We do not have income measures for all families. The village income per capita variable shown in summary Table \ref{summtwo} is from the village-head survey and not based on household incomes.} All standard errors are clustered at the village-level. Column 1 contains our focal main result, other columns contain results for robustness checks which we discuss later.

\begin{table}[t!]
\centering
\def\sym#1{\ifmmode^{#1}\else\(^{#1}\)\fi}
\caption{Effect of School Closure on Educational Attainment\label{regone}}
\begin{adjustbox}{max width=\regone\textwidth}
\begin{tabular}{m{\lablcolwidth} >{\centering\arraybackslash}m{1.5cm} >{\centering\arraybackslash}m{1.5cm} >{\centering\arraybackslash}m{1.5cm} >{\centering\arraybackslash}m{1.5cm} >{\centering\arraybackslash}m{1.5cm} >{\centering\arraybackslash}m{1.5cm}}
\toprule
& \multicolumn{6}{L{\innerheadwidth}}{Outcome: grades completed by year 2011} \\
\cmidrule(l{5pt}r{5pt}){2-7}
& \multicolumn{2}{C{3.0cm}}{\small } & \multicolumn{2}{C{3.0cm}}{\footnotesize \(10 \le 2011 \text{ Age} \le 34\)} & \multicolumn{2}{C{3.0cm}}{\footnotesize \(15 \le 2011 \text{ Age} \le 34\)} \\
\cmidrule(l{5pt}r{5pt}){2-3}\cmidrule(l{5pt}r{5pt}){4-5}\cmidrule(l{5pt}r{5pt}){6-7}
& \multicolumn{1}{C{1.5cm}}{(1)} & \multicolumn{1}{C{1.5cm}}{(2)} & \multicolumn{1}{C{1.5cm}}{(3)} & \multicolumn{1}{C{1.5cm}}{(4)} & \multicolumn{1}{C{1.5cm}}{(5)} & \multicolumn{1}{C{1.5cm}}{(6)} \\
\midrule
\formatpanelheadsupregs{7}{L{15cm}}{\baselinegroup}
\formatpanelheadregs{7}{L{8cm}}{Panel A: Regression for Females and Males}
\closeinteragestart{} 0--5 & -0.24 & -0.30 &   &   &   &    \\
& \vspace*{-2mm}{\footnotesize (0.17) } &\vspace*{-2mm}{\footnotesize (0.19) } &   &   &   &    \\
\closeinteragestart{} 6--9 & -0.29\sym{*}  & -0.38\sym{**} & -0.39\sym{**} & -0.54\sym{***}&   &    \\
& \vspace*{-2mm}{\footnotesize (0.16) } &\vspace*{-2mm}{\footnotesize (0.18) } &\vspace*{-2mm}{\footnotesize (0.18) } &\vspace*{-2mm}{\footnotesize (0.20) } &   &    \\
\closeinteragestart{} 10--13 & -0.42\sym{***} & -0.47\sym{***} & -0.46\sym{***} & -0.53\sym{***} & -0.49\sym{***} & -0.55\sym{***}\\
& \vspace*{-2mm}{\footnotesize (0.14) } &\vspace*{-2mm}{\footnotesize (0.16) } &\vspace*{-2mm}{\footnotesize (0.14) } &\vspace*{-2mm}{\footnotesize (0.15) } &\vspace*{-2mm}{\footnotesize (0.15) } &\vspace*{-2mm}{\footnotesize (0.16) }    \\
\closeinteragestart{} 22--29&    0.11 &    0.18 & 0.026 &    0.12 &     -0.0077 & 0.079    \\
& \vspace*{-2mm}{\footnotesize (0.17) } &\vspace*{-2mm}{\footnotesize (0.18) } &\vspace*{-2mm}{\footnotesize (0.17) } &\vspace*{-2mm}{\footnotesize (0.19) } &\vspace*{-2mm}{\footnotesize (0.17) } &\vspace*{-2mm}{\footnotesize (0.19) }    \\
\midrule
Observations   & 18804 & 15918 & 12072 & 10289 &    9998 &    8538    \\
\midrule
\formatpanelheadregs{7}{L{8cm}}{Panel B: \newtablepanfemale}
\closeinteragestart{} 0--5 & -0.43\sym{*}  & -0.61\sym{**} &   &   &   &    \\
& \vspace*{-2mm}{\footnotesize (0.23) } &\vspace*{-2mm}{\footnotesize (0.25) } &   &   &   &    \\
\closeinteragestart{} 6--9 & -0.49\sym{**} & -0.65\sym{***} & -0.58\sym{**} & -0.78\sym{***}&   &    \\
& \vspace*{-2mm}{\footnotesize (0.22) } &\vspace*{-2mm}{\footnotesize (0.25) } &\vspace*{-2mm}{\footnotesize (0.26) } &\vspace*{-2mm}{\footnotesize (0.29) } &   &    \\
\closeinteragestart{} 10--13 & -0.60\sym{***} & -0.69\sym{***} & -0.60\sym{**} & -0.69\sym{***} & -0.60\sym{**} & -0.65\sym{**} \\
& \vspace*{-2mm}{\footnotesize (0.23) } &\vspace*{-2mm}{\footnotesize (0.25) } &\vspace*{-2mm}{\footnotesize (0.23) } &\vspace*{-2mm}{\footnotesize (0.26) } &\vspace*{-2mm}{\footnotesize (0.28) } &\vspace*{-2mm}{\footnotesize (0.30) }    \\
\closeinteragestart{} 22--29&    0.19 &    0.27 & 0.051 &    0.12 & 0.067 &    0.11    \\
& \vspace*{-2mm}{\footnotesize (0.23) } &\vspace*{-2mm}{\footnotesize (0.23) } &\vspace*{-2mm}{\footnotesize (0.25) } &\vspace*{-2mm}{\footnotesize (0.25) } &\vspace*{-2mm}{\footnotesize (0.26) } &\vspace*{-2mm}{\footnotesize (0.26) }    \\
\midrule
Observations  &     8869 &    7466 &    5664 &    4790 &    4658 &    3946    \\
\midrule
\formatpanelheadregs{7}{L{8cm}}{Panel C: \newtablepanmale}
\closeinteragestart{} 0--5&      -0.067 &      0.0034 &   &   &   &    \\
& \vspace*{-2mm}{\footnotesize (0.20) } &\vspace*{-2mm}{\footnotesize (0.22) } &   &   &   &    \\
\closeinteragestart{} 6--9&      -0.042 &      -0.060 &      -0.096 & -0.20 &   &    \\
& \vspace*{-2mm}{\footnotesize (0.20) } &\vspace*{-2mm}{\footnotesize (0.22) } &\vspace*{-2mm}{\footnotesize (0.23) } &\vspace*{-2mm}{\footnotesize (0.25) } &   &    \\
\closeinteragestart{} 10--13 & -0.24 & -0.21 & -0.27 & -0.28 & -0.33 & -0.34    \\
& \vspace*{-2mm}{\footnotesize (0.19) } &\vspace*{-2mm}{\footnotesize (0.20) } &\vspace*{-2mm}{\footnotesize (0.19) } &\vspace*{-2mm}{\footnotesize (0.20) } &\vspace*{-2mm}{\footnotesize (0.22) } &\vspace*{-2mm}{\footnotesize (0.23) }    \\
\closeinteragestart{} 22--29&    0.15 &    0.20 &    0.14 &    0.28 & 0.051 &    0.18    \\
& \vspace*{-2mm}{\footnotesize (0.22) } &\vspace*{-2mm}{\footnotesize (0.24) } &\vspace*{-2mm}{\footnotesize (0.24) } &\vspace*{-2mm}{\footnotesize (0.27) } &\vspace*{-2mm}{\footnotesize (0.24) } &\vspace*{-2mm}{\footnotesize (0.27) }    \\
\midrule
Observations&    9935 &    8452 &    6408 &    5499 &    5340 &    4592    \\
\midrule
\exclcontrol
\exclcontrolcont
\bottomrule
\footnotegap
\multicolumn{7}{L{\footwidth}}{\footnotesize \justify \footattain} \\
\end{tabular}
\end{adjustbox}
\end{table}

The estimates in the Table \ref{regone} show that the school consolidation policy had a clear negative impact on educational attainment in terms of grades completed by 2011, but only for girls. Panel A of the first column shows that the policy decreased the educational attainment for children who were below age 6, between age 6 and 9, and between age 10 and 13 in the year of closure by 0.24 (s.e. 0.17), 0.29 (s.e. 0.16), and 0.42 (s.e. 0.14) years, respectively. The effects for these three age ranges are small and insignificant for boys, but large for girls with reductions of 0.43 (s.e. 0.23), 0.49 (s.e. 0.22), and 0.60 (s.e. 0.23) years. Results are consistent across all columns.

Different age effects reflect the different possible mechanisms by which children under different age groups are affected by the school consolidation policy. Children who were below age 6 and not enrolled in any school yet at the year of closure could be affected by delaying entry into primary schools that are much farther away due to concerns about safety. Children who were between age 6 and 9 at the year of closure faced transitioning from village school to consolidated schools and the possible disruption of school life, as well as much longer traveling distance. Finally children who were between age 10 and 13 at year of closure attended consolidated schools in their final primary school years. These individuals are on average 16.8 years old in 2011. For these individuals, closure took place during final years of primary school when school transition was potentially the most disruptive and the opportunity cost of travel time was higher. The 0.60 year reduction in grades completed for girls in this group is due to the cumulative effects of consolidation during and after primary school.

\subsection{Age and Duration Effects Results}
\label{sec:ageduraeffects}

Each age-at-closure group includes individuals of different ages in 2011 as shown in Table \ref{summtwo}, our estimates from Table \ref{regone} show the weighted average effect of children exposed to closure starting at the same age group but with different durations. The effects of school closure on educational attainment may not be homogeneous across children with different durations of exposure to the policy. The impacts of closure could amplify or weaken as children progress through school.

In order to capture the duration effects, we next allow the estimated impact of the policy to vary in the short, medium and long term. For our main results, we consider five age-at-closure groups (\(Z = 5\)), and for those under 13 at year of closure, we further divide each age group into 3 subgroups according to the number of years since closure (\(L = 3\)): 0 to 3, 4 to 6 and 7 to 12 years, representing short, medium and long-term exposure separately.\footnote{Individuals who were 10 to 13 at the time of closure and experienced 7 to 12 years of exposure are on average about 21 years old in 2011, hence we can interpret the impact here as the impact of policy on final attainment.} As before, age group 14 to 21 serves as reference group. By separating children into these \(Z\) and \(L\) groups, we are following, in principle, a similar strategy as \textcite{BehrmanParkerTodd, BehrmanParkerTodd2011} who analyze the effects of policy changes on grades completed by looking at both what they call \emph{exposure differential} as well as \emph{time since program initiation}. In our analysis, children who were at younger ages in the year of closure had greater \emph{exposure differential} to consolidated primary schools, and children from villages that experienced closure closer to 1999 have had longer \emph{time since program initiation}.\footnote{\textcite{BehrmanParkerTodd2011} focus on short and long run time-since-program-initiation effects of short policy exposure, and they also analyze the long run time-since-program-initiation effects of longer policy exposure. Here, given variation in school closure years, we have a continuous measure of time-since-program-initiation (duration-effects) which we group into short, medium and long duration sub-categories. And we analyze effects for three subsets of individuals with different lengths of policy exposures (age-effects subgroups).}

Given our earlier findings on differential gender effects, we estimate the model for girls and boys separately, with the results presented in Table \ref{regtwogirl} for girls and Table \ref{regtwoboy} for boys. All standard errors are clustered at the village-level. Column 1 presents our main results. The other columns provide robustness checks, to which we return following presentation of main results. The structure of the tables is the same as in Table \ref{regone}. We again focus on column 1.

\subsubsection{Impact on Girls}

For girls, results from Table \ref{regtwogirl} are consistent with our finding from Table \ref{regone} in that school consolidation had a large and significantly negative impact on girls. Across all affected age groups, the coefficients are all small and insignificant in the first three years after exposure, but more negative and significant after a longer duration. The first column in Table \ref{regtwogirl} shows that for the 0 to 5 age-at-closure group, the impact of policy was negative but insignificant at -0.15 (s.e. 0.26) years of education within the first 3 years after school closure. The lack of a strong impact here is expected because most of the children in this group are still too young to attend primary school. After 4 to 6 years and 7 to 12 years, the closure policy decreases average grades completed by 0.55 (s.e. 0.29) and 0.68 (s.e. 0.36) years, respectively. Most of the children in these two subgroups are still attending or just finishing up with primary school in 2011. These negative effects of school closure for girls could be driven by delayed entry into primary school due to potential safety and cost concerns given longer travel distance. Once children start primary school, continued safety and cost concerns could make regular attendance more difficult and longer travel distance could reduce time available for studying. With only 16 children in the 0 to 5 age-at-closure group at age above 15 in 2011 (see Table \ref{summthree}), we do not know the full effects of closure on final attainment for individuals in the 0 to 5 age-at-closure group, however, the overall pattern here indicates that longer duration amplifies the negative effects of closure on girl's educational attainment.

\begin{table}[t!]
\centering
\def\sym#1{\ifmmode^{#1}\else\(^{#1}\)\fi}
\caption{Effect of School Closure on Female Educational Attainment \label{regtwogirl}}
\begin{adjustbox}{max width=\regtwogirl\textwidth}
\begin{tabular}{m{\lablcolwidth} >{\centering\arraybackslash}m{1.5cm} >{\centering\arraybackslash}m{1.5cm} >{\centering\arraybackslash}m{1.5cm} >{\centering\arraybackslash}m{1.5cm} >{\centering\arraybackslash}m{1.5cm} >{\centering\arraybackslash}m{1.5cm}}
\toprule
& \multicolumn{6}{L{\innerheadwidth}}{\textbf{Female} Outcome: grades completed by year 2011} \\
\cmidrule(l{5pt}r{5pt}){2-7}
& \multicolumn{2}{C{3.0cm}}{\footnotesize } & \multicolumn{2}{C{3.0cm}}{\footnotesize \(10 \le 2011 \text{ Age} \le 34\)} & \multicolumn{2}{C{3.0cm}}{\footnotesize \(15 \le 2011 \text{ Age} \le 34\)} \\
\cmidrule(l{5pt}r{5pt}){2-3} \cmidrule(l{5pt}r{5pt}){4-5} \cmidrule(l{5pt}r{5pt}){6-7}
& \multicolumn{1}{C{1.5cm}}{(1)} & \multicolumn{1}{C{1.5cm}}{(2)} & \multicolumn{1}{C{1.5cm}}{(3)} & \multicolumn{1}{C{1.5cm}}{(4)} & \multicolumn{1}{C{1.5cm}}{(5)} & \multicolumn{1}{C{1.5cm}}{(6)} \\
\midrule
\formatpanelheadsupregs{7}{L{\subheadwidth}}{\baselinegroup}
\formatpanelheadregs{7}{L{\subheadwidth}}{Panel \newtablepanfemale}
\formatpanelheadsubregs{7}{L{\subheadwidth}}{\closeinteragestardura{} 0--5}
\vspace*{0mm}\hspace*{5mm}$\times$ (0--3 years since closure) & -0.15 & -0.29 &   &   &   &    \\
& \vspace*{-2mm}{\footnotesize (0.26) } &\vspace*{-2mm}{\footnotesize (0.30) } &   &   &   &    \\
\vspace*{0mm}\hspace*{5mm}$\times$ (4--6 years since closure) & -0.55\sym{*}  & -0.68\sym{**} &   &   &   &    \\
& \vspace*{-2mm}{\footnotesize (0.29) } &\vspace*{-2mm}{\footnotesize (0.31) } &   &   &   &    \\
\vspace*{0mm}\hspace*{5mm}$\times$ (7--12 years since closure) & -0.68\sym{*}  & -0.94\sym{**} &   &   &   &    \\
& \vspace*{-2mm}{\footnotesize (0.36) } &\vspace*{-2mm}{\footnotesize (0.38) } &   &   &   &    \\
\formatpanelheadsubregs{7}{L{\subheadwidth}}{\closeinteragestardura{} 6--9}
\vspace*{0mm}\hspace*{5mm}$\times$ (0--3 years since closure) & -0.22 & -0.38 & -0.25 & -0.64 &   &    \\
& \vspace*{-2mm}{\footnotesize (0.33) } &\vspace*{-2mm}{\footnotesize (0.41) } &\vspace*{-2mm}{\footnotesize (0.51) } &\vspace*{-2mm}{\footnotesize (0.65) } &   &    \\
\vspace*{0mm}\hspace*{5mm}$\times$ (4--6 years since closure) & -0.56\sym{*}  & -0.76\sym{**} & -0.63 & -0.83\sym{**} &   &    \\
& \vspace*{-2mm}{\footnotesize (0.34) } &\vspace*{-2mm}{\footnotesize (0.34) } &\vspace*{-2mm}{\footnotesize (0.38) } &\vspace*{-2mm}{\footnotesize (0.39) } &   &    \\
\vspace*{0mm}\hspace*{5mm}$\times$ (7--12 years since closure) & -0.77\sym{**} & -0.86\sym{**} & -0.76\sym{**} & -0.80\sym{*} &   &    \\
& \vspace*{-2mm}{\footnotesize (0.32) } &\vspace*{-2mm}{\footnotesize (0.34) } &\vspace*{-2mm}{\footnotesize (0.38) } &\vspace*{-2mm}{\footnotesize (0.42) } &   &    \\
\formatpanelheadsubregs{7}{L{\subheadwidth}}{\closeinteragestardura{} 10--13}
\vspace*{0mm}\hspace*{5mm}$\times$ (0--3 years since closure) & -0.53 & -0.71\sym{*}  & -0.53 & -0.71\sym{*} &   &    \\
& \vspace*{-2mm}{\footnotesize (0.35) } &\vspace*{-2mm}{\footnotesize (0.39) } &\vspace*{-2mm}{\footnotesize (0.38) } &\vspace*{-2mm}{\footnotesize (0.43) } &   &    \\
\vspace*{0mm}\hspace*{5mm}$\times$ (4--6 years since closure) & -0.59\sym{*}  & -0.72\sym{**} & -0.63\sym{*}  & -0.78\sym{**} & -0.66 & -0.81\sym{*}  \\
& \vspace*{-2mm}{\footnotesize (0.33) } &\vspace*{-2mm}{\footnotesize (0.33) } &\vspace*{-2mm}{\footnotesize (0.37) } &\vspace*{-2mm}{\footnotesize (0.37) } &\vspace*{-2mm}{\footnotesize (0.42) } &\vspace*{-2mm}{\footnotesize (0.41) }    \\
\vspace*{0mm}\hspace*{5mm}$\times$ (7--12 years since closure) & -0.76\sym{**} & -0.74\sym{*}  & -0.65\sym{*}  & -0.59 & -0.58 & -0.51    \\
& \vspace*{-2mm}{\footnotesize (0.35) } &\vspace*{-2mm}{\footnotesize (0.41) } &\vspace*{-2mm}{\footnotesize (0.38) } &\vspace*{-2mm}{\footnotesize (0.45) } &\vspace*{-2mm}{\footnotesize (0.40) } &\vspace*{-2mm}{\footnotesize (0.47) }    \\
\formatpanelheadsubregs{7}{L{\subheadwidth}}{\closeinteragestardura{} 22--29}
\vspace*{0mm}\hspace*{5mm}all years since closure&    0.20 &    0.28 & 0.074 &    0.12 & 0.078 &    0.10    \\
& \vspace*{-2mm}{\footnotesize (0.22) } &\vspace*{-2mm}{\footnotesize (0.23) } &\vspace*{-2mm}{\footnotesize (0.25) } &\vspace*{-2mm}{\footnotesize (0.26) } &\vspace*{-2mm}{\footnotesize (0.26) } &\vspace*{-2mm}{\footnotesize (0.27) }    \\
\midrule
Observations&    8869 &    7466 &    5664 &    4790 &    4642 &    3932    \\
\midrule
\exclcontrol
\exclcontrolcont
\bottomrule
\footnotegap
\multicolumn{7}{L{\footwidth}}{\footnotesize \justify \footattain} \\
\end{tabular}
\end{adjustbox}
\end{table}

\begin{table}[t!]
\centering
\def\sym#1{\ifmmode^{#1}\else\(^{#1}\)\fi}
\caption{Effect of School Closure on Male Educational Attainment\label{regtwoboy}}
\begin{adjustbox}{max width=\regtwoboy\textwidth}
\begin{tabular}{m{\lablcolwidth} >{\centering\arraybackslash}m{1.5cm} >{\centering\arraybackslash}m{1.5cm} >{\centering\arraybackslash}m{1.5cm} >{\centering\arraybackslash}m{1.5cm} >{\centering\arraybackslash}m{1.5cm} >{\centering\arraybackslash}m{1.5cm}}
\toprule
&    \multicolumn{6}{L{\innerheadwidth}}{\textbf{Male} Outcome: grades completed by year 2011} \\
\cmidrule(l{5pt}r{5pt}){2-7}
&    \multicolumn{2}{C{3.0cm}}{\footnotesize } & \multicolumn{2}{C{3.0cm}}{\footnotesize \(10 \le 2011 \text{Age} \le 34\)} & \multicolumn{2}{C{3.0cm}}{\footnotesize \(15 \le 2011 \text{Age} \le 34\)} \\
\cmidrule(l{5pt}r{5pt}){2-3}\cmidrule(l{5pt}r{5pt}){4-5}\cmidrule(l{5pt}r{5pt}){6-7}
&   \multicolumn{1}{C{1.5cm}}{(1)} & \multicolumn{1}{C{1.5cm}}{(2)} & \multicolumn{1}{C{1.5cm}}{(3)} & \multicolumn{1}{C{1.5cm}}{(4)} & \multicolumn{1}{C{1.5cm}}{(5)} & \multicolumn{1}{C{1.5cm}}{(6)} \\
\midrule
\formatpanelheadsupregs{7}{L{\subheadwidth}}{\baselinegroup}
\formatpanelheadregs{7}{L{\subheadwidth}}{Panel \newtablepanmale}
\formatpanelheadsubregs{7}{L{\subheadwidth}}{\closeinteragestardura{} 0--5}
\vspace*{0mm}\hspace*{5mm}$\times$ (0--3 years since closure)&      -0.030 &    0.17 &   &   &   &    \\
& \vspace*{-2mm}{\footnotesize (0.25) } &\vspace*{-2mm}{\footnotesize (0.26) } &   &   &   &    \\
\vspace*{0mm}\hspace*{5mm}$\times$ (4--6 years since closure) & -0.17 & -0.20 &   &   &   &    \\
& \vspace*{-2mm}{\footnotesize (0.27) } &\vspace*{-2mm}{\footnotesize (0.28) } &   &   &   &    \\
\vspace*{0mm}\hspace*{5mm}$\times$ (7--12 years since closure)&     -0.0044 &      0.0088 &   &   &   &    \\
& \vspace*{-2mm}{\footnotesize (0.26) } &\vspace*{-2mm}{\footnotesize (0.29) } &   &   &   &    \\
\formatpanelheadsubregs{7}{L{\subheadwidth}}{\closeinteragestardura{} 6--9}
\vspace*{0mm}\hspace*{5mm}$\times$ (0--3 years since closure) & -0.47 & -0.57\sym{*}  & -0.59 & -0.83\sym{*} &   &    \\
& \vspace*{-2mm}{\footnotesize (0.29) } &\vspace*{-2mm}{\footnotesize (0.32) } &\vspace*{-2mm}{\footnotesize (0.43) } &\vspace*{-2mm}{\footnotesize (0.46) } &   &    \\
\vspace*{0mm}\hspace*{5mm}$\times$ (4--6 years since closure)&      -0.030 &      -0.063 & -0.11 & -0.16 &   &    \\
& \vspace*{-2mm}{\footnotesize (0.31) } &\vspace*{-2mm}{\footnotesize (0.33) } &\vspace*{-2mm}{\footnotesize (0.35) } &\vspace*{-2mm}{\footnotesize (0.37) } &   &    \\
\vspace*{0mm}\hspace*{5mm}$\times$ (7--12 years since closure)&    0.28 &    0.31 &    0.22 &    0.14 &   &    \\
& \vspace*{-2mm}{\footnotesize (0.28) } &\vspace*{-2mm}{\footnotesize (0.31) } &\vspace*{-2mm}{\footnotesize (0.35) } &\vspace*{-2mm}{\footnotesize (0.38) } &   &    \\
\formatpanelheadsubregs{7}{L{\subheadwidth}}{\closeinteragestardura{} 10--13}
\vspace*{0mm}\hspace*{5mm}$\times$ (0--3 years since closure) & -0.48\sym{*}  & -0.54\sym{*}  & -0.54\sym{*}  & -0.58 &   &    \\
& \vspace*{-2mm}{\footnotesize (0.26) } &\vspace*{-2mm}{\footnotesize (0.32) } &\vspace*{-2mm}{\footnotesize (0.29) } &\vspace*{-2mm}{\footnotesize (0.36) } &   &    \\
\vspace*{0mm}\hspace*{5mm}$\times$ (4--6 years since closure) & 0.047 & 0.064 & 0.034 & 0.016 & -0.12 & -0.15    \\
& \vspace*{-2mm}{\footnotesize (0.27) } &\vspace*{-2mm}{\footnotesize (0.27) } &\vspace*{-2mm}{\footnotesize (0.29) } &\vspace*{-2mm}{\footnotesize (0.28) } &\vspace*{-2mm}{\footnotesize (0.31) } &\vspace*{-2mm}{\footnotesize (0.30) }    \\
\vspace*{0mm}\hspace*{5mm}$\times$ (7--12 years since closure) & -0.30 & -0.26 & -0.32 & -0.34 & -0.36 & -0.38    \\
& \vspace*{-2mm}{\footnotesize (0.30) } &\vspace*{-2mm}{\footnotesize (0.32) } &\vspace*{-2mm}{\footnotesize (0.32) } &\vspace*{-2mm}{\footnotesize (0.33) } &\vspace*{-2mm}{\footnotesize (0.34) } &\vspace*{-2mm}{\footnotesize (0.35) }    \\
\formatpanelheadsubregs{7}{L{\subheadwidth}}{\closeinteragestardura{} 22--29}
\vspace*{0mm}\hspace*{5mm}all years since closure&    0.14 &    0.20 &    0.12 &    0.25 & 0.038 &    0.17    \\
& \vspace*{-2mm}{\footnotesize (0.22) } &\vspace*{-2mm}{\footnotesize (0.24) } &\vspace*{-2mm}{\footnotesize (0.24) } &\vspace*{-2mm}{\footnotesize (0.27) } &\vspace*{-2mm}{\footnotesize (0.25) } &\vspace*{-2mm}{\footnotesize (0.28) }    \\
\midrule
Observations&    9935 &    8452 &    6408 &    5499 &    5321 &    4578    \\
\midrule
\exclcontrol
\exclcontrolcont
\bottomrule
\footnotegap
\multicolumn{7}{L{\footwidth}}{ \footnotesize \justify \footattain} \\
\end{tabular}
\end{adjustbox}
\end{table}

The second group of coefficients in column 1 of Table \ref{regtwogirl} shows the impact of the policy on girls who were 6 to 9 when school closure took place. For this group of children, the impact of the policy one to three years after closure is insignificant at -0.22 (s.e. 0.33) years. This indicates that girls who were in earlier primary school grades were successfully transferred from village schools to consolidated schools without immediate disruption to grade progression. It is possible that student achievement is negatively affected by school closure and transition in the short-run \autocite{hanushek_disruption_2004, sacerdote_when_2012}, but lower achievement might not have a clear impact on grade progression in the short-run. Interestingly, we do find that the impact of consolidation amplifies over the medium and long run for this subset of girls. The policy reduces attainment by 0.56 (s.e. 0.34) years in 4 to 6 years after closure, when girls are due to attend middle school; and the policy reduces attainment by 0.77 (s.e. 0.32) years in 7 to 12 years after closure, when it is about time for them to finish high school. These findings suggest that after exposure to consolidated schools in the second half of primary school, girls on average experience slower progression through middle school. Some of the long run effects are due to continued slower path of progression in high school and some are possibly due to failure to enter high school. In a separate set of regressions, we do indeed find that the high school completion rate is up to 8 percentage points lower for girls in this age-at-closure group 7 to 12 years after school closure.\footnote{Results for high school completion are available by request from the corresponding author.}

In the third group of coefficients in the first column of Table \ref{regtwogirl}, we show estimates for the impact of closure on children who were 10 to 13 years old at the time of closure. The policy reduces attainment by 0.53 (s.e. 0.35) years in the short run. The negative impact of the policy on girls is significantly amplified 4 to 6 years and 7 to 12 years after the policy, reducing attainment by 0.59 (s.e. 0.33) years and 0.76 (s.e. 0.35) years, respectively. The magnitude of the short run impact of the policy here is larger than the two younger age-at-closure groups, indicating that perhaps there is a more immediate effect of school closure for children who were in the 4th, 5th and 6th grade at the time of closure. The closure policy could be more disruptive for these children who need to learn more difficult material and prepare for middle school entry. Children in this age-at-closure cohort are on average 17 years old 4 to 6 years after the policy, and 21 years old 7 to 12 years after the policy. The similar coefficient on the medium and long run impact indicates that the negative effects of the policy persist until the end of high school age: consolidation has a potentially persistent negative effect on final attainment for girls.

Overall, the results here show that the closure policy impacts both girls who are already attending primary school and girls who were yet to enter primary school. We find a strengthening of the negative impact of closure over time for all girls' age-at-closure groups. For the two younger age-at-closure cohorts, we do not know the effect of the policy on final attainment. However, given that impacts persist even until 12 years after closure, our long run estimates are likely a lower bound for the effect of the policy on final attainment for these girls in the younger year-at-closure groups. Results do not differ significantly across columns.

\subsubsection{Impact on Boys}
\label{sec:impactboy}

We again see a large difference between boys and girls when we compare Table \ref{regtwoboy} for boys with Table \ref{regtwogirl} for girls: while we see medium and long run negative impacts of school closure on different subgroups of age-at-closure for girls, most of impacts on boys are not significant. For boys in the 0 to 6 age-at-closure group, column 1 in Table \ref{regtwoboy} shows that the impact of school consolidation is very close to zero from 0 years up to 12 years after closure. For boys in the 6 to 9 and 10 to 13 age-at-closure groups, the impacts of the policy are again mostly insignificant. The exception is for boys who were 10 to 13 at the time of closure. For this group, the closure policy reduces grades completed by 0.48 years in the short--run (s.e. 0.26).

This finding matches up with our finding from Table \ref{regtwogirl}, where the policy has similar magnitude of impact on girls in the same age subgroup. These results indicate that perhaps the policy is disruptive to both boys and girls who are in the higher grades of elementary school. This effect could be due to difficulty of completing---in the process of school transition---the relatively heavy school workload. While girls' attainment in this age-at-closure group worsens 4 to 12 years after closure, we see no impact of the policy on boys 4 to 12 years after. Perhaps boys catch up quickly afterwards, as parents continue to support their educational efforts and they eventually succeed in transitioning. Girls, however, are unable to overcome this difficulty and the negative impacts persist.

\label{rr:rtwoqthreegenderboy}Overall, our results on gender difference (as shown in panels B and C of Table \ref{regone} and in Tables \ref{regtwogirl} and \ref{regtwoboy} are not based on a comparison between girls and boys, but are based on within-gender comparisons across different age-at-closure groups that have differential exposures to closure. Specifically, Panels B and C of Table \ref{regone} show estimation results of Equation \eqref{eq:startsOnly} for the female and male samples separately, and Tables \ref{regtwogirl} and \ref{regtwoboy} show estimation results of Equation \eqref{eq:startsLengths} for the female and male samples separately as well. Our gender-specific closure effects capture the gender-specific deviations in attainments from gender-specific village fixed effects and gender-province-specific cohort (age in 2011) fixed effects. In this context, factors that have been considered as driving gender differences in educational attainment in the Chinese context, such as son preference and related issues of diluted resources for girls \autocite{lu_treiman_2008,liu_qualityquantity_2014} would not explain the gender differences in policy effects in our finding, unless shifts in these other factors line up with the village specific timing of school closure, which is unlikely.\footnote{We provide additional information on sibship sizes in Appendix Section \ref{sec:childcompo}.}

\subsection{Robustness and Common Trend}
\subsubsection{Common Trend}
\label{sec:parallel}
\begin{figure}
	\centering
	\caption{\small Effect of School Closure on Educational Attainment (Number of Grades Completed by 2011) by 8 Age-at-Closure Group.}
 	\includegraphics[width=1.0\textwidth, center]{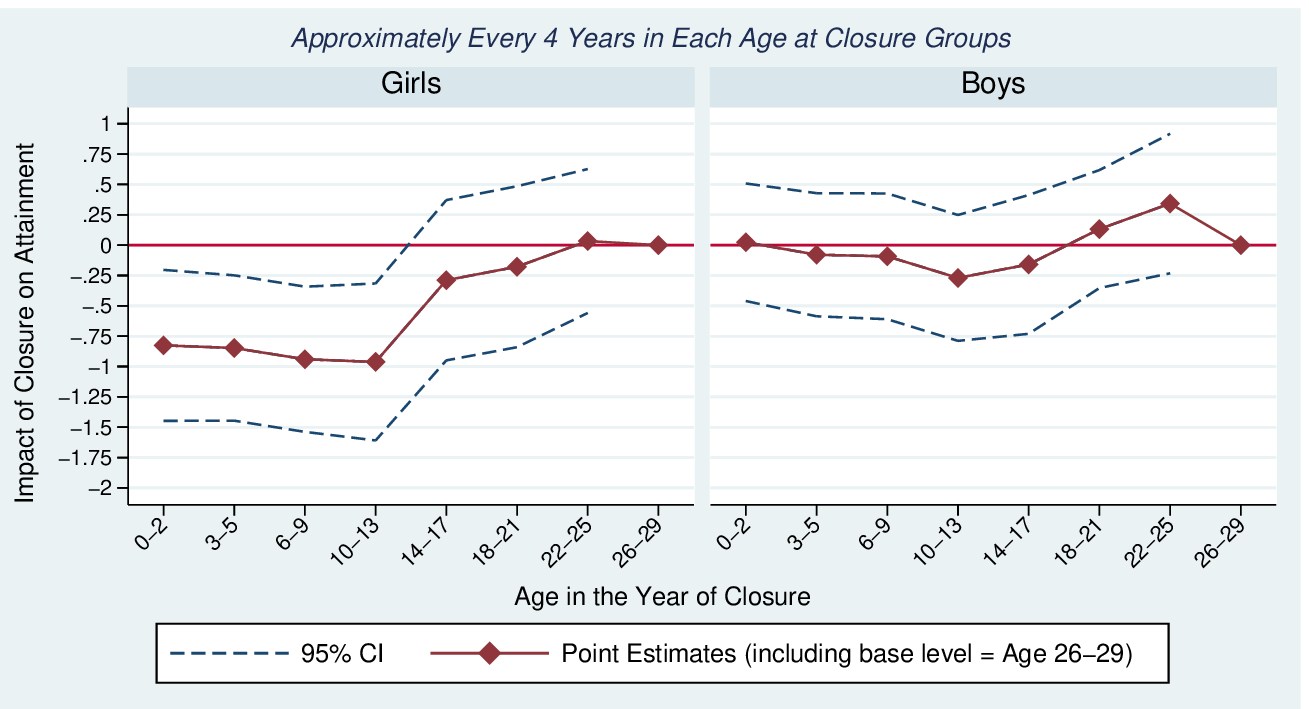}
	\captionsetup{width=1.0\textwidth}
	\caption*{\footnotesize \footgraphmain \label{figone}}
\end{figure}

As discussed earlier, our estimation relies on the ``common trend'' assumption, which assumes in the absence of the policy, the within-province cohort-trend in educational attainment are the same in the treated and control villages. We check the common trend assumption for our specification with only age effects in each regression table and in Figure \ref{figone}. In Tables \ref{regone}, \ref{regtwogirl} and \ref{regtwoboy}, we test the assumption by including another group -- those who were 22 to 30 years-of-age at the time of school closure -- and should not be affected by the policy to compare with the reference group. As shown in the final row for each table, the coefficients on this age group's interaction with the closure policy are generally slightly positive but not significantly different from the baseline group (i.e. age-at-closure between 14 and 21). These coefficients are generally closer to zero when we exclude younger and older individuals by 2011 age in columns 3 through 6 of the tables. If villages that were selected for closure were already on a significant downward trajectory compared to provincial trend, we would expect to see a positive and significant coefficient here.

Following \textcite{duflo_2001_school}, we test parallel trends with a finer set of age subgroups in Figure \ref{figone}. In Figure \ref{figone}, we show regression coefficients when we estimate Equation \eqref{eq:startsOnly} by considering eight age-at-closure groups, 0 to 2, 3 to 5, 6 to 9, 10 to 13, 14 to 17, 18 to 21, 22 to 25, and 26 to 29\label{rr:roneqtwocommontrend}.\footnote{\label{rr:roneqtwocommontrendfoot}In villages without closure, there is no ``age-at-closure'', i.e., the variable for the x-axis of Figures \ref{figone} and \ref{figoneb}. As discussed before, the policy effects are capturing the deviations in attainment from village fixed effects and province-specific cohort (age in 2011) fixed effects for different age-at-closure sub-groups from closure villages. For the regression tables and for the parallel trend figures, individuals in the villages without closure are matched to their counterparts in the treated villages by cohort (age in 2011). The correspondence between age-at-closure and current age is shown in Table \ref{summthree}.} Age-at-closure group 26-29 is the base group. The regression is the same as the one from Column 1 of Table \ref{regone} except with the addition of more age-at-closure groups and a change in the base group. For the four groups that were above 13 years of age at the time of school closure, the policy effects, with the 26-29 age-at-closure group as base group, are not significantly different from zero. For the four age groups lower than 14, the coefficients are significantly negatively deviating from the pre-existing trend. The figure also shows that for boys, the trend line is flat and not significantly different from zero for all age-at-closure ages. Figure \ref{figoneb} in the Appendix shows similar results as Figure \ref{figone} with even finer age-at-closure breakdowns.

\label{rr:aethreecommontrend}The presence of time-varying unobservables that determine both the school closure decision as well as children's educational attainment and that are not removed by village fixed effects would threaten our identification strategy \autocite{Ma2017}. For example, policy makers may have closed schools that were of poor quality and were getting even worse in teaching quality. Our parallel trend analysis above, which showed no pre-existing trends, largely addresses this concern. To investigate the issue further, we use whether a school was a ``teaching point'' before closure to proxy for schooling quality and examine whether there is heterogeneous effect of closure on students in villages with schools that were teaching-points, compared to others.\footnote{The only quality-linked characteristic available to us for closed-off schools is whether the school was a teaching-point---which goes up to grade four---or not. When quality is defined narrowly in terms of the physical facility quality and teacher qualifications, teaching-points are of lower quality compared to complete primary schools \autocite{SargentHannumJTE2009}.} Results are shown in Appendix Section \ref{sec:teachpoint} and do not show significant heterogeneous effects. Results are thus are consistent with our main results from Table \ref{regone}.

\subsubsection{Robustness Checks}
\label{sec:resrobust}

Our results are robust to several checks: 1) excluding younger and older age groups who have only completed a part of formal education at schools by 2011 or whose educational attainment might be systematically different from that of younger cohorts, 2) dropping villages that never had a school from the villages without school closure group, and 3) testing results across different age-at-closure groups and with different cutoffs ages.

First, although the inclusion of young age-at-closure groups allows us to see the impact of closure on children who started school after school closure, many of these individuals are in still early schooling years in 2011. These very young individuals and also older individuals are potentially ill-suited to be estimated jointly in an environment with village fixed effects. For Tables \ref{regone}, \ref{regtwogirl} and \ref{regtwoboy}, in columns three and four, we restrict the sample to individuals who are included inside the black dashed box as shown in Table \ref{summthree}: only individuals who are between 10 and 34 years of age in 2011 and who were at least age 6 at the year of school closure among those in villages with closure. In columns five and six of these tables, we further restrict the sample and include only those who are between 15 and 34 years of age in 2011, and who were above age 10 at the year of closure among those in closure villages. For these tables, there are no significant differences in coefficients as we go from columns 1 and 2 to the more restricted age samples of later columns, although standard errors tend to increase due to significant drops in sample size.

Second, as reported in the data section, there is a subset of villages that report never having had a school or as having only had a school at some point between 1949 and 1999. For the odd-numbered columns in the regression tables, we include these villages along with villages that currently have a school and did not experience closure. We drop them in the even-numbered columns. For Table \ref{regone}, columns 2, 4, and 6 have slightly more negative coefficients than columns 1, 3 and 5, perhaps due to a more precise comparison between villages with and without closure. We also seem to have a similar strengthening of coefficient magnitude in the even columns of Tables \ref{regtwogirl} and \ref{regtwoboy}, but the results are less clear there.

Third, we run the same regression model but with different subgroups, cutoff ages, and groupings for age-at-closure and years-since-closure. Some of these results are shown in Figure \ref{figone} discussed earlier, and also shown in Figure \ref{figoneb} in the Appendix.\footnote{Appendix Section \ref{sec:amorereg} shows some addition tables and discusses some of these results. More results are available upon request from the corresponding author.} All of these analyses show the robustness of our main estimation results.

\label{rr:moreappendixrobust}In addition to conducting robustness checks, we also run additional analysis to learn about other dimensions of effects heterogeneity. For these analyses, we continue to focus on the heterogeneity of policy effects by age-at-closure and the gender of the child, but allow for the gender specific age-at-closure effects to be different for different subsets of individuals or villages. We do not focus on the duration-since-closure effects here because the number of observations in each age-duration cell with additional interaction is often too small. Specifically, we examine the heterogeneous effects by individual's ethnic minority status (minority/Han) in Appendix Section \ref{sec:minoindi}. The results indicate that while both minority and Han girls are negatively impacted by closure, the negative effects are larger in magnitude for minority girls. We also examine the heterogeneous effects by the quality of the closed schools in Appendix Section \ref{sec:teachpoint}, where we use whether the school was a teaching point to proxy for the quality. The effects of closure on non-teaching-point schools is similar the overall effects of closure without distinguishing between these two village school types. We find weaker negative effects of closure on teaching-point schools, which indicates potentially less negative effects for children who moved to the consolidated schools from schools with closure. Finally, in Appendix Section \ref{sec:boarding}, we examine the heterogeneous effects of school closure on students in terms of boarding arrangements after school was closed. Similar to \textcite{chen_poor_2014}, we find that boarding provision when interacted with closure is associated with greater reductions in educational attainments.

\section{Mechanisms--Distance, Quality and Enrollment}
\label{sec:mechanism}

As noted earlier, school closures typically imply disruption, greater distance, and better quality school facilities for affected students. Our prior analyses showed that the impact of closure on attainment is enduring and is not simply a short-term disruption. To investigate mechanisms behind the impacts of school closure on educational attainment, we analyze how two possible factors---distance to school and quality of school---are linked to enrollment status at the time of the survey. Here, we focus only on children between 5 to 12 years of age in 2011---ages at which nearly all children attend primary school in 2011.\footnote{Although previously in regressions on educational attainment we include children who were age 13 at year of closure into the group that may still attend primary school and therefore be exposed to school closure, we only include children who are in the age range that is definitely eligible for primary school in the enrollment regression here.} For these children, we have information about distance to primary school and primary school quality.\footnote{For older individuals in the survey who are included in the earlier attainment regressions, we do not have measures for the quality of school when they were attending primary school.} We use children from both villages with and without school closure in these regressions. The data for these analyses come from columns 2 and 3 of Table \ref{summthree}.

We make use of the information from the village survey on school facilities and distance to closest primary school, which we summarized in Panel A and B in Table \ref{summtwo}. For the distance to school variable, we use both the continuous version of the variable shown in Table \ref{summtwo}, and also group the values into three categories: \(0\) kilometers to school, (\(0\) to \(3\)) kilometers to school (medium distance), and greater than \(3\) kilometers (long distance). The median distance is 2 kilometers and 7 kilometers for the second and third category, respectively. Schools of these three distance categories serve 46, 32, and 22 percent of the sample villages. We create an index for school facility quality by summing up the nine facility dummy variables: the value for this variable ranges from 0 to 9. We also divide the index value into three groups for a categorical version of the variable indicating the number of facilities that a school has: 0 to 3, 4 to 6, or 7 to 9. Schools of these three facility categories serve 19, 48, and 33 percent of the sample villages, respectively.

Primary school enrollment in 2011 is high but not full. For children at 5 years of age, the enrollment rate is 11 percent. At age 6, the enrollment rate increases to 49 percent. Enrollment peaks at 96 percent at age 9 to 10.\footnote{After main primary school ages, at age 13, 14 and 15, primary school enrollment rates drops to 45, 15, and 4 percent respectively.} In the following anlysis, we study the relationship between distance, quality and school enrollment.

We regress enrollment on distance to closest primary school and quality of these schools. Regressions control for county fixed effects, province-specific age fixed effects, village per capita income, village per capita land size, village population size, household relative wealth, the number of household members and household ethnicity. Despite the controls, the coefficients we obtain for distance to school and quality of school would not be causal if there are unobserved village-level attributes that affect enrollment and that are also correlated with distance and quality. Our inclusion of village-level controls and county fixed effects, however, seeks to reduce the risk that omitted variables bias.

\begin{table}[t!]
\centering
\caption{Linear Probability Model of School Enrollment, Age 5 to 12\label{regfive}}
\begin{adjustbox}{max width=1\textwidth}
\begin{tabular}{m{\lablcolwidthenroll} >{\centering\arraybackslash}m{1.85cm} >{\centering\arraybackslash}m{1.85cm} >{\centering\arraybackslash}m{1.85cm} >{\centering\arraybackslash}m{1.85cm} >{\centering\arraybackslash}m{1.85cm} >{\centering\arraybackslash}m{1.85cm}}
\toprule
& \multicolumn{6}{L{\innerheadwidthenroll}}{Outcome: enrolled in school or not in 2011} \\                                                 \cmidrule(l{5pt}r{5pt}){2-7}                           &   \multicolumn{2}{C{3.7cm}}{\small All Age 5 to 12} & \multicolumn{2}{C{3.7cm}}{\small Girls Age 5 to 12} & \multicolumn{2}{C{3.7cm}}{\small Boys Age 5 to 12} \\                                                  \cmidrule(l{5pt}r{5pt}){2-3} \cmidrule(l{5pt}r{5pt}){4-5} \cmidrule(l{5pt}r{5pt}){6-7}                          &   \multicolumn{1}{C{1.85cm}}{{\footnotesize all villages}} & \multicolumn{1}{C{1.85cm}}{{\footnotesize no teaching points}} & \multicolumn{1}{C{1.85cm}}{{\footnotesize all villages}} & \multicolumn{1}{C{1.85cm}}{{\footnotesize no teaching points}} & \multicolumn{1}{C{1.85cm}}{{\footnotesize all villages}} & \multicolumn{1}{C{1.85cm}}{{\footnotesize no teaching points}} \\
\midrule
\formatpanelheadregs{7}{L{\subheadwidthenroll}}{Panel A: Continuous Distance and Quality Measures}
distance (km) to primary school&     -0.0047\sym{**} &     -0.0060\sym{**} &     -0.0056\sym{*} &      -0.011\sym{***}&     -0.0035 &     -0.0019    \\
	& \vspace*{-2mm}{\footnotesize (0.0023) } &\vspace*{-2mm}{\footnotesize (0.0028) } &\vspace*{-2mm}{\footnotesize (0.0031) } &\vspace*{-2mm}{\footnotesize (0.0039) } &\vspace*{-2mm}{\footnotesize (0.0029) } &\vspace*{-2mm}{\footnotesize (0.0042) }    \\
number of primary school facilities&      0.0028 &     0.00017 &     -0.0048 &     -0.0025 &      0.0094\sym{*} &      0.0033    \\
	& \vspace*{-2mm}{\footnotesize (0.0044) } &\vspace*{-2mm}{\footnotesize (0.0047) } &\vspace*{-2mm}{\footnotesize (0.0067) } &\vspace*{-2mm}{\footnotesize (0.0078) } &\vspace*{-2mm}{\footnotesize (0.0055) } &\vspace*{-2mm}{\footnotesize (0.0057) }    \\
\midrule
Observations   &     2460 &    2033 &    1130 &     942 &    1330 &    1091    \\
\midrule
\formatpanelheadregs{7}{L{\subheadwidthenroll}}{Panel B: Categorical Distance and Quality Measures}
\formatpanelheadsubregs{7}{L{\subheadwidthenroll}}{\enrolldistbase}
\vspace*{0mm}\hspace*{5mm}$0<x\leq3\hspace{0.1cm}(median\approx2)\hspace{0.1cm}km$&      -0.022 &      -0.025 &      -0.032 &      -0.048 &      -0.028 &      -0.020    \\
	& \vspace*{-2mm}{\footnotesize (0.017) } &\vspace*{-2mm}{\footnotesize (0.019) } &\vspace*{-2mm}{\footnotesize (0.025) } &\vspace*{-2mm}{\footnotesize (0.030) } &\vspace*{-2mm}{\footnotesize (0.024) } &\vspace*{-2mm}{\footnotesize (0.026) }    \\
\vspace*{0mm}\hspace*{5mm}$3<x\leq\max\hspace{0.1cm}(median\approx7)\hspace{0.1cm}km$&      -0.063\sym{***}&      -0.082\sym{***}&      -0.084\sym{***} & -0.14\sym{***}&      -0.053\sym{*} &      -0.039    \\
	& \vspace*{-2mm}{\footnotesize (0.023) } &\vspace*{-2mm}{\footnotesize (0.030) } &\vspace*{-2mm}{\footnotesize (0.032) } &\vspace*{-2mm}{\footnotesize (0.042) } &\vspace*{-2mm}{\footnotesize (0.030) } &\vspace*{-2mm}{\footnotesize (0.041) }    \\
\formatpanelheadsubregs{7}{L{\subheadwidthenroll}}{\enrollqualbase}
\vspace*{0mm}\hspace*{5mm}4 to 6 Facilities   & 0.021 & 0.014 &     -0.0094 &      -0.013 & 0.049\sym{*}  & 0.038    \\
	& \vspace*{-2mm}{\footnotesize (0.023) } &\vspace*{-2mm}{\footnotesize (0.025) } &\vspace*{-2mm}{\footnotesize (0.028) } &\vspace*{-2mm}{\footnotesize (0.031) } &\vspace*{-2mm}{\footnotesize (0.028) } &\vspace*{-2mm}{\footnotesize (0.031) }    \\
\vspace*{0mm}\hspace*{5mm}7 to 9 Facilities   & 0.035 & 0.023 &      0.0029 &      0.0077 & 0.067\sym{**} & 0.041    \\
	& \vspace*{-2mm}{\footnotesize (0.025) } &\vspace*{-2mm}{\footnotesize (0.027) } &\vspace*{-2mm}{\footnotesize (0.033) } &\vspace*{-2mm}{\footnotesize (0.038) } &\vspace*{-2mm}{\footnotesize (0.031) } &\vspace*{-2mm}{\footnotesize (0.034) }    \\
\midrule
Observations   &     2460 &    2033 &    1130 &     942 &    1330 &    1091    \\
\bottomrule
\footnotegap
\multicolumn{7}{L{\footwidthenroll}}{\footnotesize\justify\footenroll}\\
\end{tabular}
\end{adjustbox}
\end{table}

We present the enrollment regression results for children from 5 to 12 years of age in Table \ref{regfive}\label{rr:ronemech}.\footnote{\label{rr:ronemechfoot}We focus on ages 5 to 12 during which boys, girls, minority and Han children are all predominantly enrolled in primary schools. We provided additional details on enrollment information by age, gender and ethnicity in Appendix Section \ref{sec:enrollage}.} We run similar regressions in Table \ref{regtwelve} in the Appendix, where we allow effects of distance and quality to differ for age subgroups 5 to 8 and 9 to 12. Panel A of Table \ref{regfive} presents results where we regress enrollment on the continuous distance to school variable and the continuous aggregate school facility quality variable. In panel B of Table \ref{regfive}, we show results for regressing school enrollment on the categorical variables for school distance and facility quality. In the first two columns of the table, we show results for both boys and girls between age 5 and 12, in columns three and four, we show results for girls between age 5 and 12, and in columns five and six, we show results for boys between age 5 and 12. In columns 1, 3 and 5, we show results including all villages. In columns 2, 4, and 6, we drop villages that contain a partial primary school (teaching-point) but not a complete primary school.\footnote{Village heads were asked to answer distance and quality questions for the closest full primary school to the village. But in villages with primary school teaching-point, most children between age 5 and 12 should be attending the village school and only attend the full primary school at the end of primary school age for the 5th and 6th grades. The distance and school quality variables reported for this type of schools might not, therefore, reflect the actual distance to and quality of the relevant school for the vast majority of children between age 5 and 12 in these villages.}

Overall, Panel A in Table \ref{regfive} shows that longer distance to school is linked to lower enrollment, especially for girls. School facility quality does not have an impact on girls' enrollment, but boys are more likely to attend schools with better physical facilities. For all children between age 5 and 12, columns 1 and 2 show that a kilometer increase in school distance is associated with an 0.47 (s.e. 0.23) and 0.60 (s.e. 0.28) percentage point reduction in school enrollment respectively. Columns 3 and 5 show that the reduction is -0.56 (s.e. 0.31) percentage points for girls and -0.35 (s.e. 0.29) percentage points for boys; excluding villages with primary school teaching-points, columns 4 and 6 show that the reduction is -1.1 (s.e. 0.39) percentage points for girls and -0.19 (s.e. 0.42) percentage points for boys. There is no overall significance in the impact of school facility quality on enrollment, but for boys, columns 5 and 6 show that an additional item in the school facility index is associated with an increase in the likelihood of school enrollment of 0.94 (s.e. 0.55) percentage points and 0.33 (s.e. 0.57) percentage points, respectively.

Panel B of Table \ref{regfive} uses the categorical variables for distance and quality we constructed and matches up with the results from Panel A. For all regressions, the comparison villages have 0 kilometers for distance to school and 0 to 3 school facilities. In terms of distance, from column 3, for girls, there is a 3.2 (s.e. 2.5) and an 8.4 (s.e. 3.0) percentage point reduction in enrollment associated with attending medium distance (0\textless distance\textless=3 kilometers) and long distance (distance\textgreater3 kilometers) schools. For boys, from column 5, the respective reductions are 2.8 (s.e. 2.4) and 5.3 (s.e. 3.0) percentage points. As in Panel A, the coefficients in the even columns are more negative for the girls and closer to zero for the boys. In terms of quality, columns 3 and 5 of Panel B of Table \ref{regfive} shows that better school facility scores do not impact school enrollment for girls but do for boys: having a school with 4 to 6 facilities and 7 to 9 facilities are associated with a 4.9 (s.e. 2.8) and a 6.7 (s.e. 3.1) percentage point increase in enrollment for boys, respectively. As in Panel A, the quality coefficients are still positive for the boys but less significant in column 6 of Panel B. In combination, a village with a primary school that is more than 3 kilometers away and that has 6 to 7 facilities might see a significant reduction in enrollment for girls, but minimal effects for boys, as the potential positive facility quality impact and negative distance impact for boys largely cancel out.

Village parents determine school enrollment for young boys and girls based on the costs and benefits of enrollment. The school consolidation policy potentially changed both the costs and benefits concurrently, but differently for boys and girls. Longer travel distance in difficult terrains might involve more transportation costs for both girls and boys, leading to lower school enrollment. The larger negative effects associated with girls might be due to potential additional parental concerns for girls' safety during longer travel, and higher opportunity cost of school enrollment for girls who otherwise could help out with household chores. Improved school facilities in consolidated schools could potentially increase the value of schooling for all children. However, the results here indicate that parents might have only perceived gains for boys in attending schools with better facilities, but did not value as much the gain in school quality for girls. The resistance towards school enrollment for young girls when distance to school increases is likely one of the contributing factors to the persistent negative effect of closure on girls' educational attainment that we find in earlier sections. We did not find a significant effect of the consolidation policy on boys' attainment, possibly because the positive effects from better quality and negative effects from longer distance cancel out.

\section{Conclusion}
\label{subsec:conclude}

This paper has estimated the impact of school consolidation on educational attainment in China, which is at the vanguard of a trend emerging in rural areas of many large middle--income countries. Specifically, we use multi-province data to estimate the impact of school consolidation on school progression and attainment in China. We find that children's educational outcomes are negatively affected, overall, by this policy. Our analyses indicate that children under 14 years of age at year of school closure experienced on average 0.24 to 0.42 fewer years of school attainment by 2011. This negative effect is not a temporary disruption: negative effects appear to strengthen with time since closure. Moreover, there is a striking contrast between boys and girls: while boys are less affected by this policy, girls exposed to the policy experience on average up to 0.60 fewer grades completed by 2011.

Our empirical results are consistent with certain possible mechanisms. The first mechanism, and probably the most important one, is the much greater distance to schools following closure, and the corresponding increase in travel costs. This change could impede families in sending children to school. Media outlets report families delaying entry for young children and leaving school-aged children in boarding schools or rented apartments in town centers with parents or grandparents \autocite{hui2011}. Such strategies will obviously increase financial costs for families who are exposed to school closure, and for those who have difficulty affording such strategies, children are likely to drop out of school earlier. A second possible mechanism of impact is the change of schooling environment and education quality. A core rationale for consolidation has been the expectation that more centrally located schools provide better quality education and thus improve students' performance, and some studies are consistent with this idea, though the effects may be partially offset by boarding \autocite{mo_transfer_2012, chen_poor_2014}.\footnote{However, a counter-narrative has emerged in news reports highlighting that schools were overcrowded and underprepared for an influx of children \autocite{yu_yi_2010, wu_tingting_2014}, and academic studies have provided inconsistent evidence \autocite{liu_effect_2010, chen_poor_2014}.} As families and youth make decisions about educational continuation, the greater costs and risk associated with attending school at a distance must be weighed against the potential benefits of attending a better-resourced school. These calculations may differ for girls and boys.

Underlying the observed patterns of gender difference is a decision process in which parents weigh the benefits and costs of attending school for children--and may do so differently for boys and girls. Even though the cost of attending school increases for households who lose access to within-village primary education, the perceived benefit of boys attending school appears large enough to counteract the extra cost brought by school consolidation. On the other hand, the instrumental benefit associated with educating daughters is likely perceived to be less than that for sons by parents, on average. Although previous studies estimate higher rates of return to schooling for girls than boys in the 1980s-1990s \autocite{zhang_economic_2005}, a tradition of patrilocal marriage in most parts of rural China means that girls' returns may be viewed as unlikely to flow to natal households. In addition, attending schools outside the villages is likely more risky and costly for girls than for boys. As a result, the school-consolidation policy may have pushed a fraction of households over the margin from sending girls to school to not sending them.

Beyond China, in an age of global population aging and large-scale migration, sparse school-aged populations in rural communities are common. Designing education supply policies that appropriately balance efficiency and equity concerns in such contexts is a difficult challenge. School consolidation initiatives are emerging as a common response in many middle-income countries with large rural populations, with recent media reports describing closure initiatives in a number of countries \autocite[for example, see][]{chowdhury_cramped_2017, harun_schools_2017, saengpassa_ministrys_2017, setiawati_schools_2010, tawie_sarawak_2017}.

Yet, the likely implications of these initiatives for educational access and inequality are poorly understood. Results presented here indicate that the school consolidation policy in rural China has had a negative impact on girls' educational attainment, but not boys'. Strictly speaking, these findings pertain to areas of China where the CHES sample was drawn, which are areas with large minority populations. While the impact, and gender differences in the impact, of consolidation may differ across contexts and nations, these findings highlight a significant case where consolidation has affected access and inequality and suggest the need for further scholarly attention to an emerging policy response to global demographic change.
\clearpage 

\begingroup
\setstretch{1.0}
\setlength\bibitemsep{3pt}
\printbibliography[title=References]

@article{grau_school_2018,
	title = {School closure and educational attainment: {Evidence} from a market-based system},
	volume = {65},
	issn = {02727757},
	shorttitle = {School closure and educational attainment},
	url = {https://www.sciencedirect.com/science/article/pii/S0272775717303953},
	doi = {https://doi.org/10.1016/j.econedurev.2018.05.003},
	language = {en},
	urldate = {2021-07-19},
	journal = {Economics of Education Review},
	author = {Grau, Nicolas and Hojman, Daniel and Mizala, Alejandra},
	month = aug,
	year = {2018},
	pages = {1--17},
}

@article{post_district_1999,
	title = {District consolidation and rural school closure: {E} pluribus unum?},
	volume = {15},
	issn = {1551-0670},
	number = {2},
	journal = {Journal of Research in Rural Education},
	author = {Post, David and Stambach, Amy},
	year = {1999},
	note = {Publisher: THE UNIVERSITY OF MAINE},
	pages = {106--117},
}

@article{lee_impact_2017,
	title = {The {Impact} of {School} {Closures} on {Equity} of {Access} in {Chicago}},
	volume = {49},
	issn = {0013-1245, 1552-3535},
	url = {http://journals.sagepub.com/doi/10.1177/0013124516630601},
	doi = {10.1177/0013124516630601},
	language = {en},
	number = {1},
	urldate = {2021-02-25},
	journal = {Education and Urban Society},
	author = {Lee, Jin and Lubienski, Christopher},
	month = jan,
	year = {2017},
	pages = {53--80},
}

@online{united_states_department_of_education_closed_2021,
	type = {Fast {Facts}},
	title = {Closed Schools},
	url = {https://nces.ed.gov/fastfacts/display.asp?id=619},
	language = {EN},
	urldate = {2021-07-18},
	author = {{United States Department of Education}},
	year = {2021},
	note = {Publisher: National Center for Education Statistics},
}

@online{ceic_cn_2021,
	type = {Data {Portal}},
	title = {{CN}: {No} of {School}: {Primary} ({ID}: 4967601 {SR} {Code}: {SR255140})},
	url = {https://info.ceicdata.com/en-products-china-premium-database},
	language = {en},
	urldate = {2021-07-18},
	journal = {CEIC Data - China Premium Database},
	author = {CEIC},
	month = jun,
	year = {2021},
}

@online{textor_china_2021,
	type = {Data {Portal}},
	title = {China: urbanization 2020},
	shorttitle = {China},
	url = {http://www.statista.com/statistics/270162/urbanization-in-china/},
	language = {en},
	urldate = {2021-07-18},
	journal = {Statista},
	author = {Textor, C.},
	month = may,
	year = {2021},
}

@book{national_bureau_of_statistics_of_china_china_2012,
	location = {Beijing},
	title = {China Rural Statistical Yearboook 2012},
	url = {http://cdi.cnki.net/Titles/SingleNJ?NJCode=N2012120246},
	publisher = {China Statistics Press},
	author = {{National Bureau of Statistics of China}},
	date = {2012}
}

@book{gustafsson_ethnicity_2019,
	location = {New York; Oxford},
	title = {Ethnicity and Inequality in China},
	publisher = {Routledge},
	date = {2019},
	author = {Gustafsson, Björn and Hasmath, Reza and Ding, Sai}
}

@article{lei_sibling_2017,
	title = {Sibling gender composition’s effect on education: evidence from {China}},
	volume = {30},
	issn = {1432-1475},
	shorttitle = {Sibling gender composition’s effect on education},
	url = {https://doi.org/10.1007/s00148-016-0614-z},
	doi = {10.1007/s00148-016-0614-z},
	language = {en},
	number = {2},
	journal = {Journal of Population Economics},
	author = {Lei, Xiaoyan and Shen, Yan and Smith, James P. and Zhou, Guangsu},
	month = apr,
	year = {2017},
	pages = {569--590}
}

@article{liu_qualityquantity_2014,
	title = {The quality–quantity trade-off: evidence from the relaxation of {China}’s one-child policy},
	volume = {27},
	issn = {1432-1475},
	shorttitle = {The quality–quantity trade-off},
	url = {https://doi.org/10.1007/s00148-013-0478-4},
	doi = {10.1007/s00148-013-0478-4},
	language = {en},
	number = {2},
	journal = {Journal of Population Economics},
	author = {Liu, Haoming},
	month = apr,
	year = {2014},
	pages = {565--602}
}

@article{KingBehrman2009WBRO,
    author = {King, Elizabeth M. and Behrman, Jere R.},
    title = "{Timing and Duration of Exposure in Evaluations of Social Programs}",
    journal = {The World Bank Research Observer},
    volume = {24},
    number = {1},
    pages = {55-82},
    year = {2009},
    month = {02},
    issn = {0257-3032},
    doi = {10.1093/wbro/lkn009},
    url = {https://doi.org/10.1093/wbro/lkn009},
    eprint = {http://oup.prod.sis.lan/wbro/article-pdf/24/1/55/4692681/lkn009.pdf},
}

@article{SargentHannumJTE2009,
author = {Tanja C. Sargent and Emily Hannum},
title ={Doing More With Less: Teacher Professional Learning Communities in Resource-Constrained Primary Schools in Rural China},
journal = {Journal of Teacher Education},
volume = {60},
number = {3},
pages = {258-276},
year = {2009},
doi = {10.1177/0022487109337279},
    note ={PMID: 21191452}
}

@article{lu_treiman_2008,
    author = {Yao Lu and Donald J. Treiman},
    title ={The Effect of Sibship Size on Educational Attainment in China: Period Variations},
    journal = {American Sociological Review},
    volume = {73},
    number = {5},
    pages = {813-834},
    year = {2008},
    doi = {10.1177/000312240807300506}
}

@article{harun_schools_2017,
	title = {Schools with low enrolment of students to be merged into one},
	location = {Kuala Lumpur, Malaysia},
	url = {https://www.nst.com.my/news/government-public-policy/2017/07/262570/schools-low-enrolment-students-be-merged-one},
	journaltitle = {New Straits Times},
	type = {Newspaper},
	author = {Harun, Hana Nax and Yunus, Arfa and Yusof, Teh Athira},
	date = {2017-07-31}
}

@article{hui2011,
	location = {Beijing, China},
	title = {School Closure and Consolidation does not Match Local Needs in Western China},
	url = {https://finance.qq.com/a/20090630/000162.htm},
	journaltitle = {21st Century Business Herald},
	author = {Hui, Ma},
	urldate = {2017-07-05},
	date = {2009-06-30},
}

@article{Zhang2018,
author = {Zhang, Shuang},
title = {Effects of High School Closure on Education and Labor Market Outcomes in Rural China},
journal = {Economic Development and Cultural Change},
volume = {67},
number = {1},
pages = {171-191},
year = {2018},
doi = {10.1086/697564},
URL = { 
        https://doi.org/10.1086/697564
},
eprint = { 
        https://doi.org/10.1086/697564
}
}

@incollection{BehrmanParkerTodd,
	location = {Cambridge, MA},
	title = {Medium-Term Impacts of the Oportunidades Conditional Cash Transfer Program on Rural Youth in Mexico},
	pages = {219--270},
	booktitle = {Poverty, Inequality, and Policy in Latin America},
	publisher = {MIT Press},
	author = {Behrman, Jere R. and Parker, Susan W. and Todd, Petra E.},
	editor = {Klasen, Stephan and Nowak-Lehmann, Felicitas},
	date = {2009}
}

@article{BehrmanParkerTodd2011,
 ISSN = {0022166X},
 URL = {http://www.jstor.org/stable/25764805},
 author = {Jere R. Behrman and Susan W. Parker and Petra E. Todd},
 journal = {The Journal of Human Resources},
 number = {1},
 pages = {93-122},
 publisher = {[University of Wisconsin Press, Board of Regents of the University of Wisconsin System]},
 title = {Do Conditional Cash Transfers for Schooling Generate Lasting Benefits? A Five-Year Followup of PROGRESA/Oportunidades},
 volume = {46},
 year = {2011}
}

@article{ministry_of_education_compulsory_1986,
	title = {Compulsory Education Law of the People's Republic of China (Adopted at the Fourth Session of the Sixth National People's Congress April 12, 1986 and promulgated by Order No. 38 of the President of the People's Republic of China on April 12, 1986)},
	url = {http://www.china.org.cn/government/laws/2007-04/17/content_1207402.htm},
	location = {Beijing, China},
	type = {Government and Law Information Website},
	author = {{Ministry of Education}},
	urldate = {2017-12-22},
	date = {1986}
}

@article{state_council_2012,
	title = {Directive 48 (2012) of the General Office of the State Council of the People's Republic of China on Rural Compulsory Education and School Reorganization},
	location = {Beijing, China},
	url = {http://www.gov.cn/zwgk/2012-09/07/content_2218779.htm},
	type = {Government and Law Information Website},
	author = {{General Office of the State Council}},
	date = {2012},
}

@article{state_council_state_2001,
	author = {{General Office of the State Council}},
	location = {Beijing, China},
	title = {Directive 21 (2011) of the General Office of the State Council of the People's Republic of China on Basic Education Reform and Development},
	url = {http://old.moe.gov.cn//publicfiles/business/htmlfiles/moe/moe_16/200105/132.html},
	type = {Government and Law Information Website},
	date = {2001},
}

@article{huisman_effects_2009,
	title = {Effects of Household- and District-Level Factors on Primary School Enrollment in 30 Developing Countries},
	volume = {37},
	issn = {0305750X},
	url = {http://linkinghub.elsevier.com/retrieve/pii/S0305750X08001666},
	doi = {10.1016/j.worlddev.2008.01.007},
	pages = {179--193},
	number = {1},
	journaltitle = {World Development},
	author = {Huisman, Janine and Smits, Jeroen},
	urldate = {2017-02-11},
	date = {2009-01},
	langid = {english}
}

@article{hanushek_students_2008,
	title = {Do Students Care about School Quality? Determinants of Dropout Behavior in Developing Countries},
	volume = {2},
	issn = {1932-8575, 1932-8664},
	url = {http://www.journals.uchicago.edu/doi/10.1086/529446},
	doi = {10.1086/529446},
	shorttitle = {Do Students Care about School Quality?},
	pages = {69--105},
	number = {1},
	journaltitle = {Journal of Human Capital},
	author = {Hanushek, Eric A. and Lavy, Victor and Hitomi, Kohtaro},
	urldate = {2017-02-11},
	date = {2008-03},
	langid = {english}
}

@report{howley_consolidation_2011,
	location = {Boulder, {CO}},
	title = {Consolidation of Schools and Districts: What the Research Says and What It Means.},
	url = {https://eric.ed.gov/?id=ED515900},
	shorttitle = {Consolidation of Schools and Districts},
	institution = {National Education Policy Center},
    author={Howley, Craig and Johnson, Jerry and Petrie, Jennifer},
	urldate = {2017-02-03},
	date = {2011-02},
	langid = {english}
}

@article{Muralidharan_2017,
Author = {Muralidharan, Karthik and Prakash, Nishith},
Title = {Cycling to School: Increasing Secondary School Enrollment for Girls in India},
Journal = {American Economic Journal: Applied Economics},
Volume = {9},
Number = {3},
Year = {2017},
Month = {7},
Pages = {321-50},
DOI = {10.1257/app.20160004},
URL = {http://www.aeaweb.org/articles?id=10.1257/app.20160004}}

@article{liu_migrate_2016,
	title = {Migrate for education: An unintended effect of school district combination in rural China},
	volume = {40},
	pages = {192--206},
	journaltitle = {China Economic Review},
	author = {Liu, Jing and Xing, Chunbing},
	date = {2016}
}

@article{YangWang_2013,
	title = {From Extending Teaching to Remote Areas and Running School in Various Forms to Canceling Teaching Centers and Merging Schools -- A trade-off between equity and efficiency of the policy on rural compulsory education},
	volume = {34},
	pages = {25--34},
	number = {5},
	journaltitle = {Tsinghua Journal of Education},
	author = {Yang, Dong-ping and Wang, Shuai},
	date = {2013-10}
}

@article{chen_poor_2014,
	title = {Do poor students benefit from China's Merger Program? Transfer path and educational performance},
	volume = {34},
	issn = {0218-8791, 1742-6855},
	url = {http://www.tandfonline.com/doi/abs/10.1080/02188791.2013.790781},
	doi = {10.1080/02188791.2013.790781},
	shorttitle = {Do poor students benefit from China's Merger Program?},
	pages = {15--35},
	number = {1},
	journaltitle = {Asia Pacific Journal of Education},
	author = {Chen, Xinxin and Yi, Hongmei and Zhang, Linxiu and Mo, Di and Chu, James and Rozelle, Scott},
	urldate = {2016-06-29},
	date = {2014-01-02},
	langid = {english}
}

@article{liu_effect_2010,
	title = {The effect of primary school mergers on academic performance of students in rural China},
	volume = {30},
	issn = {07380593},
	url = {http://linkinghub.elsevier.com/retrieve/pii/S0738059310000659},
	doi = {10.1016/j.ijedudev.2010.05.003},
	pages = {570--585},
	number = {6},
	journaltitle = {International Journal of Educational Development},
	author = {Liu, Chengfang and Zhang, Linxiu and Luo, Renfu and Rozelle, Scott and Loyalka, Prashant},
	urldate = {2016-06-29},
	date = {2010-11},
	langid = {english}
}

@article{bartl_economisation_2013,
	title = {Economisation of the Education System in Shrinking Regions? The Demographic Responsiveness of Education Demand and Supply at Different Levels of the Education System},
	volume = {39},
	url = {http://www.comparativepopulationstudies.de/index.php/CPoS/article/viewFile/130/151},
	number = {2},
	journaltitle = {Comparative Population Studies},
	author = {Bartl, Walter},
	date = {2013}
}

@article{slee_school_2015,
	title = {School Closures as a Driver of Rural Decline in Scotland: A Problem in Pursuit of Some Evidence?},
	volume = {131},
	issn = {1470-2541, 1751-665X},
	url = {http://www.tandfonline.com/doi/full/10.1080/14702541.2014.988288},
	doi = {10.1080/14702541.2014.988288},
	shorttitle = {School Closures as a Driver of Rural Decline in Scotland},
	pages = {78--97},
	number = {2},
	journaltitle = {Scottish Geographical Journal},
	author = {Slee, Bill and Miller, Dave},
	urldate = {2016-06-29},
	date = {2015-04-03},
	langid = {english}
}

@article{mo_transfer_2012,
	title = {Transfer paths and academic performance: The primary school merger program in China},
	volume = {32},
	issn = {0738-0593},
	url = {http://www.sciencedirect.com/science/article/pii/S0738059311001581},
	doi = {10.1016/j.ijedudev.2011.11.001},
	pages = {423--431},
	number = {3},
	journaltitle = {International Journal of Educational Development},
	shortjournal = {International Journal of Educational Development},
	author = {Mo, Di and Yi, Hongmei and Zhang, Linxiu and Shi, Yaojiang and Rozelle, Scott and Medina, Alexis},
	date = {2012-05}
}

@article{mei_school_2015,
	title = {School Consolidation: Whither China's Rural Education?},
	issn = {17531403},
	url = {http://doi.wiley.com/10.1111/aswp.12053},
	doi = {10.1111/aswp.12053},
	shorttitle = {School Consolidation},
	pages = {138–150},
	journaltitle = {Asian Social Work and Policy Review},
	author = {Mei, Hong and Jiang, Quanbao and Xiang, Yuanyuan and Song, Xiaoping},
	urldate = {2015-06-04},
	date = {2015-03},
	langid = {english}
}

@article{duflo_2001_school,
	author = {Duflo, Esther},
	title = {Schooling and Labor Market Consequences of School Construction in Indonesia: Evidence from an Unusual Policy Experiment},
	journaltitle = {American Economic Review},
	volume = {91},
	number = {4},
	date = {2001},
	month = {9},
	pages = {795-813},
	doi = {10.1257/aer.91.4.795},
	url = {http://www.aeaweb.org/articles?id=10.1257/aer.91.4.795}
	}

@article{zhang_economic_2005,
	title = {Economic returns to schooling in urban China, 1988 to 2001},
	volume = {33},
	issn = {0147-5967},
	url = {http://www.sciencedirect.com/science/article/pii/S014759670500048X},
	doi = {10.1016/j.jce.2005.05.008},
	series = {Symposium: Poverty and Labor Markets in China},
	pages = {730--752},
	number = {4},
	journaltitle = {Journal of Comparative Economics},
	shortjournal = {Journal of Comparative Economics},
	author = {Zhang, Junsen and Zhao, Yaohui and Park, Albert and Song, Xiaoqing},
	date = {2005-12-01}
}

@article{yu_yi_2010,
	title = {An American Chinese Teacher Investigates Chinese School Consolidation Program},
	url = {http://news.163.com/10/0222/03/603LT0B6000120GR.html},
	journaltitle = {China Youth Daily},
	location = {Beijing, China},
	author = {Yu, Lei and Jiang, Shuai},
	date = {2010-02-22}
}

@article{wu_tingting_2014,
	title = {Tingting Shang Xue Neng Bu Neng Jin Yi Dian [Can Tingting go to a closer school--An investigation from Inner Mongolia]},
	url = {http://cpc.people.com.cn/n/2014/0625/c83083-25197357.html},
	journaltitle = {People's Daily},
	location = {Beijing, China},
	author = {Wu, Yong},
	date = {2014-06-25},
	pages = {23}
}

@article{ding2015dismantling,
  title={Dismantling teaching points: integrate education resources or reduce investment in education},
  author={Ding, Dong and Zheng, Fengtian},
  journal={China Economic Quarterly},
  volume={14},
  number={2},
  pages={603--622},
  year={2015}
}

@article{cai_has_2017,
	title = {Has the compulsory school merger program reduced the welfare of rural residents in {China}?},
	volume = {46},
	issn = {1043-951X},
	url = {https://www.sciencedirect.com/science/article/pii/S1043951X17301013},
	doi = {10.1016/j.chieco.2017.07.010},
	language = {en},
	urldate = {2021-07-19},
	journal = {China Economic Review},
	author = {Cai, Weixian and Chen, Gong and Zhu, Feng},
	month = dec,
	year = {2017},
	pages = {123--141},
}

@article{tawie_sarawak_2017,
	title = {Sarawak to start merging low enrolment schools next year},
	url = {http://www.themalaymailonline.com/print/malaysia/sarawak-to-start-merging-low-enrolment-schools-next-year-says-minister},
	journal = {Malay Mail},
	location = {Kuala Lumpur, Malaysia},
	type = {Newspaper},
	author = {Tawie, Sulok},
	urldate = {2017-11-13},
	date = {2017-09-06}
}

@article{setiawati_schools_2010,
	title = {Schools merger to boost efficiency},
	url = {https://www.pressreader.com/indonesia/the-jakarta-post/20100301/282024733423092},
	journal = {The Jakarta Post},
	location = {Jakarta, Indonesia},
	type = {Press Reader},
	author = {Setiawati, Indah},
	urldate = {2017-11-13},
	date = {2010-03-01}
}

@article{zhao_increasingly_2015,
	title = {The increasingly long road to school in rural China: the impacts of education network consolidation on broadly defined schooling distance in Xinfeng County of rural China},
	volume = {16},
	issn = {1598-1037, 1876-407X},
	url = {http://link.springer.com/10.1007/s12564-015-9380-y},
	doi = {10.1007/s12564-015-9380-y},
	shorttitle = {The increasingly long road to school in rural China},
	pages = {413--431},
	number = {3},
	journaltitle = {Asia Pacific Education Review},
	author = {Zhao, Dan and Barakat, Bilal},
	urldate = {2017-10-25},
	date = {2015-09},
	langid = {english}
}

@article{fan_reasons_2013,
	title = {The Reasons, Motivation, and Selection of Approach to the Consolidation of Primary and Secondary Schools in Rural Areas},
	volume = {46},
	issn = {1061-1932},
	url = {http://www.tandfonline.com/doi/full/10.2753/CED1061-1932460501},
	doi = {10.2753/CED1061-1932460501},
	pages = {9--20},
	number = {5},
	journaltitle = {Chinese Education \& Society},
	author = {Fan, Xianzuo},
	urldate = {2017-10-25},
	date = {2013-09-01}
}

@article{xie_consolidating_2013,
	title = {Consolidating Rural Schools in China: Policy, Issues, and Debates: Guest Editors' Introduction},
	volume = {46},
	issn = {1061-1932},
	url = {http://www.tandfonline.com/doi/full/10.2753/CED1061-1932460500},
	doi = {10.2753/CED1061-1932460500},
	shorttitle = {Consolidating Rural Schools in China},
	pages = {3--8},
	number = {5},
	journaltitle = {Chinese Education \& Society},
	author = {Xie, Ailei and Wu, Zhihui},
	urldate = {2017-10-25},
	date = {2013-09-01}
}

@article{liu_closures_2013,
	title = {Closures and Consolidation: The Effective Path to Improving the Overall Quality of Basic Education},
	volume = {46},
	issn = {1061-1932},
	url = {http://www.tandfonline.com/doi/full/10.2753/CED1061-1932460503},
	doi = {10.2753/CED1061-1932460503},
	shorttitle = {Closures and Consolidation},
	pages = {36--55},
	number = {5},
	journaltitle = {Chinese Education \& Society},
	author = {Liu, Haiming and Gaowa, Naren and Wang, Shu},
	urldate = {2017-10-25},
	date = {2013-09-01}
}

@article{dai_cost_2017,
	title = {The Cost of School Consolidation Policy: Implications from Decomposing School Commuting Distances in Yanqing, Beijing},
	issn = {1874-4621},
	url = {https://doi.org/10.1007/s12061-017-9238-2},
	doi = {10.1007/s12061-017-9238-2},
	shorttitle = {The Cost of School Consolidation Policy},
	pages = {1--14},
	issue = {September},
	journaltitle = {Applied Spatial Analysis and Policy},
	author = {Dai, Te-qi and Wang, Liang and Zhang, Yu-chao and Liao, Cong and Liu, Zheng-bing},
	date = {2017},
	langid = {english}
}

@article{chowdhury_cramped_2017,
	title = {Cramped classrooms, long commutes, dropouts: How Rajasthan's school mergers have hurt students},
	location = {Delhi, India},
	url = {https://scroll.in/article/835687/cramped-classrooms-long-commutes-dropouts-the-impact-of-rajasthans-school-mergers},
	shorttitle = {Cramped classrooms, long commutes, dropouts},
	titleaddon = {the Scroll},
	type = {Text},
	author = {Chowdhury, Shreya Roy},
	urldate = {2017-10-24},
	date = {2017-05-03},
	langid = {american}
}

@article{saengpassa_ministrys_2017,
	title = {Ministry's plan to merge schools criticised: Ministry predicts improved quality of education, but NGO fears policy failure.},
	url = {http://www.nationmultimedia.com/detail/national/30303271},
	location = {Bangkok, Thailand},
	journal = {The Nation},
	author = {Saengpassa, Chularat},
	urldate = {2017-10-24},
	date = {2017-01-02},
}

@article{chicago_tribune_editorial_board_case_2017,
	location = {Chicago, {IL}},
	edition = {Online: 1:06 {PM}},
	title = {The case for a new Englewood High School},
	url = {http://www.chicagotribune.com/news/opinion/editorials/ct-englewood-high-emanuel-chicago-edit-0616-md-20170615-story.html},
	journaltitle = {Chicago Tribune},
	author = {{Chicago Tribune Editorial Board}},
	urldate = {2017-07-05},
	date = {2017-06-15}
}

@article{kremer_challenge_2013,
	title = {The Challenge of Education and Learning in the Developing World},
	volume = {340},
	issn = {0036-8075, 1095-9203},
	url = {http://www.sciencemag.org/cgi/doi/10.1126/science.1235350},
	doi = {10.1126/science.1235350},
	pages = {297--300},
	number = {6130},
	journaltitle = {Science},
	author = {Kremer, Michael and Brannen, Conner and Glennerster, Rachel},
	urldate = {2017-07-05},
	date = {2013-04-19},
	langid = {english}
}

@article{gershenson_effect_2015,
	title = {The Effect of Primary School Size on Academic Achievement},
	volume = {37},
	issn = {0162-3737, 1935-1062},
	url = {http://epa.sagepub.com/cgi/doi/10.3102/0162373715576075},
	doi = {10.3102/0162373715576075},
	pages = {135S--155S},
	number = {1},
	journaltitle = {Educational Evaluation and Policy Analysis},
	author = {Gershenson, Seth and Langbein, Laura},
	urldate = {2017-07-05},
	date = {2015-05-01},
	langid = {english}
}

@article{berry_growing_2010,
	title = {Growing Pains: The School Consolidation Movement and Student Outcomes},
	volume = {26},
	issn = {8756-6222, 1465-7341},
	url = {https://academic.oup.com/jleo/article-lookup/doi/10.1093/jleo/ewn015},
	doi = {10.1093/jleo/ewn015},
	shorttitle = {Growing Pains},
	pages = {1--29},
	number = {1},
	journaltitle = {Journal of Law, Economics, and Organization},
	author = {Berry, Christopher and West, Martin},
	urldate = {2017-07-05},
	date = {2010-04-01},
	langid = {english}
}

@article{sacerdote_when_2012,
	title = {When the Saints Go Marching Out: Long-Term Outcomes for Student Evacuees from Hurricanes Katrina and Rita},
	volume = {4},
	issn = {1945-7782, 1945-7790},
	url = {http://pubs.aeaweb.org/doi/10.1257/app.4.1.109},
	doi = {10.1257/app.4.1.109},
	shorttitle = {When the Saints Go Marching Out},
	pages = {109--135},
	number = {1},
	journaltitle = {American Economic Journal: Applied Economics},
	author = {Sacerdote, Bruce},
	urldate = {2017-05-30},
	date = {2012-01},
	langid = {english}
}

@incollection{united_states_department_of_education_chapter_2016,
	location = {Washington, D.C.},
	title = {Elementary and Secondary Education},
	url = {https://nces.ed.gov/programs/digest/d15/ch_2.asp},
	booktitle = {Digest of Education Statistics, 2015},
	publisher = {National Center for Education Statistics},
	author = {Snyder, Thomas D. and de Brey, Cristobal  and Dillow, Sally A.},
% 	author = { {United States Department of Education} },
	urldate = {2017-05-26},
	date = {2016},
	file = {Digest of Education Statistics, 2015 - Chapter 2\: Elementary and Secondary Education:C\:\\Users\\fan\\Zotero\\storage\\TPXJ57VK\\ch_2.html:text/html}
}

@article{schwartz_small_2013,
	title = {Do small schools improve performance in large, urban districts? Causal evidence from New York City},
	volume = {77},
	issn = {00941190},
	url = {http://linkinghub.elsevier.com/retrieve/pii/S0094119013000326},
	doi = {10.1016/j.jue.2013.03.008},
	shorttitle = {Do small schools improve performance in large, urban districts?},
	pages = {27--40},
	journaltitle = {Journal of Urban Economics},
	author = {Schwartz, Amy Ellen and Stiefel, Leanna and Wiswall, Matthew},
	urldate = {2017-05-26},
	date = {2013-09},
	langid = {english}
}

@article{kearns_status_2009,
	title = {`The status quo is not an option': Community impacts of school closure in South Taranaki, New Zealand},
	volume = {25},
	issn = {07430167},
	url = {http://linkinghub.elsevier.com/retrieve/pii/S0743016708000594},
	doi = {10.1016/j.jrurstud.2008.08.002},
	shorttitle = {`The status quo is not an option'},
	pages = {131--140},
	number = {1},
	journaltitle = {Journal of Rural Studies},
	author = {Kearns, Robin A. and Lewis, Nicolas and {McCreanor}, Tim and Witten, Karen},
	urldate = {2017-05-25},
	date = {2009-01},
	langid = {english}
}

@article{blauwkamp_school_2011,
	title = {School consolidation in Nebraska: Economic efficiency vs. rural community life},
	volume = {6},
	pages = {1},
	number = {1},
	journaltitle = {Online Journal of Rural Research \& Policy},
	author = {Blauwkamp, Joan M. and Longo, Peter J. and Anderson, John},
	date = {2011}
}

@article{de_haan_school_2016,
	title = {School Consolidation and Student Achievement},
	volume = {32},
	issn = {8756-6222, 1465-7341},
	url = {https://academic.oup.com/jleo/article-lookup/doi/10.1093/jleo/eww006},
	doi = {10.1093/jleo/eww006},
	pages = {816--839},
	number = {4},
	journaltitle = {Journal of Law, Economics, and Organization},
	author = {De Haan, Monique and Leuven, Edwin and Oosterbeek, Hessel},
	urldate = {2017-05-25},
	date = {2016-11},
	langid = {english}
}

@report{beuchert_short_term_2016,
	title = {The Short-Term Effects of School Consolidation on Student Achievement: Evidence of Disruption?},
	url = {http://EconPapers.repec.org/RePEc:iza:izadps:dp10195},
	number = {10195},
	institution = {Institute for the Study of Labor ({IZA})},
	type = {{IZA} Discussion Papers},
	author = {Beuchert, Louise Voldby and Humlum, Maria and Nielsen, Helena and Smith, Nina},
	date = {2016}
}

@article{de_witte_influence_2014,
	title = {The influence of closing poor performing primary schools on the educational attainment of students},
	volume = {20},
	issn = {13803611 ({ISSN})},
	url = {https://www.scopus.com/inward/record.uri?eid=2-s2.0-84906251136&doi=10.1080%2f13803611.2014.940979&partnerID=40&md5=13a1ab26dbc037dfcbe3a30c4107ccd3},
	doi = {10.1080/13803611.2014.940979},
	pages = {290--307},
	number = {4},
	journaltitle = {Educational Research and Evaluation},
	shortjournal = {Educ. Res. Eval.},
	author = {De Witte, Kristof and Van Klaveren, Chris P.B.J.},
	date = {2014}
}

@article{deeds_organizational_2015,
	title = {Organizational ``Failure'' and Institutional Pluralism: A Case Study of an Urban School Closure},
	volume = {50},
	issn = {00420859 ({ISSN})},
	url = {https://www.scopus.com/inward/record.uri?eid=2-s2.0-84930702594&doi=10.1177%2f0042085913519337&partnerID=40&md5=c73afac5deaa0fa3cb35235cab261d6f},
	doi = {10.1177/0042085913519337},
	pages = {474--504},
	number = {4},
	journaltitle = {Urban Education},
	shortjournal = {Urban Educ.},
	author = {Deeds, Vontrese and Pattillo, Mary E.},
	date = {2015}
}

@article{carlson_charter_2016,
	title = {Charter school closure and student achievement: Evidence from Ohio},
	volume = {95},
	issn = {00941190 ({ISSN})},
	url = {https://www.scopus.com/inward/record.uri?eid=2-s2.0-84979502162&doi=10.1016%2fj.jue.2016.07.001&partnerID=40&md5=a378f43856795b032fb2e7c8d5cf83e5},
	doi = {10.1016/j.jue.2016.07.001},
	pages = {31--48},
	journaltitle = {Journal of Urban Economics},
	shortjournal = {J. Urban Econ.},
	author = {Carlson, Deven E. and Lavertu, Stephane},
	date = {2016}
}

@article{chiu_effects_20161,
	title = {The effects of school closure threats on student performance: Evidence from a natural experiment},
	volume = {16},
	issn = {19351682 ({ISSN})},
	url = {https://www.scopus.com/inward/record.uri?eid=2-s2.0-85012887821&doi=10.1515%2fbejeap-2015-0149&partnerID=40&md5=1e9a614e1d9c415db36d5a1b1fb8ef22},
	doi = {10.1515/bejeap-2015-0149},
	number = {4},
	journaltitle = {B.E. Journal of Economic Analysis and Policy},
	shortjournal = {B.E. J. Econ. Anal. Policy},
	author = {Chiu, Mingming and Joh, Sungwook and Khoo, Lawrence},
	date = {2016}
}

@article{brummet_effect_2014,
	title = {The effect of school closings on student achievement},
	volume = {119},
	issn = {00472727},
	url = {http://linkinghub.elsevier.com/retrieve/pii/S0047272714001509},
	doi = {10.1016/j.jpubeco.2014.06.010},
	pages = {108--124},
	journaltitle = {Journal of Public Economics},
	author = {Brummet, Quentin},
	urldate = {2017-03-16},
	date = {2014-11},
	langid = {english}
}

@ARTICLE{Howell_2017,
	title = {Impacts of Migration and Remittances on Ethnic Income Inequality in Rural China},
	author = {Howell, Anthony},
	year = {2017},
	journal = {{World Development}},
	volume = {94},
	number = {C},
	pages = {200-211},
	url = {https://EconPapers.repec.org/RePEc:eee:wdevel:v:94:y:2017:i:c:p:200-211}
}

@online{ChinaEduYearBook,
	title = {China Education Statistics Year Book},
	url = {http://en.moe.gov.cn/Resources/Statistics/},
	type = {Database},
	author = {{Ministry of Education}},
	urldate = {2018-01-25},
	year = {1998--2015}
}

@online{UScensus,
	title = {International Data Base},
	url = {https://www.census.gov/programs-surveys/international-programs/data/tools.html},
	type = {Database},
	author = {{United States Bureau of the Census}},
	urldate = {2018-01-25},
	year = {2017}
}

@article{Conley_Taber_2011,
	author = {Timothy G. Conley and Christopher R. Taber},
	title = {Inference with ``Difference in Differences'' with a Small Number of Policy Changes},
	journal = {The Review of Economics and Statistics},
	volume = {93},
	number = {1},
	pages = {113-125},
	year = {2011},
	doi = {10.1162/REST\_a\_00049},
	URL = {https://doi.org/10.1162/REST_a_00049}
}

@article{liu2017early,
  title={Early poverty exposure predicts young adult educational outcomes in China},
  author={Liu, Xiaoying and Hannum, Emily},
  journal={China Economic Review},
  volume={44},
  pages={79--97},
  year={2017},
  publisher={Elsevier}
}

@report{Ma2017,
  author      = {Ma, Xiang},
  title       = {Economies of Scale and Heterogeneity in the Provision of Public Goods: Evidence from School Consolidation in China},
  institution = {Jinan University, Department of Economics},
  date = {2017-03-15}
}

@article{BurdeLinden_2013,
Author = {Burde, Dana and Linden, Leigh L.},
Title = {Bringing Education to Afghan Girls: A Randomized Controlled Trial of Village-Based Schools},
Journal = {American Economic Journal: Applied Economics},
Volume = {5},
Number = {3},
Year = {2013},
Month = {7},
Pages = {27-40},
DOI = {10.1257/app.5.3.27},
URL = {http://www.aeaweb.org/articles?id=10.1257/app.5.3.27}}

@article{ANDRABI20131,
title = "Students today, teachers tomorrow: Identifying constraints on the provision of education",
journal = "Journal of Public Economics",
volume = "100",
number = "Supplement C",
pages = "1 - 14",
year = "2013",
issn = "0047-2727",
doi = "https://doi.org/10.1016/j.jpubeco.2012.12.003",
url = "http://www.sciencedirect.com/science/article/pii/S0047272712001387",
author = "Tahir Andrabi and Jishnu Das and Asim Ijaz Khwaja"
}

@article{Kazianga_2013,
Author = {Kazianga, Harounan and Levy, Dan and Linden, Leigh L. and Sloan, Matt},
Title = {The Effects of ``Girl-Friendly'' Schools: Evidence from the BRIGHT School Construction Program in Burkina Faso},
Journal = {American Economic Journal: Applied Economics},
Volume = {5},
Number = {3},
Year = {2013},
Month = {7},
Pages = {41-62},
DOI = {10.1257/app.5.3.41},
URL = {http://www.aeaweb.org/articles?id=10.1257/app.5.3.41}}

@article{kirshner_tracing_2010,
	title = {Tracing Transitions: The Effect of High School Closure on Displaced Students},
	volume = {32},
	issn = {0162-3737, 1935-1062},
	url = {http://epa.sagepub.com/cgi/doi/10.3102/0162373710376823},
	doi = {10.3102/0162373710376823},
	shorttitle = {Tracing Transitions},
	pages = {407--429},
	number = {3},
	journaltitle = {Educational Evaluation and Policy Analysis},
	author = {Kirshner, Ben and Gaertner, Matthew and Pozzoboni, Kristen},
	urldate = {2015-04-23},
	date = {2010-09-01},
	langid = {english}
}

@article{hanushek_disruption_2004,
	title = {Disruption versus Tiebout improvement: the costs and benefits of switching schools},
	volume = {88},
	issn = {00472727},
	url = {http://linkinghub.elsevier.com/retrieve/pii/S004727270300063X},
	doi = {10.1016/S0047-2727(03)00063-X},
	shorttitle = {Disruption versus Tiebout improvement},
	pages = {1721--1746},
	number = {9},
	journaltitle = {Journal of Public Economics},
	author = {Hanushek, Eric A. and Kain, John F. and Rivkin, Steven G.},
	urldate = {2017-05-04},
	date = {2004-08},
	langid = {english}
}

@article{engberg_closing_2012,
	title = {Closing schools in a shrinking district: Do student outcomes depend on which schools are closed?},
	volume = {71},
	issn = {00941190},
	url = {http://linkinghub.elsevier.com/retrieve/pii/S009411901100060X},
	doi = {10.1016/j.jue.2011.10.001},
	shorttitle = {Closing schools in a shrinking district},
	pages = {189--203},
	number = {2},
	journaltitle = {Journal of Urban Economics},
	author = {Engberg, John and Gill, Brian and Zamarro, Gema and Zimmer, Ron},
	urldate = {2017-03-16},
	date = {2012-03},
	langid = {english}
}

@online{brazil_ministry_of_education_sinopses_2020,
	title = {Sinopses Estatísticas da Educação Básica  (Synopses of Basic Education Statistics)},
	url = {http://portal.inep.gov.br/web/guest/sinopses-estatisticas-da-educacao-basica},
	titleaddon = {National Institute of Educational Studies and Research ({INEP})},
	type = {Government Statistics},
	author = {{Brazil Ministry of Education}},
	urldate = {2020-02-01},
	date = {2020}
}
\endgroup
\pagebreak

\processdelayedfloats

\appendix

\makeatletter
\efloat@restorefloats
\makeatother

\setlength{\footnotemargin}{5.75mm}
\begingroup
\doublespacing
\centering
\Large ONLINE APPENDIX \\
\Large\begin{singlespace}\href{\PAPERDOIURL}{\PAPERTITLE}\end{singlespace}
\large \AUTHORHANNUM, \AUTHORLIU, and \AUTHORWANG\\[1.0em]
\endgroup


\section{Additional Data Details \label{sec:amoredata}}
\renewcommand{\thefigure}{A.\arabic{figure}}
\setcounter{figure}{0}
\renewcommand{\thetable}{A.\arabic{table}}
\setcounter{table}{0}
\renewcommand{\theequation}{A.\arabic{equation}}
\setcounter{equation}{0}
\renewcommand{\thefootnote}{A.\arabic{footnote}}
\setcounter{footnote}{0}

\vspace*{-0.5\baselineskip}
\begin{figure}[H]
\centering
\caption[Survey Prefecture Map]{Survey Prefectures Containing Survey Counties\footnotemark \label{map}}
\includegraphics[width=1.2\textwidth, center]{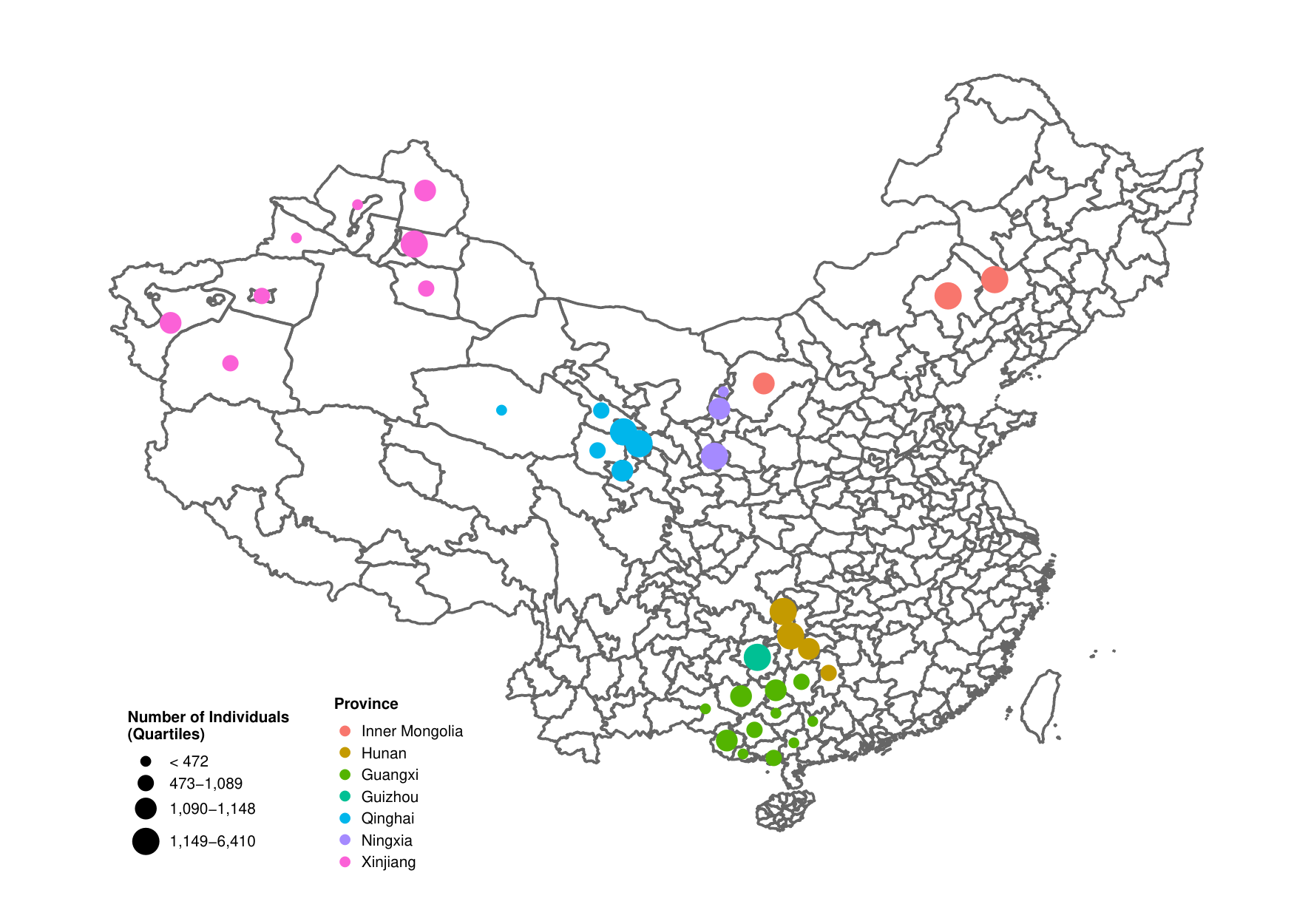}

\captionsetup{width=1.3\textwidth}
\end{figure}
\footnotetext{Reprinted with permission from \autocite{Howell_2017}.}
\vspace*{-15mm}

\subsection{Location of Closures}
\label{sec:aloc}
\begin{spacing}{1.7}
This paper utilizes data from the rural sample of the China Household Ethnic Survey (CHES 2011), which covers households and villages from 728 villages\footnote{There are 751 unique village IDs in the survey, but 17 villages do not have school closure information, and 6 villages report closure without a closure year.} in 81 counties of 7 provinces with substantial minority populations in western China: Qinghai Province (119 villages surveyed); Ningxia Hui Autonomous Region (97 villages surveyed); Xinjiang Uygur Autonomous Region (94 villages surveyed); Inner Mongolia Autonomous Region (100 villages surveyed); Qiandongnan Miao and Dong Autonomous Prefecture in Guizhou Province (120 villages surveyed); Hunan Province (101 villages surveyed); and Guangxi Zhuang Autonomous Region (103 villages surveyed).
\end{spacing}

\vspace*{-0.5\baselineskip}
\subsection{Sampling Procedure}
\label{sec:asampling}
\begin{spacing}{1.7}
The CHES survey is the largest-scale cross-province survey ever gathered to study the socio-economic conditions of minorities in ethnically diverse regions of China. It was designed by China's Academy of Social Sciences and the Central Nationalities University, and it was administrated by local offices of the National Bureau of Statistics (NBS). The survey villages were selected based on a subset of the NBS's Rural Household Survey (RHS) \autocite{national_bureau_of_statistics_of_china_china_2012}. The villages selected are not representative of their respective provinces and autonomous regions overall, but are from minority rich prefectures in order for the survey to capture socio-economic conditions of minorities and Han individuals in ethnically diverse areas of each province and autonomous regions. Households in villages were selected by systematic sampling based on their agricultural census address codes. The survey was implemented in early 2012 and asked households to report information from the end of 2011. Household surveys are complemented by surveys of villages heads from the sampled villages. \textcite{gustafsson_ethnicity_2019} provide more information on the sampling procedure and other information related to the survey.
\end{spacing}

\vspace*{-0.5\baselineskip} \subsection{Additional Summary Statistics}

\begin{spacing}{\aptxspc}
To complement Table \ref{summtwo}, in Panels A, B, C and D of Appendix Table \ref{summtwob}, we test how villages with and without school closure differ along several other dimensions. Panel A shows that non-closure villages are more likely to have someone originally from the village working at the county or higher level of government. The first variable shows if the village has any contacts in governments above the county level, and there is no statistical differences between closure and non-closure villages for this measure controlling for provincial fixed effects. For the second variable, however, we find that 45 percent of villages with closure have someone from the village at county or above county level governments, but 61 percent of the villages without closure do. The difference is significant.

In Panel B, we do not find any statistically significant differences between closure and non-closure villages for income and labor market variables. Villages that have closure have 2 percent higher average net annual income (4503 Yuan vs 4421 Yuan in 2011), and 4 percent higher average daily wage than villages without closure (85.24 Yuan vs 82.19 Yuan).\footnote{1 Dollar = 6.5 Yuan in 2011.} These differences are not statistically different. Additionally, villages with closure have 26 percent of the population working as migrant workers, compared to 23 percent in villages without closure. Males migrant workers account for 16 and 15 percent of the population in closure and non-closure villages respectively. The vast majority of those not working as migrant workers are agricultural workers/farmers, accounting for 65.3 percent and 63.7 percent of the labor force in villages with and without closure respectively.

\begin{table}[htbp]
\centering
\def\sym#1{\ifmmode^{#1}\else\(^{#1}\)\fi}
\caption{Additional Summary Statistics for Village Characteristics\label{summtwob}}
\begin{adjustbox}{max width=1\textwidth}
\begin{tabular}{m{10cm} >{\centering\arraybackslash}m{1.5cm} >{\centering\arraybackslash}m{1.5cm} >{\centering\arraybackslash}m{1.5cm} >{\centering\arraybackslash}m{1.5cm} >{\centering\arraybackslash}m{1.5cm}}
\toprule
& \multicolumn{5}{L{8.5cm}}{Villages with and without school closures} \\
\cmidrule(l{5pt}r{5pt}){2-6}
& \multicolumn{1}{C{1.5cm}}{\small \textbf{all}} & \multicolumn{2}{C{3cm}}{\small \textbf{group averages}} & \multicolumn{2}{C{3cm}}{\small \textbf{p-values testing}} \\
\cmidrule(l{5pt}r{5pt}){2-2} \cmidrule(l{5pt}r{5pt}){3-4} \cmidrule(l{5pt}r{5pt}){5-6}
& \multicolumn{1}{C{1.5cm}}{\textit{\small mean}} & \multicolumn{1}{C{1.5cm}}{\textit{\small non-closure}} & \multicolumn{1}{C{1.5cm}}{\textit{\small closure}} & \multicolumn{1}{C{1.5cm}}{\textit{\small closure vs non-closure\dag}} & \multicolumn{1}{C{1.5cm}}{\textit{\small years of closure trend\ddag}} \\
\midrule
\formatpanelheadsumm{6}{L{20cm}}{Panel A: Leadership}
Village has contact at above county level&    0.62&    0.63&    0.58&    0.40&    0.86\\
There are villagers who became officials at county level&    0.56&    0.61&    0.45&    0.00&    0.06\\
\midrule
\formatpanelheadsumm{6}{L{20cm}}{Panel B: Income, Wage, Migration}
Per capita annual net income (Yuan) 2011&     4445.89&     4421.79&     4503.87&    0.97&    0.16\\
Local temp work wage (Yuan) 2011 & 83.08 & 82.19 & 85.24&    0.32&    0.34\\
Fraction of village labor force in agriculture & 64.16 & 63.71 & 65.28&    0.17&    0.39\\
Fraction of migrant worker in total population&    0.24&    0.23&    0.26&    0.13&    0.40\\
Fraction of male migrant worker in total population&    0.15&    0.15&    0.16&    0.20&    0.14\\
\midrule
\formatpanelheadsumm{6}{L{20cm}}{Panel C: Village Expenditures}
Per capita village budget spending (Yuan) 2011 & 69.03 & 63.13 & 83.95&    0.20&    0.47\\
Per capita spending on education (Yuan) 2011&    1.26&    1.62&    0.38&    0.40&    0.38\\
If village has education spending in 2011&    0.17&    0.18&    0.15&    0.21&    0.39\\
\midrule
\formatpanelheadsumm{6}{L{20cm}}{Panel D: Other Policies}
Village implemented Grain for Green&    0.69&    0.66&    0.74&    0.35&    0.54\\
Village has been consolidated since 1999&    0.14&    0.14&    0.14&    0.19&    0.22\\
Village has collective-owned medical station&    0.65&    0.66&    0.64&    0.64&    0.50\\
Village started rural medical insurance scheme after 2006&    0.50&    0.52&    0.45&    0.08&    0.99\\
\midrule
\bottomrule
\footnotegap
\multicolumn{6}{L{20cm}}{\footnotesize\justify\footsummstatsmain} \\
\end{tabular}\end{adjustbox}
\end{table}

In Panel D, we check whether there is a relationship between the implementation of the closure policy and three other village-level policies. The fraction of villages with collectively-owned medical stations among closure and non-closure villages is 64 percent and 66 percent. 74 percent of villages with closure implemented the Grain for Green (Grain for Green) policy,\footnote{This was a policy to convert cultivated land back into forest, as these lands were converted from forest to cultivated land before, which led to decrease in forest coverage, and caused flood, and soil erosion.} while 66 percent of villages without closure had. 45 percent of villages with closure had implemented cooperative medical insurance after 2006\footnote{The cooperative medical insurance policy started in 2004. By the end of 2006, 51 percent of villages in our sample had this insurance program. 99 percent of our sample villages had the program by the end of 2009.}, and 52 percent of villages without closure had. Controlling for provincial fixed effects, the differences between these variables in closure and non-closure villages are not significant. Overall, in our sample of villages, it seems that the closure decision is unrelated to the level of village economic development and other socio-economic policies, but might be partly driven by the size of villages (or school size) as discussed in Section \ref{sec:closecomp} of the paper.

\end{spacing}

\vfill
\pagebreak
 \subsection{Grades Completed by 2011 for Males and Females}
\label{sec:agrpedu}

\begin{spacing}{\aptxspc}
Table \ref{summthree} shows average grades completed by 2011 for all individuals. Appendix Table \ref{summfive} presents information on grades completed by 2011 for males and females separately.
\end{spacing}

\begin{spacing}{1.0}
\begin{table}[H]
\centering
\def\sym#1{\ifmmode^{#1}\else\(^{#1}\)\fi}
\caption{Summary Statistics for Educational Attainment\label{summfive}}
\begin{adjustbox}{max width=1\textwidth}
\begin{tabular}{m{\lablcolwidthsummeight} >{\centering\arraybackslash}m{1.5cm} >{\centering\arraybackslash}m{1.5cm} >{\centering\arraybackslash}m{1.5cm} >{\centering\arraybackslash}m{1.5cm} >{\centering\arraybackslash}m{1.5cm} >{\centering\arraybackslash}m{1.5cm} >{\centering\arraybackslash}m{1.5cm} >{\centering\arraybackslash}m{1.5cm}}
\toprule
& \multicolumn{8}{L{\innerheadwidthsummeight}}{Age at village-specific year of closure and 2011 age}\\
\cmidrule(l{5pt}r{5pt}){2-9} & \multicolumn{2}{C{3.0cm}}{\small\textit{Age in 2011}} & \multicolumn{2}{C{3.0cm}}{\small\textit{Age in 2011}} & \multicolumn{2}{C{3.0cm}}{\small \textit{Age in 2011}} & \multicolumn{2}{C{3.0cm}}{\small\textit{Age in 2011}} \\
\cmidrule(l{5pt}r{5pt}){2-3} \cmidrule(l{5pt}r{5pt}){4-5} \cmidrule(l{5pt}r{5pt}){6-7} \cmidrule(l{5pt}r{5pt}){8-9}
& \multicolumn{1}{C{1.5cm}}{\textbf{\small 0-4}} & \multicolumn{1}{C{1.5cm}}{\textbf{\small 5-9}} & \multicolumn{1}{C{1.5cm}}{\textbf{\small 10-14}} & \multicolumn{1}{C{1.5cm}}{\textbf{\small 15-19}} & \multicolumn{1}{C{1.5cm}}{\textbf{\small 20-24}} & \multicolumn{1}{C{1.5cm}}{\textbf{\small 25-29}} & \multicolumn{1}{C{1.5cm}}{\textbf{\small 30-34}} & \multicolumn{1}{C{1.5cm}}{\textbf{\small 35-44}} \\
\midrule
\formatpanelheadsumm{9}{L{\subheadwidthsummeight}}{Group A: age 1 to 5 at year of closure}
Number of grades completed: \textit{Female} & 0.083&    1.00&    5.37&    9.64&   &   &   &    \\
Number of grades completed: \textit{Male} & 0.082&    1.20 & 5&    9.80&   &   &   &    \\
Fraction completed middle school: \textit{Female} & 0 & 0 & 0.032&    0.91&   &   &   &    \\
Fraction completed middle school: \textit{Male}&      0.0058 & 0 & 0.031&    0.80&   &   &   &    \\
\midrule
Observations   &      303&     333&     126&      16&  &  &   & \\
\midrule
\formatpanelheadsumm{9}{L{\subheadwidthsummeight}}{Group B: age 6 to 9 at year of closure}
Number of grades completed: \textit{Female}&   &    2.24&    5.58&      10&    8.80&   &   &    \\
Number of grades completed: \textit{Male}&   &    1.66&    5.43&    9.60&    11.5&   &   &    \\
Fraction completed middle school: \textit{Female}&&  0 & 0.018&    0.74&    0.80&   &   &    \\
Fraction completed middle school: \textit{Male}&&  0 & 0.010&    0.69&    0.83&   &   &    \\
\midrule
Observations   &   &      98&     211&      69&      17&  &   & \\
\midrule
\formatpanelheadsumm{9}{L{\subheadwidthsummeight}}{Group C: age 10 to 13 at year of closure}
Number of grades completed: \textit{Female}&   &   &    5.85&    9.27&    10.4 & 8&   &    \\
Number of grades completed: \textit{Male}&   &   &    6.09&    9.49&    10.3 & 9&   &    \\
Fraction completed middle school: \textit{Female}&   &&  0.081&    0.68&    0.80 & 0&   &    \\
Fraction completed middle school: \textit{Male}&   &&  0.042&    0.71&    0.84 & 1&   &    \\
\midrule
Observations   &   &  &     133&     224&     117 & 2&   & \\
\midrule
\formatpanelheadsumm{9}{L{\subheadwidthsummeight}}{Group D: age 14 to 21 at year of closure}
Number of grades completed: \textit{Female}&   &   &    7.33&    9.90&    10.1&    8.97&    7.69&    \\
Number of grades completed: \textit{Male}&   &   &    6.57&    9.71&    10.4&    9.11&    9.21&    \\
Fraction completed middle school: \textit{Female}&   &   &    0.22&    0.80&    0.78&    0.66&    0.46&    \\
Fraction completed middle school: \textit{Male}&   &   &    0.14&    0.75&    0.82&    0.71&    0.68&    \\
\midrule
Observations   &   &  &      16&     275&     590&     241&      32 & \\
\midrule
\formatpanelheadsumm{9}{L{\subheadwidthsummeight}}{Group E: age 22 to 29 at year of closure}
Number of grades completed: \textit{Female}&   &   &   &   &    9.75&    8.41&    7.81&    7.82\\
Number of grades completed: \textit{Male}&   &   &   &   &    10.2&    9.11&    8.41&    8.13\\
Fraction completed middle school: \textit{Female}&   &   &   &   &    0.75&    0.59&    0.48&    0.42\\
Fraction completed middle school: \textit{Male}&   &   &   &   &    0.84&    0.68&    0.61&    0.51\\
\midrule
Observations   &   &  &  &  &     101&     460&     322&      98\\
\midrule
\formatpanelheadsumm{9}{L{\subheadwidthsummeight}}{Group F: individuals from non-closure villages}
Number of grades completed: \textit{Female}&    0.11&    1.47&    5.58&    9.31&    9.48&    8.01&    6.64&    5.88\\
Number of grades completed: \textit{Male} & 0.070&    1.28&    5.52&    9.35&    9.92&    8.80&    7.98&    7.71\\
Fraction completed middle school: \textit{Female}&      0.0057 & 0 & 0.048&    0.71&    0.71&    0.52&    0.41&    0.27\\
Fraction completed middle school: \textit{Male}&      0.0047 & 0 & 0.040&    0.73&    0.75&    0.63&    0.54&    0.47\\
\midrule
Observations   &      783&    1227&    1521&    1774&    2237&    1713&    1420&    1569\\
\bottomrule
\footnotegap
\multicolumn{9}{L{\footwidthsummeight}}{\small \justify \footeduattain} \\
\end{tabular}
\end{adjustbox}
\end{table}
 \end{spacing}
 \clearpage
\pagebreak

\section{Age Grouping and Compositions \label{sec:amoreage}}
\renewcommand{\thefigure}{B.\arabic{figure}}
\setcounter{figure}{0}
\renewcommand{\thetable}{B.\arabic{table}}
\setcounter{table}{0}
\renewcommand{\theequation}{B.\arabic{equation}}
\setcounter{equation}{0}
\renewcommand{\thefootnote}{B.\arabic{footnote}}
\setcounter{footnote}{0}

\subsection{Age of Enrollment}
\label{sec:enrollage}
\begin{spacing}{\aptxspc}
Table \ref{summenroll} shows the age pattern of primary school enrollment in 2011 in the CHES data.\footnote{School closure took place in the decade preceding 2011. In this study we define age cutoffs for the educational attainment and school enrollment regressions based on primary school enrollment age patterns in our data from 2011, the only year in which we have enrollment data.} We summarize the enrollment rates in primary school and lower middle school at each age separately for all children (Panel A), for children by gender (Panels B and C) and ethnicity subgroups (Panels D and E). Overall enrollment rates into primary school jumps from 11 percent at age 5 to 49 percent at age 6, and are on average 76 percent between ages 6 to 8 as shown in Panel A. The primary enrollment rates decrease from 79 percent at age 12 to less than 50 percent at age 13, and further decrease to 15 percent age age 14. We observe higher enrollment rates for boys than girls at age 5, but otherwise similar enrollment rates across genders between ages 6 and 15. Across ethnic groups, we see a lower proportion of 5 year old ethnic minorities enrolled in primary schools (compared to Han children), and a higher proportion of Han children between ages 12 and 15 enrolled in middle school and beyond. Despite these subgroup differences, overall, children are predominantly enrolled in primary schools between 6 and 13 years of age in 2011.

Given this pattern, we consider children who were between 6 and 13 years of age at the year of closure in their respective villages as individuals who were exposed to school closure during years in which they were most likely to have been enrolled in primary schools.  Additionally, we consider children age 5 or lower at year of closure as individuals who most likely did not experience the previously closed village primary school. Finally, we consider individuals age 14 or above at year of closure as those who were most likely to have been too old to be directly influenced by school closure. Given that there is still a small fraction of children 14 and 15 years of age in 2011 who were enrolled in primary schools, classifying these individuals as above the closure impact cut-off might lead to a dampening of the estimates in Table \ref{regone} and related attainment tables. To test the robustness of our results against different age cutoffs, we re-estimate Equation \eqref{eq:startsOnly} and show attainment results for finer age-at-closure subgroups in Figure \ref{figoneb}, with individuals who were 28--29 years old at year of closure as the baseline group. We see a jump between the 12--13 and 14--15 year old age-at-closure sub-groups in the figure for girls. There is no statistically significant impact of closure on the 14--15 year old age-at-closure subgroup compared to the baseline group, while the effects of closure on attainment are significant and negative for the 12--13 year old (female) age-at-closure subgroup.

\begin{table}[t]
\centering
\caption{Enrollment in Primary and Lower Middle School by Age in 2011\label{summenroll}}
\begin{adjustbox}{max width=1\textwidth}
\begin{tabular}{m{\lablcolwidthsummeight} >{\centering\arraybackslash}m{1.5cm} >{\centering\arraybackslash}m{1.5cm} >{\centering\arraybackslash}m{1.5cm} >{\centering\arraybackslash}m{1.5cm} >{\centering\arraybackslash}m{1.5cm} >{\centering\arraybackslash}m{1.5cm} >{\centering\arraybackslash}m{1.5cm} >{\centering\arraybackslash}m{1.5cm}}
\toprule
& \multicolumn{8}{L{\innerheadwidthsummeight}}{Outcome: fraction of children enrolled (\%)} \\
\cmidrule(l{5pt}r{5pt}){2-9}
& \multicolumn{2}{C{3cm}}{\small Entry Ages} & \multicolumn{2}{C{3cm}}{\small Primary Ages} & \multicolumn{4}{C{6cm}}{\small Primary End Ages} \\
\cmidrule(l{5pt}r{5pt}){2-3} \cmidrule(l{5pt}r{5pt}){4-5} \cmidrule(l{5pt}r{5pt}){6-9}     &  \multicolumn{1}{C{1.5cm}}{{\small 4}} & \multicolumn{1}{C{1.5cm}}{{\small 5}} & \multicolumn{1}{C{1.5cm}}{{\small 6-8}} & \multicolumn{1}{C{1.5cm}}{{\small 9-11}} & \multicolumn{1}{C{1.5cm}}{{\small 12}} & \multicolumn{1}{C{1.5cm}}{{\small 13}} & \multicolumn{1}{C{1.5cm}}{{\small 14}} & \multicolumn{1}{C{1.5cm}}{{\small 15}} \\
\midrule
\formatpanelheadsumm{9}{L{\subheadwidthsummeight}}{Panel A: All Children}
Enrolled in primary school&    0.04&    0.11&    0.76&    0.93&    0.79&    0.45&    0.15&    0.04\\
Enrolled in lower middle school and above&    0.00&    0.00&    0.01&    0.03&    0.15&    0.51&    0.75&    0.84\\
\midrule
Observations   &      306&     335&     889&     920&     316&     366&     341&     357\\
\midrule
\formatpanelheadsumm{9}{L{\subheadwidthsummeight}}{Panel B: Female Children}
Enrolled in primary school&    0.01&    0.08&    0.75&    0.95&    0.78&    0.42&    0.15&    0.04\\
Enrolled in lower middle school and above&    0.00&    0.00&    0.01&    0.02&    0.18&    0.55&    0.75&    0.86\\
\midrule
Observations   &      141&     129&     422&     432&     147&     166&     184&     180\\
\midrule
\formatpanelheadsumm{9}{L{\subheadwidthsummeight}}{Panel C: Male Children}
Enrolled in primary school&    0.05&    0.13&    0.77&    0.92&    0.80&    0.47&    0.15&    0.05\\
Enrolled in lower middle school and above&    0.00&    0.00&    0.01&    0.04&    0.13&    0.47&    0.76&    0.83\\
\midrule
Observations   &      165&     206&     467&     488&     169&     200&     157&     177\\
\midrule
\formatpanelheadsumm{9}{L{\subheadwidthsummeight}}{Panel D: Minority Children}
Enrolled in primary school&    0.04&    0.08&    0.77&    0.93&    0.84&    0.49&    0.16&    0.05\\
Enrolled in lower middle school and above&    0.00&    0.00&    0.01&    0.03&    0.11&    0.47&    0.72&    0.82\\
\midrule
Observations   &      210&     241&     621&     632&     218&     239&     231&     239\\
\midrule
\formatpanelheadsumm{9}{L{\subheadwidthsummeight}}{Panel E: Han Children}
Enrolled in primary school&    0.03&    0.18&    0.74&    0.94&    0.69&    0.39&    0.12&    0.03\\
Enrolled in lower middle school and above&    0.00&    0.00&    0.01&    0.03&    0.24&    0.57&    0.82&    0.88\\
\midrule
Observations    &  96 & 94&      268&      288 & 98&      127&      110&      118\\
\bottomrule
\footnotegap
\multicolumn{9}{p{\footwidthsummeight}}{\small\justify}\\
\end{tabular}
\end{adjustbox}
\end{table}

In Section \ref{sec:mechanism}, given our interests in studying the association between distance to primary school and the quality of primary school facility on primary school enrollment, we focus on school enrollment rates between ages 5 and 12 in 2011. Based on Table \ref{summenroll} statistics across gender and ethnicity, age 5 is included because that is the first age in which some children begin attending primary school, and age 13 is excluded because a substantial portion of children are in lower middle school at age 13.

\end{spacing}
\clearpage
\pagebreak

\subsection{Age Effects with Finer Age-at-closure Breakdowns \label{sec:amorereg}}
\begin{spacing}{\aptxspc}
Appendix Figure \ref{figoneb} shows the same style of graph as Figure \ref{figone} except now with even finer age breakdowns, grouping every 2 closure-year ages together on the x-axis. There is more instability in the coefficients due to the smaller sample size for each group, but we see the same pattern as in Figure \ref{figone}. For girls, there is a significant drop in the trend line at closure-age 12 to 13, coefficients are insignificant from 0 above this closure-age, but significantly negative below this closure-age group. For boys, there is a flat trend along the x-axis.

\vspace*{+2mm}
\begin{figure}[H]
	\centering
	\caption{\small Effect of School Closure on Educational Attainment (Number of Grades Completed by 2011) by 15 Age-at-Closure Groups.}
	\includegraphics[width=1.0\textwidth, center]{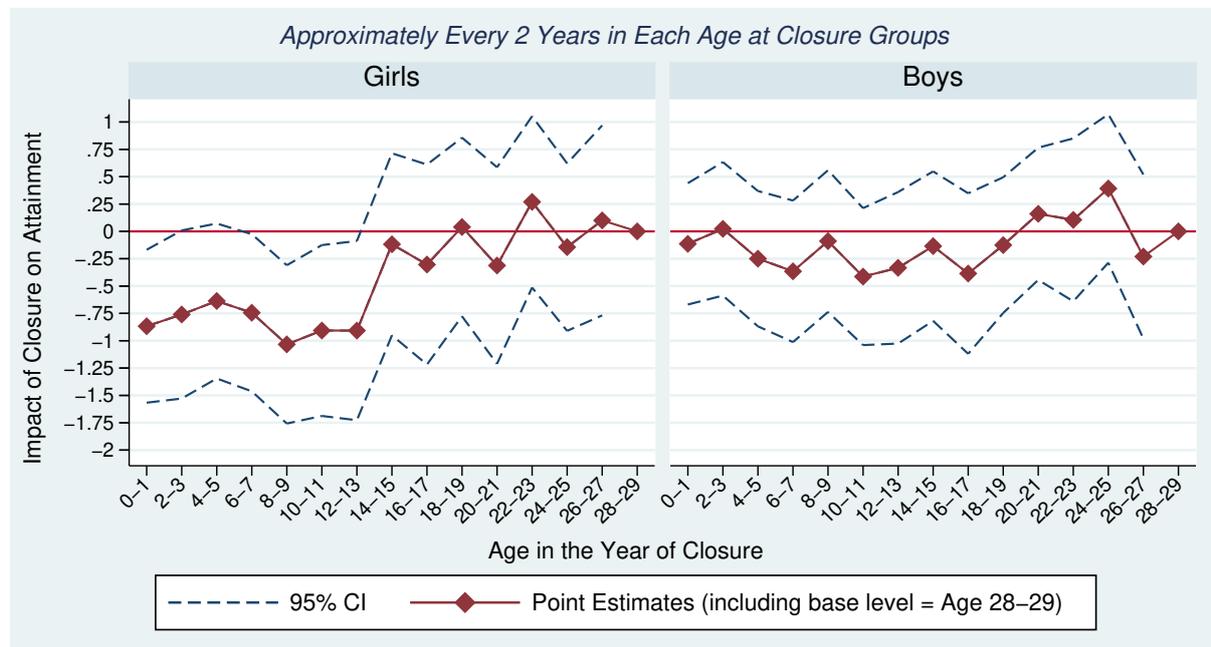}
	\captionsetup{width=1.0\textwidth}
	\caption*{\footnotesize \footgraphfiner \label{figoneb}}
\end{figure}
\end{spacing}
\clearpage

\subsection{Attainment Regression for Girls in Younger Cohorts}
\begin{spacing}{\aptxspc}
The panels of Appendix Table \ref{regyoung} show results for regression samples that only include women below 35, 30 and 25 years of age in 2011. We also exclude individuals who were 0 to 5 or 22 to 30 years-of-age at the time of school closure from villages with closure. Results are similar across panels.
\end{spacing}
\vspace*{+4mm}
\begin{spacing}{1.0}
\begin{table}[H]
\centering
\def\sym#1{\ifmmode^{#1}\else\(^{#1}\)\fi}
\caption{{\small Effect of School Closure on Educational Attainment for Restricted Age Cohorts\label{regyoung}}}
\begin{adjustbox}{max width=\regone\textwidth}
\begin{tabular}{m{\lablcolwidth} >{\centering\arraybackslash}m{1.5cm} >{\centering\arraybackslash}m{1.5cm} >{\centering\arraybackslash}m{1.5cm} >{\centering\arraybackslash}m{1.5cm} >{\centering\arraybackslash}m{1.5cm} >{\centering\arraybackslash}m{1.5cm}}
\toprule
&    \multicolumn{6}{L{\innerheadwidth}}{Outcome: grades completed by year 2011} \\
\cmidrule(l{5pt}r{5pt}){2-7}
&    \multicolumn{2}{C{3.0cm}}{\footnotesize } & \multicolumn{2}{C{3.0cm}}{\footnotesize \(10 \le 2011 \text{ Age} \le X\)} & \multicolumn{2}{C{3.0cm}}{\footnotesize \(15 \le 2011 \text{ Age} \le X\)} \\
\cmidrule(l{5pt}r{5pt}){2-3}\cmidrule(l{5pt}r{5pt}){4-5}\cmidrule(l{5pt}r{5pt}){6-7}
&   \multicolumn{1}{C{1.5cm}}{(1)} & \multicolumn{1}{C{1.5cm}}{(2)} & \multicolumn{1}{C{1.5cm}}{(3)} & \multicolumn{1}{C{1.5cm}}{(4)} & \multicolumn{1}{C{1.5cm}}{(5)} & \multicolumn{1}{C{1.5cm}}{(6)} \\
\midrule
\formatpanelheadsupregs{7}{L{15cm}}{\baselinegroup}
\formatpanelheadregs{7}{L{15cm}}{Panel A: Female Below 35 in 2011}
\closeinteragestart{} 6--9 & -0.54\sym{**} & -0.62\sym{**} & -0.54\sym{**} & -0.65\sym{**} &   &    \\
& \vspace*{-2mm}{\footnotesize (0.25) } &\vspace*{-2mm}{\footnotesize (0.27) } &\vspace*{-2mm}{\footnotesize (0.28) } &\vspace*{-2mm}{\footnotesize (0.30) } &   &    \\
\closeinteragestart{} 10--13 & -0.55\sym{**} & -0.61\sym{**} & -0.54\sym{**} & -0.60\sym{**} & -0.54\sym{*}  & -0.59\sym{*}  \\
& \vspace*{-2mm}{\footnotesize (0.24) } &\vspace*{-2mm}{\footnotesize (0.27) } &\vspace*{-2mm}{\footnotesize (0.25) } &\vspace*{-2mm}{\footnotesize (0.28) } &\vspace*{-2mm}{\footnotesize (0.30) } &\vspace*{-2mm}{\footnotesize (0.33) }    \\
\midrule
Observations  &     6485 &    5472 &    5316 &    4492 &    4310 &    3648    \\
\midrule
\formatpanelheadregs{7}{L{15cm}}{Panel B: Female Below 30 in 2011}
\closeinteragestart{} 6--9 & -0.59\sym{**} & -0.66\sym{**} & -0.58\sym{**} & -0.66\sym{**} &   &    \\
& \vspace*{-2mm}{\footnotesize (0.25) } &\vspace*{-2mm}{\footnotesize (0.28) } &\vspace*{-2mm}{\footnotesize (0.28) } &\vspace*{-2mm}{\footnotesize (0.31) } &   &    \\
\closeinteragestart{} 10--13 & -0.59\sym{**} & -0.62\sym{**} & -0.58\sym{**} & -0.62\sym{**} & -0.63\sym{**} & -0.62\sym{*}  \\
& \vspace*{-2mm}{\footnotesize (0.24) } &\vspace*{-2mm}{\footnotesize (0.27) } &\vspace*{-2mm}{\footnotesize (0.25) } &\vspace*{-2mm}{\footnotesize (0.28) } &\vspace*{-2mm}{\footnotesize (0.30) } &\vspace*{-2mm}{\footnotesize (0.33) }    \\
\midrule
Observations  &     5769 &    4879 &    4758 &    4029 &    3752 &    3185    \\
\midrule
\formatpanelheadregs{7}{L{15cm}}{Panel C: Female Below 25 in 2011}
\closeinteragestart{} 6--9 & -0.59\sym{**} & -0.73\sym{**} & -0.58\sym{*}  & -0.74\sym{**} &   &    \\
& \vspace*{-2mm}{\footnotesize (0.28) } &\vspace*{-2mm}{\footnotesize (0.31) } &\vspace*{-2mm}{\footnotesize (0.32) } &\vspace*{-2mm}{\footnotesize (0.35) } &   &    \\
\closeinteragestart{} 10--13 & -0.51\sym{**} & -0.63\sym{**} & -0.50\sym{*}  & -0.62\sym{**} & -0.49 & -0.59    \\
& \vspace*{-2mm}{\footnotesize (0.26) } &\vspace*{-2mm}{\footnotesize (0.29) } &\vspace*{-2mm}{\footnotesize (0.27) } &\vspace*{-2mm}{\footnotesize (0.30) } &\vspace*{-2mm}{\footnotesize (0.34) } &\vspace*{-2mm}{\footnotesize (0.36) }    \\
\midrule
Observations&    4900 &    4138 &    3889 &    3288 &    2883 &    2444    \\
\midrule
\exclcontrol
\exclcontrolcont
\bottomrule
\footnotegap
\multicolumn{7}{L{18.5cm}}{\footnotesize \justify  \footattainmoreage} \\
\end{tabular}
\end{adjustbox}
\end{table}
\end{spacing}
\clearpage

\subsection{Enrollment Results with Smaller Age Groups}
\begin{spacing}{\aptxspc}
Table \ref{regtwelve} shows results from enrollment regressions using more fine-grained age groups compared to Table \ref{regfive}. We find that for 5 to 8 year old girls, the presence of medium and long distance schools are associated with larger reductions in enrollment than for boys. We also find that for 5 to 8 year old boys, better facility schools are associated with increases in enrollment. For children between 9 and 12, there are no enrollment differences for girls when schools are less than 3 kilometers away, but more distant schools are associated with a reduction in girls' enrollment. For this age group, schools with better facilities are associated with small increases in enrollment for both boys and girls, but not significantly so.
\end{spacing}

\vspace*{-3mm}
\begin{spacing}{1.0}
\begin{table}[htbp]
\centering
\caption{Linear Probability Model of School Enrollment by Age Subgroups\label{regtwelve}}
\begin{adjustbox}{max width=1\textwidth}
\begin{tabular}{m{\lablcolwidthenroll} >{\centering\arraybackslash}m{1.85cm} >{\centering\arraybackslash}m{1.85cm} >{\centering\arraybackslash}m{1.85cm} >{\centering\arraybackslash}m{1.85cm} >{\centering\arraybackslash}m{1.85cm} >{\centering\arraybackslash}m{1.85cm}}
\toprule
& \multicolumn{6}{L{\innerheadwidthenroll}}{Outcome: enrolled in school or not in 2011} \\
\cmidrule(l{5pt}r{5pt}){2-7}
& \multicolumn{2}{C{3.7cm}}{\small All Age 5 to 12} & \multicolumn{2}{C{3.7cm}}{\small Girls Age 5 to 12} & \multicolumn{2}{C{3.7cm}}{\small Boys Age 5 to 12} \\                                                  \cmidrule(l{5pt}r{5pt}){2-3} \cmidrule(l{5pt}r{5pt}){4-5} \cmidrule(l{5pt}r{5pt}){6-7}                          &   \multicolumn{1}{C{1.85cm}}{{\footnotesize all villages}} & \multicolumn{1}{C{1.85cm}}{{\footnotesize no teaching points}} & \multicolumn{1}{C{1.85cm}}{{\footnotesize all villages}} & \multicolumn{1}{C{1.85cm}}{{\footnotesize no teaching points}} & \multicolumn{1}{C{1.85cm}}{{\footnotesize all villages}} & \multicolumn{1}{C{1.85cm}}{{\footnotesize no teaching points}} \\
\midrule
\formatpanelheadsupregs{7}{L{\subheadwidthenroll}}{categorical distance and quality measures}
\formatpanelheadregs{7}{L{\subheadwidthenroll}}{Age 5--8 (2011) $\times$}
\formatpanelheadsubregs{7}{L{\subheadwidthenroll}}{\vspace*{0mm}\hspace*{0mm}\enrolldistbase}
\vspace*{0mm}\hspace*{5mm}$0<x\leq3\hspace{0.1cm}(median\approx2)\hspace{0.1cm}km$&      -0.048\sym{*} &      -0.061\sym{**} &      -0.055 &      -0.095\sym{**} &      -0.070\sym{*} &      -0.065\sym{*}  \\
 & \vspace*{-2mm}{\footnotesize (0.027) } &\vspace*{-2mm}{\footnotesize (0.029) } &\vspace*{-2mm}{\footnotesize (0.039) } &\vspace*{-2mm}{\footnotesize (0.045) } &\vspace*{-2mm}{\footnotesize (0.036) } &\vspace*{-2mm}{\footnotesize (0.039) }    \\
\vspace*{0mm}\hspace*{5mm}$3<x\leq\max\hspace{0.1cm}(median\approx7)\hspace{0.1cm}km$&      -0.076\sym{**} &      -0.094\sym{**} &      -0.084\sym{*}  & -0.15\sym{**} &      -0.075 &      -0.046    \\
 & \vspace*{-2mm}{\footnotesize (0.034) } &\vspace*{-2mm}{\footnotesize (0.042) } &\vspace*{-2mm}{\footnotesize (0.045) } &\vspace*{-2mm}{\footnotesize (0.058) } &\vspace*{-2mm}{\footnotesize (0.046) } &\vspace*{-2mm}{\footnotesize (0.060) }    \\
\formatpanelheadsubregs{7}{L{\subheadwidthenroll}}{\vspace*{0mm}\hspace*{0mm}\enrollqualbase}
\vspace*{0mm}\hspace*{5mm}4 to 6 Facilities & 0.027 & 0.014 &      -0.035 &      -0.059 & 0.085\sym{**} & 0.073\sym{*}  \\
 & \vspace*{-2mm}{\footnotesize (0.031) } &\vspace*{-2mm}{\footnotesize (0.032) } &\vspace*{-2mm}{\footnotesize (0.042) } &\vspace*{-2mm}{\footnotesize (0.047) } &\vspace*{-2mm}{\footnotesize (0.040) } &\vspace*{-2mm}{\footnotesize (0.043) }    \\
\vspace*{0mm}\hspace*{5mm}7 to 9 Facilities & 0.033 & 0.013 &      -0.027 &      -0.029 & 0.098\sym{**} & 0.066    \\
 & \vspace*{-2mm}{\footnotesize (0.038) } &\vspace*{-2mm}{\footnotesize (0.039) } &\vspace*{-2mm}{\footnotesize (0.053) } &\vspace*{-2mm}{\footnotesize (0.059) } &\vspace*{-2mm}{\footnotesize (0.047) } &\vspace*{-2mm}{\footnotesize (0.049) }    \\
\formatpanelheadregs{7}{L{\subheadwidthenroll}}{Age 9--12 (2011) $\times$}
\formatpanelheadsubregs{7}{L{\subheadwidthenroll}}{\vspace*{0mm}\hspace*{0mm}\enrolldistbase}
\vspace*{0mm}\hspace*{5mm}$0<x\leq3\hspace{0.1cm}(median\approx2)\hspace{0.1cm}km$&      0.0035 & 0.014 &      -0.012 &     -0.0011 & 0.013 & 0.028    \\
 & \vspace*{-2mm}{\footnotesize (0.018) } &\vspace*{-2mm}{\footnotesize (0.019) } &\vspace*{-2mm}{\footnotesize (0.027) } &\vspace*{-2mm}{\footnotesize (0.032) } &\vspace*{-2mm}{\footnotesize (0.024) } &\vspace*{-2mm}{\footnotesize (0.025) }    \\
\vspace*{0mm}\hspace*{5mm}$3<x\leq\max\hspace{0.1cm}(median\approx7)\hspace{0.1cm}km$&      -0.049\sym{**} &      -0.066\sym{**} &      -0.082\sym{**} & -0.13\sym{**} &      -0.026 &      -0.025    \\
 & \vspace*{-2mm}{\footnotesize (0.024) } &\vspace*{-2mm}{\footnotesize (0.034) } &\vspace*{-2mm}{\footnotesize (0.037) } &\vspace*{-2mm}{\footnotesize (0.050) } &\vspace*{-2mm}{\footnotesize (0.030) } &\vspace*{-2mm}{\footnotesize (0.044) }    \\
\formatpanelheadsubregs{7}{L{\subheadwidthenroll}}{\vspace*{0mm}\hspace*{0mm}\enrollqualbase}
\vspace*{0mm}\hspace*{5mm}4 to 6 Facilities & 0.013 & 0.013 & 0.016 & 0.034 &      0.0093 &     -0.0045    \\
 & \vspace*{-2mm}{\footnotesize (0.026) } &\vspace*{-2mm}{\footnotesize (0.030) } &\vspace*{-2mm}{\footnotesize (0.030) } &\vspace*{-2mm}{\footnotesize (0.035) } &\vspace*{-2mm}{\footnotesize (0.036) } &\vspace*{-2mm}{\footnotesize (0.043) }    \\
\vspace*{0mm}\hspace*{5mm}7 to 9 Facilities & 0.037 & 0.032 & 0.030 & 0.045 & 0.037 & 0.017    \\
 & \vspace*{-2mm}{\footnotesize (0.025) } &\vspace*{-2mm}{\footnotesize (0.028) } &\vspace*{-2mm}{\footnotesize (0.030) } &\vspace*{-2mm}{\footnotesize (0.037) } &\vspace*{-2mm}{\footnotesize (0.036) } &\vspace*{-2mm}{\footnotesize (0.043) }    \\
\midrule
Observations   &     2460 &    2033 &    1130 &     942 &    1330 &    1091    \\
\bottomrule
\footnotegap
\multicolumn{7}{L{\footwidthenroll}}{\footnotesize\justify\footenroll}\\
\end{tabular}
\end{adjustbox}
\end{table}
 \end{spacing}
 \clearpage
\pagebreak

\subsection{Child Composition}
\label{sec:childcompo}
\begin{spacing}{\aptxspc}
\begin{table}[htbp]
\centering
\caption{Number of Children in the Household for Children of Heads and for All Children by Sex and Age in 2011, Children Ages 5 to 12.\label{summsibs}}
\begin{adjustbox}{max width=1\textwidth}
\begin{tabular}{m{\lablcolwidthsummeight} >{\centering\arraybackslash}m{1.5cm} >{\centering\arraybackslash}m{1.5cm} >{\centering\arraybackslash}m{1.5cm} >{\centering\arraybackslash}m{1.5cm} >{\centering\arraybackslash}m{1.5cm} >{\centering\arraybackslash}m{1.5cm} >{\centering\arraybackslash}m{1.5cm} >{\centering\arraybackslash}m{1.5cm}}
\toprule
& \multicolumn{8}{L{\innerheadwidthsummeight}}{Outcome: Number of children} \\
\cmidrule(l{5pt}r{5pt}){2-9}
& \multicolumn{4}{C{6cm}}{\small Early Primary Ages} & \multicolumn{4}{C{6cm}}{\small Later Primary Ages} \\
\cmidrule(l{5pt}r{5pt}){2-5} \cmidrule(l{5pt}r{5pt}){6-9}
& \multicolumn{1}{C{1.5cm}}{{\small 5}} & \multicolumn{1}{C{1.5cm}}{{\small 6}} & \multicolumn{1}{C{1.5cm}}{{\small 7}} & \multicolumn{1}{C{1.5cm}}{{\small 8}} & \multicolumn{1}{C{1.5cm}}{{\small 9}} & \multicolumn{1}{C{1.5cm}}{{\small 10}} & \multicolumn{1}{C{1.5cm}}{{\small 11}} & \multicolumn{1}{C{1.5cm}}{{\small 12}} \\
\midrule
\formatpanelheadsumm{9}{L{\subheadwidthsummeight}}{Panel A: Number of Siblings for a Daughter of the Household Head}
Number of siblings (self-included) 5 to 12&    1.62&    1.56&    1.59&    1.55&    1.60&    1.56&    1.61&    1.44\\
Number of siblings (self-included) 0 to 18&    2.18&    2.20&    2.31&    2.30&    2.20&    2.32&    2.33&    2.13\\
\midrule
Observations   &       74&      90&     107&      94&     109&     138&     126&     158\\
\midrule
\formatpanelheadsumm{9}{L{\subheadwidthsummeight}}{Panel B: Number of Siblings for a Son of the Household Head}
Number of siblings (self-included) 5 to 12&    1.58&    1.54&    1.57&    1.40&    1.39&    1.30&    1.34&    1.40\\
Number of siblings (self-included) 0 to 18&    2.10&    2.01&    2.23&    2.01&    2.09&    2.05&    2.08&    2.05\\
\midrule
Observations   &      100&      95&     117&     109&     117&     150&     158&     164\\
\midrule
\formatpanelheadsumm{9}{L{\subheadwidthsummeight}}{Panel C: Number of Siblings/Cousins/etc. (self-included) for a Girl in Household}
Number of children 5 to 12&    1.52&    1.53&    1.46&    1.61&    1.62&    1.58&    1.63&    1.58\\
Number of children 0 to 18&    2.14&    2.23&    2.13&    2.35&    2.27&    2.33&    2.41&    2.28\\
\midrule
Observations   &      133&     145&     168&     155&     158&     180&     174&     186\\
\midrule
\formatpanelheadsumm{9}{L{\subheadwidthsummeight}}{Panel D: Number of Siblings/Cousins/etc. (self-included) for a Boy in Household}
Number of children 5 to 12&    1.60&    1.47&    1.59&    1.46&    1.44&    1.37&    1.38&    1.44\\
Number of children 0 to 18&    2.08&    1.92&    2.19&    2.04&    2.08&    2.06&    2.12&    2.10\\
\midrule
Observations   &       191&      165&      195&      166&      170&      196&      206&      204\\
\bottomrule
\footnotegap
\multicolumn{9}{p{\footwidthsummeight}}{\small\justify In Panels A and B, We calculate the number of children of household heads between age ranges in 2011. In Panels C and D, we calculate the number of children overall in the household between age ranges in 2011}\\
\end{tabular}
\end{adjustbox}
\end{table}

Panels A and B of Table \ref{summsibs} show that the average number of siblings (including self) between age 0 and 18 varies between 2.01 to 2.10 for boys and 2.13 to 2.32 for girls.\footnote{As a comparison, \textcite{lu_treiman_2008} find that from 1978 to 1998, there are on average 2.5 siblings for males and 2.9 siblings for females considering all siblings of an individual when that individual is at age 14.} Panels C and D---where we count all children including siblings, cousins and other individuals between ages 0 and 18---show similar results. The slightly larger sibship size for girls is likely due to son preference. In the urban Chinese context, sibship size has been shown to be negatively correlated with educational attainment due to more constrained family resources. However, in rural areas, \textcite{lu_treiman_2008} find no statistically significant relationship between sibship size and educational attainment partly due to the flexible implementation of the one child policy in rural areas. Exploiting exogenous variations in birth-control policies, \textcite{liu_qualityquantity_2014} also finds a generally insignificant relationship between schooling outcomes and sibship size.

\begin{table}[t]
\centering
\caption{School Enrollment, Age 5 to 12, Controls for Family Characteristics \label{regfamilyenrollcontrols}}
\begin{adjustbox}{max width=1\textwidth}
\begin{tabular}{m{\lablcolwidthenroll} >{\centering\arraybackslash}m{1.85cm} >{\centering\arraybackslash}m{1.85cm} >{\centering\arraybackslash}m{1.85cm} >{\centering\arraybackslash}m{1.85cm} >{\centering\arraybackslash}m{1.85cm} >{\centering\arraybackslash}m{1.85cm}}
\toprule
& \multicolumn{6}{L{\innerheadwidthenroll}}{Outcome: enrolled in school or not in 2011} \\
\cmidrule(l{5pt}r{5pt}){2-7} &  \multicolumn{2}{L{3.7cm}}{\small All Age 5 to 12} & \multicolumn{2}{L{3.7cm}}{\small Girls Age 5 to 12} & \multicolumn{2}{L{3.7cm}}{\small Boys Age 5 to 12} \\      \cmidrule(l{5pt}r{5pt}){2-3} \cmidrule(l{5pt}r{5pt}){4-5} \cmidrule(l{5pt}r{5pt}){6-7} &  \multicolumn{1}{C{1.85cm}}{{\footnotesize All Villages}} & \multicolumn{1}{C{1.85cm}}{{\footnotesize No Teaching Points}} & \multicolumn{1}{C{1.85cm}}{{\footnotesize All Villages}} & \multicolumn{1}{C{1.85cm}}{{\footnotesize No Teaching Points}} & \multicolumn{1}{C{1.85cm}}{{\footnotesize All Villages}} & \multicolumn{1}{C{1.85cm}}{{\footnotesize No Teaching Points}} \\
\midrule
\formatpanelheadregs{7}{L{\subheadwidthenroll}}{Family Member Counts}
\vspace*{0mm}\hspace*{0mm}Number of males 0--18 in household \textit{}&    -0.00022 &      -0.020 &      -0.018 &      -0.050\sym{**} &      0.0069 &     -0.0045    \\
	& \vspace*{-2mm}{\footnotesize (0.012) } &\vspace*{-2mm}{\footnotesize (0.013) } &\vspace*{-2mm}{\footnotesize (0.019) } &\vspace*{-2mm}{\footnotesize (0.020) } &\vspace*{-2mm}{\footnotesize (0.019) } &\vspace*{-2mm}{\footnotesize (0.020) }    \\
\vspace*{0mm}\hspace*{0mm}Number of females 0--18 in household \textit{} & 0.011 &     -0.0017 &    -0.00018 &      -0.015 & 0.032\sym{**} & 0.020    \\
	& \vspace*{-2mm}{\footnotesize (0.0099) } &\vspace*{-2mm}{\footnotesize (0.011) } &\vspace*{-2mm}{\footnotesize (0.016) } &\vspace*{-2mm}{\footnotesize (0.018) } &\vspace*{-2mm}{\footnotesize (0.014) } &\vspace*{-2mm}{\footnotesize (0.015) }    \\
\vspace*{0mm}\hspace*{0mm}Total number of household members &     -0.0059 &   -0.000013 &     -0.0022 &      0.0061 &     -0.0092 &     -0.0052    \\
	& \vspace*{-2mm}{\footnotesize (0.0061) } &\vspace*{-2mm}{\footnotesize (0.0063) } &\vspace*{-2mm}{\footnotesize (0.0083) } &\vspace*{-2mm}{\footnotesize (0.0088) } &\vspace*{-2mm}{\footnotesize (0.0079) } &\vspace*{-2mm}{\footnotesize (0.0085) }    \\
\formatpanelheadregs{7}{L{\subheadwidthenroll}}{Categorical Distance and Quality Measures}
\formatpanelheadsubregs{7}{L{\subheadwidthenroll}}{\enrolldistbase}
\vspace*{0mm}\hspace*{5mm}$0<x\leq3\hspace{0.1cm}(median\approx2)\hspace{0.1cm}km$&      -0.021 &      -0.024 &      -0.031 &      -0.046 &      -0.027 &      -0.019    \\
	& \vspace*{-2mm}{\footnotesize (0.017) } &\vspace*{-2mm}{\footnotesize (0.019) } &\vspace*{-2mm}{\footnotesize (0.025) } &\vspace*{-2mm}{\footnotesize (0.030) } &\vspace*{-2mm}{\footnotesize (0.023) } &\vspace*{-2mm}{\footnotesize (0.026) }    \\
\vspace*{0mm}\hspace*{5mm}$3<x\leq\max\hspace{0.1cm}(median\approx7)\hspace{0.1cm}km$&      -0.062\sym{***}&      -0.082\sym{***}&      -0.084\sym{***} & -0.14\sym{***}&      -0.050\sym{*} &      -0.037    \\
	& \vspace*{-2mm}{\footnotesize (0.023) } &\vspace*{-2mm}{\footnotesize (0.030) } &\vspace*{-2mm}{\footnotesize (0.032) } &\vspace*{-2mm}{\footnotesize (0.042) } &\vspace*{-2mm}{\footnotesize (0.030) } &\vspace*{-2mm}{\footnotesize (0.041) }    \\
\formatpanelheadsubregs{7}{L{\subheadwidthenroll}}{\enrollqualbase}
\vspace*{0mm}\hspace*{5mm}4 to 6 Facilities & 0.022 & 0.017 &     -0.0050 &     -0.0061 & 0.048\sym{*}  & 0.038    \\
	& \vspace*{-2mm}{\footnotesize (0.023) } &\vspace*{-2mm}{\footnotesize (0.025) } &\vspace*{-2mm}{\footnotesize (0.028) } &\vspace*{-2mm}{\footnotesize (0.031) } &\vspace*{-2mm}{\footnotesize (0.028) } &\vspace*{-2mm}{\footnotesize (0.030) }    \\
\vspace*{0mm}\hspace*{5mm}7 to 9 Facilities & 0.037 & 0.024 &      0.0055 &      0.0094 & 0.069\sym{**} & 0.044    \\
	& \vspace*{-2mm}{\footnotesize (0.025) } &\vspace*{-2mm}{\footnotesize (0.027) } &\vspace*{-2mm}{\footnotesize (0.033) } &\vspace*{-2mm}{\footnotesize (0.037) } &\vspace*{-2mm}{\footnotesize (0.032) } &\vspace*{-2mm}{\footnotesize (0.034) }    \\
\midrule
Observations   &     2457 &    2030 &    1128 &     940 &    1329 &    1090    \\
\bottomrule
\footnotegap
\multicolumn{7}{L{\footwidthenroll}}{\footnotesize\justify\footenrollinter}\\
\end{tabular}
\end{adjustbox}
\end{table}

In Table \ref{regfamilyenrollcontrols}, we re-estimate the enrollment regressions. We include, in addition to the total number of household members, also the number of females and males between 0 and 18 years of age in a household. These controls have almost no effects on the distance to school and school facility quality variables' coefficients in the enrollment regressions.\footnote{Results are similar for discretized and continuous measures, Table \ref{regfamilyenrollcontrols} show the discretized results.} This indicates that while there might be effects of sibships sizes on educational outcomes, it does not seem to be a key dimension that interacts with closure mechanisms to generate the school closure effects that we find. Interestingly, Table \ref{regfamilyenrollcontrols} indicates that having more male family members between 0 and 18 is negatively associated with enrollment rates for girls (columns 3 and 4). In contrast, the number of female household members between 0 and 18 is positively associated with enrollment rate for boys (columns 5 and 6). These echo the heterogeneous household composition effects on educational outcomes that \textcite{lei_sibling_2017} find using the China Family Panel Survey. Additional analysis of household structure is out of the scope of this paper.

\end{spacing}
\clearpage
\pagebreak

\section{Minority Status\label{sec:amoremino}}
\renewcommand{\thefigure}{C.\arabic{figure}}
\setcounter{figure}{0}
\renewcommand{\thetable}{C.\arabic{table}}
\setcounter{table}{0}
\renewcommand{\theequation}{C.\arabic{equation}}
\setcounter{equation}{0}
\renewcommand{\thefootnote}{C.\arabic{footnote}}
\setcounter{footnote}{0}

\subsection{Han and non-Han Villages}
\label{sec:minovil}
\begin{spacing}{\aptxspc}

Table \ref{regeleven} shows results from estimating Equation \eqref{eq:startsOnly} for females in villages where Han individuals are in the minority or majority. Standard errors are larger when we divide females into separate village groups. The panels of Table \ref{regeleven} generally show similar results across panels. In our dataset, there are more villages where Han individuals are in the minority, which gives us slightly tighter standard errors for the estimates in the top panel of Table \ref{regeleven}.
\end{spacing}
\vspace*{+4mm}
\begin{spacing}{1.0}
\begin{table}[H]
\centering
\def\sym#1{\ifmmode^{#1}\else\(^{#1}\)\fi}
\caption{Effect of Closure on Educational Attainment by Ethnicity\label{regeleven}}
\begin{adjustbox}{max width=\regone\textwidth}
\begin{tabular}{m{\lablcolwidth} >{\centering\arraybackslash}m{1.5cm} >{\centering\arraybackslash}m{1.5cm} >{\centering\arraybackslash}m{1.5cm} >{\centering\arraybackslash}m{1.5cm} >{\centering\arraybackslash}m{1.5cm} >{\centering\arraybackslash}m{1.5cm}}
\toprule
& \multicolumn{6}{L{\innerheadwidth}}{ Outcome: grades completed by year 2011} \\
\cmidrule(l{5pt}r{5pt}){2-7}
&  \multicolumn{2}{C{3.0cm}}{\footnotesize } & \multicolumn{2}{C{3.0cm}}{\footnotesize \(10 \le 2011 \text{ Age} \le 34\)} & \multicolumn{2}{C{3.0cm}}{\footnotesize \(15 \le 2011 \text{ Age} \le 34\)} \\                                                 \cmidrule(l{5pt}r{5pt}){2-3}                                                 \cmidrule(l{5pt}r{5pt}){4-5}                                                 \cmidrule(l{5pt}r{5pt}){6-7}                         &   \multicolumn{1}{C{1.5cm}}{(1)} & \multicolumn{1}{C{1.5cm}}{(2)} & \multicolumn{1}{C{1.5cm}}{(3)} & \multicolumn{1}{C{1.5cm}}{(4)} & \multicolumn{1}{C{1.5cm}}{(5)} & \multicolumn{1}{C{1.5cm}}{(6)} \\
\midrule
\formatpanelheadregs{7}{L{15cm}}{Panel A: Female in Villages where Han are in the Minority}
\closeinteragestart{} 0--5&      -0.013 & -0.14 &   &   &   &    \\
& \vspace*{-2mm}{\footnotesize (0.29) } &\vspace*{-2mm}{\footnotesize (0.34) } &   &   &   &    \\
\closeinteragestart{} 6--9 & -0.40 & -0.45 & -0.47 & -0.61 &   &    \\
& \vspace*{-2mm}{\footnotesize (0.30) } &\vspace*{-2mm}{\footnotesize (0.35) } &\vspace*{-2mm}{\footnotesize (0.37) } &\vspace*{-2mm}{\footnotesize (0.43) } &   &    \\
\closeinteragestart{} 10--13 & -0.56\sym{*}  & -0.60\sym{*}  & -0.60\sym{*}  & -0.64\sym{*}  & -0.63\sym{*}  & -0.70\sym{*}  \\
& \vspace*{-2mm}{\footnotesize (0.30) } &\vspace*{-2mm}{\footnotesize (0.35) } &\vspace*{-2mm}{\footnotesize (0.33) } &\vspace*{-2mm}{\footnotesize (0.38) } &\vspace*{-2mm}{\footnotesize (0.36) } &\vspace*{-2mm}{\footnotesize (0.41) }    \\
\closeinteragestart{} 22--29&    0.16 &    0.28 &    0.15 &    0.19 &    0.15 &    0.21    \\
& \vspace*{-2mm}{\footnotesize (0.30) } &\vspace*{-2mm}{\footnotesize (0.30) } &\vspace*{-2mm}{\footnotesize (0.33) } &\vspace*{-2mm}{\footnotesize (0.32) } &\vspace*{-2mm}{\footnotesize (0.34) } &\vspace*{-2mm}{\footnotesize (0.33) }    \\
\midrule
Observations  &     5677 &    4758 &    3631 &    3032 &    2984 &    2490    \\

\midrule
\formatpanelheadregs{7}{L{15cm}}{Panel B: Female in Han Majority Villages}
\closeinteragestart{} 0--5 & -0.37 & -0.37 &   &   &   &    \\
& \vspace*{-2mm}{\footnotesize (0.32) } &\vspace*{-2mm}{\footnotesize (0.36) } &   &   &   &    \\
\closeinteragestart{} 6--9 & -0.26 & -0.38 & -0.36 & -0.52 &   &    \\
& \vspace*{-2mm}{\footnotesize (0.31) } &\vspace*{-2mm}{\footnotesize (0.34) } &\vspace*{-2mm}{\footnotesize (0.35) } &\vspace*{-2mm}{\footnotesize (0.40) } &   &    \\
\closeinteragestart{} 10--13 & -0.51 & -0.49 & -0.45 & -0.48 & -0.58 & -0.54    \\
& \vspace*{-2mm}{\footnotesize (0.34) } &\vspace*{-2mm}{\footnotesize (0.38) } &\vspace*{-2mm}{\footnotesize (0.36) } &\vspace*{-2mm}{\footnotesize (0.41) } &\vspace*{-2mm}{\footnotesize (0.45) } &\vspace*{-2mm}{\footnotesize (0.48) }    \\
\closeinteragestart{} 22--29&    0.17 &    0.25 & -0.11 &      -0.031 &      -0.098 & -0.13    \\
& \vspace*{-2mm}{\footnotesize (0.34) } &\vspace*{-2mm}{\footnotesize (0.34) } &\vspace*{-2mm}{\footnotesize (0.40) } &\vspace*{-2mm}{\footnotesize (0.40) } &\vspace*{-2mm}{\footnotesize (0.42) } &\vspace*{-2mm}{\footnotesize (0.43) }    \\
\midrule
Observations&    3192 &    2708 &    2033 &    1758 &    1674 &    1456    \\
\midrule
\exclcontrol
\exclcontrolcont
\bottomrule
\footnotegap
\multicolumn{7}{L{\footwidth}}{\footnotesize\justify\footattain} \\
\end{tabular}
\end{adjustbox}
\end{table}
 \end{spacing}
\clearpage

\subsection{Han and non-Han Individuals}
\label{sec:minoindi}
\newcommand{\footminoindione}{In Table \ref{regone}, when we differentiate between the effects of the policy for females and males, we estimate Equation \eqref{eq:startsOnly} separately for females and males. This means that there are gender specific village fixed effects $\beta_{v}$, provincial specific age fixed effects $\rho_{pa}$, covariate effects $\gamma$, as well as policy effects $\tilde{\lambda_{z}}$. When we differentiate the effects between minority and Han individuals, it is no longer possible to allow for all coefficients to be different because the fraction of individuals who are Han in some villages is small.}

\begin{spacing}{\aptxspc}
Following Equation \eqref{eq:startsOnly}, we differentiate the effects of school closure for minority and Han children by interacting the schooling closure variable by whether a child is minority or Han:
\begin{singlespace}\vspace*{-\baselineskip}
	\begin{eqnarray}
	\label{eq:mino}
	E_{pvia} & = & \phi + \beta_{v}  + \rho_{pa} + \rho^{\text{m}}_{a} \cdot m_i
	\nonumber \\
	& & + \sum_{\mu \in \left\{0, 1\right\}}
	\left(
	\sum_{z=1}^Z \tilde{\lambda^{\mu}_{z}} \cdot \bm{1} \left\{ l_{z}\leq t_{i}\leq u_{z} \right\}
	\cdot c_{v}
	\right)
	\cdot \bm{1} \left\{ m_i = \mu \right\}
	\\
	&  &
	+ X_i \cdot \gamma + X_i \cdot m_i \cdot \gamma^m \nonumber\\
	& & + \epsilon_{i}  \nonumber
	\end{eqnarray}
\end{singlespace}\noindent\ignorespaces

In Equation \eqref{eq:mino}, $m_i$ indicates Han ($m_i=1$) or minority ($m_i=0$) status. We allow age-specific minority fixed effects $\rho_{a}^m$ and minority-specific effect $\gamma^m$.\footnote{\footminoindione} Table \ref{tab:minotwo} presents the minority- and Han-specific effects of policy $\tilde{\lambda^{\mu}_{z}}$, both compared against the respective base groups.

\begin{table}[htbp]
\centering\caption{Effect of School Closure on Educational Attainment (Han and Minority) \label{tab:minotwo}}
\begin{adjustbox}{max width=\reginteract\textwidth}
\begin{tabular}{m{\lablcolwidth} >{\centering\arraybackslash}m{1.5cm} >{\centering\arraybackslash}m{1.5cm} >{\centering\arraybackslash}m{1.5cm} >{\centering\arraybackslash}m{1.5cm} >{\centering\arraybackslash}m{1.5cm} >{\centering\arraybackslash}m{1.5cm}}
\toprule
& \multicolumn{6}{L{\innerheadwidth}}{Outcome: grades completed by year 2011} \\                                 \cmidrule(l{5pt}r{5pt}){2-7}          &   \multicolumn{2}{L{3.0cm}}{\footnotesize } & \multicolumn{2}{L{3.0cm}}{\footnotesize 10 $\le$ 2011 Age $\le$ 34} & \multicolumn{2}{L{3.0cm}}{\footnotesize  15 $\le$ 2011 Age $\le$ 34} \\                                  \cmidrule(l{5pt}r{5pt}){2-3} \cmidrule(l{5pt}r{5pt}){4-5} \cmidrule(l{5pt}r{5pt}){6-7}          &   \multicolumn{1}{C{1.5cm}}{1} & \multicolumn{1}{C{1.5cm}}{2} & \multicolumn{1}{C{1.5cm}}{3} & \multicolumn{1}{C{1.5cm}}{4} & \multicolumn{1}{C{1.5cm}}{5} & \multicolumn{1}{C{1.5cm}}{6} \\
\midrule
\formatpanelheadsupregs{7}{L{15cm}}{\baselinegroup}
\formatpanelheadregs{7}{L{\subheadwidth}}{Panel A: \newtablepanfemale}
\formatpanelheadsubregs{7}{L{\subheadwidth}}{Minority $\times$}
\vspace*{0mm}\hspace*{5mm}\closeinteragestart{} 0--5 & -0.28 & -0.50\sym{*} &   &   &   &    \\
	& \vspace*{-2mm}{\footnotesize (0.25) } &\vspace*{-2mm}{\footnotesize (0.28) } &   &   &   &    \\
\vspace*{0mm}\hspace*{5mm}\closeinteragestart{} 6--9 & -0.55\sym{**} & -0.71\sym{**} & -0.54 & -0.73\sym{*} &   &    \\
	& \vspace*{-2mm}{\footnotesize (0.27) } &\vspace*{-2mm}{\footnotesize (0.31) } &\vspace*{-2mm}{\footnotesize (0.34) } &\vspace*{-2mm}{\footnotesize (0.37) } &   &    \\
\vspace*{0mm}\hspace*{5mm}\closeinteragestart{} 10--13 & -0.64\sym{**} & -0.77\sym{**} & -0.62\sym{**} & -0.74\sym{**} & -0.62\sym{*}  & -0.70\sym{**} \\
	& \vspace*{-2mm}{\footnotesize (0.27) } &\vspace*{-2mm}{\footnotesize (0.30) } &\vspace*{-2mm}{\footnotesize (0.29) } &\vspace*{-2mm}{\footnotesize (0.32) } &\vspace*{-2mm}{\footnotesize (0.33) } &\vspace*{-2mm}{\footnotesize (0.36) }    \\
\vspace*{0mm}\hspace*{5mm}\closeinteragestart{} 22--29&    0.20 &    0.30 &    0.20 &    0.23 &    0.23 &    0.26    \\
	& \vspace*{-2mm}{\footnotesize (0.28) } &\vspace*{-2mm}{\footnotesize (0.29) } &\vspace*{-2mm}{\footnotesize (0.30) } &\vspace*{-2mm}{\footnotesize (0.31) } &\vspace*{-2mm}{\footnotesize (0.31) } &\vspace*{-2mm}{\footnotesize (0.32) }    \\
\formatpanelheadsubregs{7}{L{\subheadwidth}}{Han $\times$}
\vspace*{0mm}\hspace*{5mm}\closeinteragestart{} 0--5 & -0.26 & -0.28 &   &   &   &    \\
	& \vspace*{-2mm}{\footnotesize (0.29) } &\vspace*{-2mm}{\footnotesize (0.32) } &   &   &   &    \\
\vspace*{0mm}\hspace*{5mm}\closeinteragestart{} 6--9 & -0.21 & -0.30 & -0.33 & -0.48 &   &    \\
	& \vspace*{-2mm}{\footnotesize (0.29) } &\vspace*{-2mm}{\footnotesize (0.32) } &\vspace*{-2mm}{\footnotesize (0.33) } &\vspace*{-2mm}{\footnotesize (0.36) } &   &    \\
\vspace*{0mm}\hspace*{5mm}\closeinteragestart{} 10--13 & -0.44 & -0.43 & -0.43 & -0.43 & -0.54 & -0.54    \\
	& \vspace*{-2mm}{\footnotesize (0.32) } &\vspace*{-2mm}{\footnotesize (0.37) } &\vspace*{-2mm}{\footnotesize (0.35) } &\vspace*{-2mm}{\footnotesize (0.39) } &\vspace*{-2mm}{\footnotesize (0.45) } &\vspace*{-2mm}{\footnotesize (0.49) }    \\
\vspace*{0mm}\hspace*{5mm}\closeinteragestart{} 22--29 & 0.092 &    0.12 & -0.29 & -0.24 & -0.29 & -0.30    \\
	& \vspace*{-2mm}{\footnotesize (0.27) } &\vspace*{-2mm}{\footnotesize (0.29) } &\vspace*{-2mm}{\footnotesize (0.33) } &\vspace*{-2mm}{\footnotesize (0.35) } &\vspace*{-2mm}{\footnotesize (0.34) } &\vspace*{-2mm}{\footnotesize (0.37) }    \\
\midrule
Observations   &     8869 &    7466 &    5664 &    4790 &    4658 &    3946    \\
\midrule
\formatpanelheadregs{7}{L{\subheadwidth}}{Panel B: \newtablepanmale}
\formatpanelheadsubregs{7}{L{\subheadwidth}}{Minority $\times$}
\vspace*{0mm}\hspace*{5mm}\closeinteragestart{} 0--5&      -0.056 & 0.025 &   &   &   &    \\
	& \vspace*{-2mm}{\footnotesize (0.24) } &\vspace*{-2mm}{\footnotesize (0.25) } &   &   &   &    \\
\vspace*{0mm}\hspace*{5mm}\closeinteragestart{} 6--9 & -0.24 & -0.31 & -0.21 & -0.40 &   &    \\
	& \vspace*{-2mm}{\footnotesize (0.21) } &\vspace*{-2mm}{\footnotesize (0.24) } &\vspace*{-2mm}{\footnotesize (0.26) } &\vspace*{-2mm}{\footnotesize (0.28) } &   &    \\
\vspace*{0mm}\hspace*{5mm}\closeinteragestart{} 10--13 & -0.46\sym{*}  & -0.43\sym{*}  & -0.47\sym{*}  & -0.45\sym{*}  & -0.52\sym{*}  & -0.48\sym{*}  \\
	& \vspace*{-2mm}{\footnotesize (0.23) } &\vspace*{-2mm}{\footnotesize (0.25) } &\vspace*{-2mm}{\footnotesize (0.24) } &\vspace*{-2mm}{\footnotesize (0.25) } &\vspace*{-2mm}{\footnotesize (0.28) } &\vspace*{-2mm}{\footnotesize (0.28) }    \\
\vspace*{0mm}\hspace*{5mm}\closeinteragestart{} 22--29&    0.25 &    0.23 &    0.18 &    0.20 & 0.099 &    0.12    \\
	& \vspace*{-2mm}{\footnotesize (0.27) } &\vspace*{-2mm}{\footnotesize (0.30) } &\vspace*{-2mm}{\footnotesize (0.29) } &\vspace*{-2mm}{\footnotesize (0.33) } &\vspace*{-2mm}{\footnotesize (0.30) } &\vspace*{-2mm}{\footnotesize (0.34) }    \\
\formatpanelheadsubregs{7}{L{\subheadwidth}}{Han $\times$}
\vspace*{0mm}\hspace*{5mm}\closeinteragestart{} 0--5 & 0.063 &    0.10 &   &   &   &    \\
	& \vspace*{-2mm}{\footnotesize (0.28) } &\vspace*{-2mm}{\footnotesize (0.32) } &   &   &   &    \\
\vspace*{0mm}\hspace*{5mm}\closeinteragestart{} 6--9&    0.34 &    0.41 &    0.24 &    0.27 &   &    \\
	& \vspace*{-2mm}{\footnotesize (0.32) } &\vspace*{-2mm}{\footnotesize (0.35) } &\vspace*{-2mm}{\footnotesize (0.39) } &\vspace*{-2mm}{\footnotesize (0.42) } &   &    \\
\vspace*{0mm}\hspace*{5mm}\closeinteragestart{} 10--13 & 0.065 &    0.10 & 0.036 &      0.0049 &      -0.085 & -0.16    \\
	& \vspace*{-2mm}{\footnotesize (0.25) } &\vspace*{-2mm}{\footnotesize (0.28) } &\vspace*{-2mm}{\footnotesize (0.27) } &\vspace*{-2mm}{\footnotesize (0.30) } &\vspace*{-2mm}{\footnotesize (0.32) } &\vspace*{-2mm}{\footnotesize (0.33) }    \\
\vspace*{0mm}\hspace*{5mm}\closeinteragestart{} 22--29&      -0.049 & 0.066 &     -0.0054 &    0.31 & -0.13 &    0.13    \\
	& \vspace*{-2mm}{\footnotesize (0.29) } &\vspace*{-2mm}{\footnotesize (0.33) } &\vspace*{-2mm}{\footnotesize (0.35) } &\vspace*{-2mm}{\footnotesize (0.40) } &\vspace*{-2mm}{\footnotesize (0.36) } &\vspace*{-2mm}{\footnotesize (0.42) }    \\
\midrule
Observations   &     9935 &    8452 &    6408 &    5499 &    5340 &    4592    \\
\midrule
\exclcontrol
\exclcontrolcont
\bottomrule
\footnotegap
\multicolumn{7}{L{\footwidth}}{\footnotesize\justify\footattaininter}\\
\end{tabular}
\end{adjustbox}
\end{table}
Similar to what we showed previously, we find that girls are more vulnerable to school closure. Table \ref{tab:minotwo} shows that the policy decreased the educational attainment for minority girls who were below age 6, between age 6 and 9, and between age 10 and 13 in the year of closure by 0.28 (s.e. 0.25), 0.55 (s.e. 0.27), and 0.64 (s.e. 0.27) years by 2011, respectively. The effects for these three age ranges for Han girls are negative as well, but smaller in magnitude and weaker in significance. Results are consistent across all columns. For boys, as before, policy effects are less negative than effects for girls and generally insignificant. We also find stronger and more negative policy impacts of closure for minority boys, compared to Han boys.

To analyze the underlying mechanisms, we show in Table \ref{tab:tabsixbhan} enrollment regression results where we interact distance to school and school facility variables with a child's minority status. Table \ref{tab:tabsixbhan} shows overall the same story as Table \ref{regfive}. While longer distance to school is linked to lower enrollment, the association is greater for minority than Han girls. Better school facilities are associated with an increase in boys enrollment, but the increase is larger in magnitude for Han boys compared to minority boys. There is a significant reduction in sub-group sample size as we disaggregate, leading to a general weakening of statistical significance.

\begin{table}[t]
\centering
\caption{Linear Probability Model of School Enrollment, Age 5 to 12 (Han and Minority)\label{tab:tabsixbhan}}
\begin{adjustbox}{max width=1\textwidth}
\begin{tabular}{m{\lablcolwidthenroll} >{\centering\arraybackslash}m{1.85cm} >{\centering\arraybackslash}m{1.85cm} >{\centering\arraybackslash}m{1.85cm} >{\centering\arraybackslash}m{1.85cm} >{\centering\arraybackslash}m{1.85cm} >{\centering\arraybackslash}m{1.85cm}}
\toprule
& \multicolumn{6}{L{\innerheadwidthenroll}}{Outcome: enrolled in school or not in 2011} \\                                         \cmidrule(l{5pt}r{5pt}){2-7}                  &   \multicolumn{2}{L{3.7cm}}{\small All Age 5 to 12} & \multicolumn{2}{L{3.7cm}}{\small Girls Age 5 to 12} & \multicolumn{2}{L{3.7cm}}{\small Boys Age 5 to 12} \\                                          \cmidrule(l{5pt}r{5pt}){2-3} \cmidrule(l{5pt}r{5pt}){4-5} \cmidrule(l{5pt}r{5pt}){6-7}                  &   \multicolumn{1}{C{1.85cm}}{{\small All Villages}} & \multicolumn{1}{C{1.85cm}}{{\small No Teaching Points}} & \multicolumn{1}{C{1.85cm}}{{\small All Villages}} & \multicolumn{1}{C{1.85cm}}{{\small No Teaching Points}} & \multicolumn{1}{C{1.85cm}}{{\small All Villages}} & \multicolumn{1}{C{1.85cm}}{{\small No Teaching Points}} \\
\midrule
\formatpanelheadsupregs{7}{L{\subheadwidthenroll}}{categorical and quality and distance measures}
\formatpanelheadregs{7}{L{\subheadwidthenroll}}{Minority $\times$}
\formatpanelheadsubregs{7}{L{\subheadwidthenroll}}{\enrolldistbase}
\vspace*{0mm}\hspace*{5mm} $0<x\leq3\hspace{0.1cm}(median\approx2)\hspace{0.1cm}km$&      -0.015 &      -0.013 &      -0.028 &      -0.048 &      -0.031 &      -0.015    \\
	& \vspace*{-2mm}{\footnotesize (0.018) } &\vspace*{-2mm}{\footnotesize (0.020) } &\vspace*{-2mm}{\footnotesize (0.028) } &\vspace*{-2mm}{\footnotesize (0.034) } &\vspace*{-2mm}{\footnotesize (0.025) } &\vspace*{-2mm}{\footnotesize (0.028) }    \\
\vspace*{0mm}\hspace*{5mm} $3<x\leq\max\hspace{0.1cm}(median\approx7)\hspace{0.1cm}km$&      -0.053\sym{**} &      -0.068\sym{**} &      -0.084\sym{**} & -0.15\sym{***}&      -0.044 &      -0.024    \\
	& \vspace*{-2mm}{\footnotesize (0.023) } &\vspace*{-2mm}{\footnotesize (0.031) } &\vspace*{-2mm}{\footnotesize (0.033) } &\vspace*{-2mm}{\footnotesize (0.046) } &\vspace*{-2mm}{\footnotesize (0.034) } &\vspace*{-2mm}{\footnotesize (0.045) }    \\
\formatpanelheadsubregs{7}{L{\subheadwidthenroll}}{\enrollqualbase}
\vspace*{0mm}\hspace*{5mm} \textit4 to 6 Facilities & 0.010 & 0.012 &      -0.016 &      -0.018 & 0.031 & 0.035    \\
	& \vspace*{-2mm}{\footnotesize (0.023) } &\vspace*{-2mm}{\footnotesize (0.027) } &\vspace*{-2mm}{\footnotesize (0.030) } &\vspace*{-2mm}{\footnotesize (0.035) } &\vspace*{-2mm}{\footnotesize (0.029) } &\vspace*{-2mm}{\footnotesize (0.034) }    \\
\vspace*{0mm}\hspace*{5mm} \textit7 to 9 Facilities & 0.050\sym{*}  & 0.043 & 0.040 & 0.047 & 0.065\sym{*}  & 0.045    \\
	& \vspace*{-2mm}{\footnotesize (0.026) } &\vspace*{-2mm}{\footnotesize (0.030) } &\vspace*{-2mm}{\footnotesize (0.034) } &\vspace*{-2mm}{\footnotesize (0.040) } &\vspace*{-2mm}{\footnotesize (0.033) } &\vspace*{-2mm}{\footnotesize (0.038) }    \\
\formatpanelheadregs{7}{L{\subheadwidthenroll}}{Han $\times$}
\formatpanelheadsubregs{7}{L{\subheadwidthenroll}}{\enrolldistbase}
\vspace*{0mm}\hspace*{5mm} $0<x\leq3\hspace{0.1cm}(median\approx2)\hspace{0.1cm}km$&      -0.033 &      -0.051 &      -0.044 &      -0.061 &     -0.0074 &      -0.023    \\
	& \vspace*{-2mm}{\footnotesize (0.037) } &\vspace*{-2mm}{\footnotesize (0.039) } &\vspace*{-2mm}{\footnotesize (0.050) } &\vspace*{-2mm}{\footnotesize (0.056) } &\vspace*{-2mm}{\footnotesize (0.047) } &\vspace*{-2mm}{\footnotesize (0.051) }    \\
\vspace*{0mm}\hspace*{5mm} $3<x\leq\max\hspace{0.1cm}(median\approx7)\hspace{0.1cm}km$&      -0.058 &      -0.092\sym{*} &      -0.062 &      -0.095 &      -0.042 &      -0.075    \\
	& \vspace*{-2mm}{\footnotesize (0.042) } &\vspace*{-2mm}{\footnotesize (0.050) } &\vspace*{-2mm}{\footnotesize (0.058) } &\vspace*{-2mm}{\footnotesize (0.067) } &\vspace*{-2mm}{\footnotesize (0.053) } &\vspace*{-2mm}{\footnotesize (0.066) }    \\
\formatpanelheadsubregs{7}{L{\subheadwidthenroll}}{\enrollqualbase}
\vspace*{0mm}\hspace*{5mm} 4 to 6 Facilities & 0.028 & 0.018 &      -0.015 &      0.0058 & 0.093\sym{*}  & 0.062    \\
	& \vspace*{-2mm}{\footnotesize (0.043) } &\vspace*{-2mm}{\footnotesize (0.042) } &\vspace*{-2mm}{\footnotesize (0.060) } &\vspace*{-2mm}{\footnotesize (0.062) } &\vspace*{-2mm}{\footnotesize (0.052) } &\vspace*{-2mm}{\footnotesize (0.051) }    \\
\vspace*{0mm}\hspace*{5mm} 7 to 9 Facilities&     0.00053 &      -0.011 &      -0.077 &      -0.051 & 0.080 & 0.050    \\
	& \vspace*{-2mm}{\footnotesize (0.043) } &\vspace*{-2mm}{\footnotesize (0.046) } &\vspace*{-2mm}{\footnotesize (0.066) } &\vspace*{-2mm}{\footnotesize (0.075) } &\vspace*{-2mm}{\footnotesize (0.055) } &\vspace*{-2mm}{\footnotesize (0.059) }    \\
\midrule
Observations   &     2444 &    2017 &    1125 &     937 &    1319 &    1080    \\
\bottomrule
\footnotegap
\multicolumn{7}{L{\footwidthenroll}}{\footnotesize\justify\footenrollinter}\\
\end{tabular}
\end{adjustbox}
\end{table}

Both the attainment and enrollment results show more significant negative consequences of school closure for minority children. This might be partly be due to the larger sample size for minorities--66 and 67 percent of the sample in the female and male attainment regression samples are minorities. Minorities might also be affected more negatively by the policy because of language disadvantage at the consolidated schools or greater difficulty of accessing consolidated schools by minorities. Overall, for both minorities and Han, however, we observe a broadly similar gender pattern for the effects of school closure.
\end{spacing}
\clearpage
\clearpage
\pagebreak

\section{Heterogeneity by Village Attributes \label{sec:amorevilvars}}
\renewcommand{\thefigure}{D.\arabic{figure}}
\setcounter{figure}{0}
\renewcommand{\thetable}{D.\arabic{table}}
\setcounter{table}{0}
\renewcommand{\theequation}{D.\arabic{equation}}
\setcounter{equation}{0}
\renewcommand{\thefootnote}{D.\arabic{footnote}}
\setcounter{footnote}{0}

\subsection{Richer and Poorer villages}
\begin{spacing}{\aptxspc}
With an average per capita income of 4446 Yuan in 2011, the CHES survey villages are all poor by Chinese standards. The panels of Appendix Table \ref{regten} show results for females in villages with below and above 4000 yuan (1 Dollar = 6.5 Yuan in 2011) per-capita income in 2011. Standard errors are larger when we divide females into separate village groups. The two panels of Appendix Table \ref{regten} both show negative effects of closure on grades completed by 2011. The negative effects are generally weaker for females from relatively richer villages.
\end{spacing}
\vspace*{-4mm}
\begin{spacing}{1.0}
\begin{table}[H]                                         \centering                                         \def\sym#1{\ifmmode^{#1}\else\(^{#1}\)\fi}
\caption{Effect of School Closure on Educational Attainment by Village Income\label{regten}}                                         \begin{adjustbox}{max width=0.95\textwidth}                                         \begin{tabular}{m{\lablcolwidth} >{\centering\arraybackslash}m{1.5cm} >{\centering\arraybackslash}m{1.5cm} >{\centering\arraybackslash}m{1.5cm} >{\centering\arraybackslash}m{1.5cm} >{\centering\arraybackslash}m{1.5cm} >{\centering\arraybackslash}m{1.5cm}}                                         \toprule                                                                                          &    \multicolumn{6}{L{\innerheadwidth}}{ Outcome: grades completed by year 2011} \\                                                 \cmidrule(l{5pt}r{5pt}){2-7}                         &    \multicolumn{2}{C{3.0cm}}{\footnotesize } & \multicolumn{2}{C{3.0cm}}{\footnotesize \(10 \le 2011 \text{ Age} \le 34\)} & \multicolumn{2}{C{3.0cm}}{\footnotesize \(15 \le 2011 \text{ Age} \le 34\)} \\                                                 \cmidrule(l{5pt}r{5pt}){2-3}                                                 \cmidrule(l{5pt}r{5pt}){4-5}                                                 \cmidrule(l{5pt}r{5pt}){6-7}                         &   \multicolumn{1}{C{1.5cm}}{(1)} & \multicolumn{1}{C{1.5cm}}{(2)} & \multicolumn{1}{C{1.5cm}}{(3)} & \multicolumn{1}{C{1.5cm}}{(4)} & \multicolumn{1}{C{1.5cm}}{(5)} & \multicolumn{1}{C{1.5cm}}{(6)} \\
\midrule
\formatpanelheadregs{7}{L{15cm}}{Panel A: Female in Villages with Per Capita Income Below 4000 Yuan}
\closeinteragestart{} 0--5 & -0.42 & -0.60\sym{*} &   &   &   &    \\
& \vspace*{-2mm}{\footnotesize (0.30) } &\vspace*{-2mm}{\footnotesize (0.33) } &   &   &   &    \\
\closeinteragestart{} 6--9 & -0.72\sym{**} & -0.74\sym{**} & -0.83\sym{**} & -0.90\sym{**} &   &    \\
& \vspace*{-2mm}{\footnotesize (0.29) } &\vspace*{-2mm}{\footnotesize (0.33) } &\vspace*{-2mm}{\footnotesize (0.38) } &\vspace*{-2mm}{\footnotesize (0.42) } &   &    \\
\closeinteragestart{} 10--13 & -0.60\sym{*}  & -0.61\sym{*}  & -0.75\sym{**} & -0.79\sym{**} & -0.70\sym{*}  & -0.75\sym{*}  \\
& \vspace*{-2mm}{\footnotesize (0.31) } &\vspace*{-2mm}{\footnotesize (0.35) } &\vspace*{-2mm}{\footnotesize (0.33) } &\vspace*{-2mm}{\footnotesize (0.38) } &\vspace*{-2mm}{\footnotesize (0.42) } &\vspace*{-2mm}{\footnotesize (0.44) }    \\
\closeinteragestart{} 22--29&    0.13 &    0.26 & 0.015 &    -0.0002 &      -0.018 &      -0.021    \\
& \vspace*{-2mm}{\footnotesize (0.31) } &\vspace*{-2mm}{\footnotesize (0.29) } &\vspace*{-2mm}{\footnotesize (0.34) } &\vspace*{-2mm}{\footnotesize (0.32) } &\vspace*{-2mm}{\footnotesize (0.34) } &\vspace*{-2mm}{\footnotesize (0.33) }    \\
\midrule
Observations  &     4495 &    4066 &    2910 &    2620 &    2403 &    2164    \\
\midrule
\formatpanelheadregs{7}{L{15cm}}{Panel B: Female in Villages with Per Capita Income Above 4000 Yuan}
\closeinteragestart{} 0--5 & -0.33 & -0.48 &   &   &   &    \\
& \vspace*{-2mm}{\footnotesize (0.35) } &\vspace*{-2mm}{\footnotesize (0.39) } &   &   &   &    \\
\closeinteragestart{} 6--9 & -0.35 & -0.67\sym{*}  & -0.44 & -0.84\sym{*} &   &    \\
& \vspace*{-2mm}{\footnotesize (0.36) } &\vspace*{-2mm}{\footnotesize (0.40) } &\vspace*{-2mm}{\footnotesize (0.39) } &\vspace*{-2mm}{\footnotesize (0.43) } &   &    \\
\closeinteragestart{} 10--13 & -0.56 & -0.76\sym{**} & -0.45 & -0.67\sym{*}  & -0.47 & -0.60    \\
& \vspace*{-2mm}{\footnotesize (0.34) } &\vspace*{-2mm}{\footnotesize (0.38) } &\vspace*{-2mm}{\footnotesize (0.35) } &\vspace*{-2mm}{\footnotesize (0.39) } &\vspace*{-2mm}{\footnotesize (0.39) } &\vspace*{-2mm}{\footnotesize (0.44) }    \\
\closeinteragestart{} 22--29&    0.29 &    0.25 &    0.17 &    0.26 &    0.27 &    0.25    \\
& \vspace*{-2mm}{\footnotesize (0.32) } &\vspace*{-2mm}{\footnotesize (0.35) } &\vspace*{-2mm}{\footnotesize (0.36) } &\vspace*{-2mm}{\footnotesize (0.39) } &\vspace*{-2mm}{\footnotesize (0.39) } &\vspace*{-2mm}{\footnotesize (0.42) }    \\
\midrule
Observations&    4374 &    3400 &    2754 &    2170 &    2255 &    1782    \\
\midrule
\exclcontrol
\exclcontrolcont
\bottomrule
\footnotegap
\multicolumn{7}{L{\footwidth}}{\footnotesize\justify\footattain} \\
\end{tabular}
\end{adjustbox}
\end{table}
 \end{spacing}
\clearpage

\subsection{Teaching-Points Closure}
\label{sec:teachpoint}
\begin{spacing}{\aptxspc}
In Section \ref{sec:data}, we categorized villages by their school closure status. Among the 193 villages that closed schools between 1999 and 2010, 45 of them (i.e. 23.3\%) had teaching-points. On the contrary only 44 villages (i.e. 10.2\%) among the 430 villages that had not closed existing schools by 2011 have teaching-points. This shows that both teaching-points and non-teaching-points schools experienced closures, although teaching-points schools were more likely to have been closed. Teaching-points in villages offer up to 4 years of within-village primary education partly with the aim of reducing travel distance for students. While teaching-points may offer proximity and small-school benefits for young village students, they are usually of lower quality compared to other primary schools, when quality is defined narrowly in terms of the physical facility quality and teacher qualifications \autocite{SargentHannumJTE2009}. In this section, we analyze the heterogeneity of closure effects by teaching-points status.\footnote{While our attainment results have not distinguished teaching-points, our enrollment regressions from Table \ref{regfive} and elsewhere exclude teaching-point villages in the even columns as discussed in Section \ref{sec:mechanism}.}

Following Equation \eqref{eq:mino} from Appendix Section \ref{sec:minoindi}, we estimate:
\begin{singlespace}\vspace*{-\baselineskip}
	\begin{eqnarray}
	\label{eq:tp}
	E_{pvia} & = & \phi + \beta_{v}  + \rho_{pa} + \rho^{\tau}_{a} \cdot \text{TP}_{v}
	\nonumber \\
	& & + \sum_{\tau \in \left\{0, 1\right\}}
	\left(
	\sum_{z=1}^Z \tilde{\lambda^{\tau}_{z}} \cdot \bm{1} \left\{ l_{z}\leq t_{i}\leq u_{z} \right\}
	\cdot c_{v}
	\right)
	\cdot \bm{1} \left\{ \text{TP}_{v} = \tau \right\}
	\\
	&  &
	+ X_i \cdot \gamma + X_i \cdot \text{TP}_{v} \cdot \gamma^{\tau} \nonumber\\
	& & + \epsilon_{i}  \nonumber
	\end{eqnarray}
\end{singlespace}\noindent\ignorespaces
where \(\text{TP}_v = 1\) if the village had a teaching-point school in 2011 or had a teaching-point school that was closed before 2001. Equation \eqref{eq:tp} allows teaching-point villages to have differential cohort (age in 2011) attainment patterns. This allows for differential pre-trends for teaching-points schools where educational quality might have been on a different cohort trajectory. In contrast to Equation \eqref{eq:mino}, the interactions are at the village level in Equation \eqref{eq:tp}. Table \ref{tab:teachone} presents teaching-point and non-teaching-point interacted closure specific policy effects compared against the respective base group of children between 14 and 21 years of age in the year of closure.

\begin{table}[htbp]
\centering\caption{Effect of School Closure on Educational Attainment (Teaching-Point)\label{tab:teachone}}
\begin{adjustbox}{max width=\reginteract\textwidth}
\begin{tabular}{m{\lablcolwidth} >{\centering\arraybackslash}m{1.5cm} >{\centering\arraybackslash}m{1.5cm} >{\centering\arraybackslash}m{1.5cm} >{\centering\arraybackslash}m{1.5cm} >{\centering\arraybackslash}m{1.5cm} >{\centering\arraybackslash}m{1.5cm}}
\toprule
& \multicolumn{6}{L{\innerheadwidth}}{Outcome: grades completed by year 2011} \\                                 \cmidrule(l{5pt}r{5pt}){2-7}          &   \multicolumn{2}{L{3.0cm}}{\footnotesize } & \multicolumn{2}{L{3.0cm}}{\footnotesize 10 $\le$ 2011 Age $\le$ 34} & \multicolumn{2}{L{3.0cm}}{\footnotesize  15 $\le$ 2011 Age $\le$ 34} \\                                  \cmidrule(l{5pt}r{5pt}){2-3} \cmidrule(l{5pt}r{5pt}){4-5} \cmidrule(l{5pt}r{5pt}){6-7}          &   \multicolumn{1}{C{1.5cm}}{1} & \multicolumn{1}{C{1.5cm}}{2} & \multicolumn{1}{C{1.5cm}}{3} & \multicolumn{1}{C{1.5cm}}{4} & \multicolumn{1}{C{1.5cm}}{5} & \multicolumn{1}{C{1.5cm}}{6} \\
\midrule
\formatpanelheadsupregs{7}{L{15cm}}{\baselinegroup}
\formatpanelheadregs{7}{L{\subheadwidth}}{Panel A: \newtablepanfemale}
\formatpanelheadsubregs{7}{L{\subheadwidth}}{Non-teach-point $\times$}
\vspace*{0mm}\hspace*{5mm}\closeinteragestart{} 0--5 & -0.38 & -0.49\sym{*} &   &   &   &    \\
	& \vspace*{-2mm}{\footnotesize (0.24) } &\vspace*{-2mm}{\footnotesize (0.26) } &   &   &   &    \\
\vspace*{0mm}\hspace*{5mm}\closeinteragestart{} 6--9 & -0.50\sym{**} & -0.57\sym{**} & -0.58\sym{**} & -0.70\sym{**} &   &    \\
	& \vspace*{-2mm}{\footnotesize (0.24) } &\vspace*{-2mm}{\footnotesize (0.26) } &\vspace*{-2mm}{\footnotesize (0.29) } &\vspace*{-2mm}{\footnotesize (0.32) } &   &    \\
\vspace*{0mm}\hspace*{5mm}\closeinteragestart{} 10--13 & -0.65\sym{**} & -0.70\sym{**} & -0.65\sym{**} & -0.71\sym{**} & -0.61\sym{*}  & -0.67\sym{**} \\
	& \vspace*{-2mm}{\footnotesize (0.25) } &\vspace*{-2mm}{\footnotesize (0.29) } &\vspace*{-2mm}{\footnotesize (0.26) } &\vspace*{-2mm}{\footnotesize (0.30) } &\vspace*{-2mm}{\footnotesize (0.31) } &\vspace*{-2mm}{\footnotesize (0.34) }    \\
\vspace*{0mm}\hspace*{5mm}\closeinteragestart{} 22--29&    0.28 &    0.40 &    0.14 &    0.21 &    0.15 &    0.21    \\
	& \vspace*{-2mm}{\footnotesize (0.25) } &\vspace*{-2mm}{\footnotesize (0.24) } &\vspace*{-2mm}{\footnotesize (0.28) } &\vspace*{-2mm}{\footnotesize (0.27) } &\vspace*{-2mm}{\footnotesize (0.28) } &\vspace*{-2mm}{\footnotesize (0.28) }    \\
\formatpanelheadsubregs{7}{L{\subheadwidth}}{Teach-point  $\times$}
\vspace*{0mm}\hspace*{5mm}\closeinteragestart{} 0--5 & -0.28 & -0.92 &   &   &   &    \\
	& \vspace*{-2mm}{\footnotesize (0.61) } &\vspace*{-2mm}{\footnotesize (0.68) } &   &   &   &    \\
\vspace*{0mm}\hspace*{5mm}\closeinteragestart{} 6--9&      -0.096 & -0.62 & -0.15 & -0.69 &   &    \\
	& \vspace*{-2mm}{\footnotesize (0.57) } &\vspace*{-2mm}{\footnotesize (0.64) } &\vspace*{-2mm}{\footnotesize (0.65) } &\vspace*{-2mm}{\footnotesize (0.70) } &   &    \\
\vspace*{0mm}\hspace*{5mm}\closeinteragestart{} 10--13 & -0.29 & -0.54 & -0.17 & -0.40 & -0.42 & -0.44    \\
	& \vspace*{-2mm}{\footnotesize (0.49) } &\vspace*{-2mm}{\footnotesize (0.54) } &\vspace*{-2mm}{\footnotesize (0.54) } &\vspace*{-2mm}{\footnotesize (0.60) } &\vspace*{-2mm}{\footnotesize (0.64) } &\vspace*{-2mm}{\footnotesize (0.69) }    \\
\vspace*{0mm}\hspace*{5mm}\closeinteragestart{} 22--29 & 0.055 & 0.029 & -0.16 & -0.17 &     -0.0067 &      -0.090    \\
	& \vspace*{-2mm}{\footnotesize (0.52) } &\vspace*{-2mm}{\footnotesize (0.58) } &\vspace*{-2mm}{\footnotesize (0.57) } &\vspace*{-2mm}{\footnotesize (0.61) } &\vspace*{-2mm}{\footnotesize (0.60) } &\vspace*{-2mm}{\footnotesize (0.64) }    \\
\midrule
Observations   &     8869 &    7466 &    5664 &    4790 &    4658 &    3946    \\
\midrule
\formatpanelheadregs{7}{L{\subheadwidth}}{Panel B: \newtablepanmale}
\formatpanelheadsubregs{7}{L{\subheadwidth}}{Non-teach-point $\times$}
\vspace*{0mm}\hspace*{5mm}\closeinteragestart{} 0--5&     -0.0053 & 0.084 &   &   &   &    \\
	& \vspace*{-2mm}{\footnotesize (0.22) } &\vspace*{-2mm}{\footnotesize (0.23) } &   &   &   &    \\
\vspace*{0mm}\hspace*{5mm}\closeinteragestart{} 6--9 & 0.016 & 0.010 & 0.051 &     -0.0028 &   &    \\
	& \vspace*{-2mm}{\footnotesize (0.22) } &\vspace*{-2mm}{\footnotesize (0.24) } &\vspace*{-2mm}{\footnotesize (0.26) } &\vspace*{-2mm}{\footnotesize (0.28) } &   &    \\
\vspace*{0mm}\hspace*{5mm}\closeinteragestart{} 10--13 & -0.41\sym{**} & -0.43\sym{*}  & -0.40\sym{*}  & -0.44\sym{*}  & -0.48\sym{*}  & -0.52\sym{**} \\
	& \vspace*{-2mm}{\footnotesize (0.21) } &\vspace*{-2mm}{\footnotesize (0.22) } &\vspace*{-2mm}{\footnotesize (0.21) } &\vspace*{-2mm}{\footnotesize (0.23) } &\vspace*{-2mm}{\footnotesize (0.25) } &\vspace*{-2mm}{\footnotesize (0.26) }    \\
\vspace*{0mm}\hspace*{5mm}\closeinteragestart{} 22--29&    0.35 &    0.41 &    0.32 &    0.45 &    0.25 &    0.37    \\
	& \vspace*{-2mm}{\footnotesize (0.24) } &\vspace*{-2mm}{\footnotesize (0.26) } &\vspace*{-2mm}{\footnotesize (0.27) } &\vspace*{-2mm}{\footnotesize (0.30) } &\vspace*{-2mm}{\footnotesize (0.28) } &\vspace*{-2mm}{\footnotesize (0.31) }    \\
\formatpanelheadsubregs{7}{L{\subheadwidth}}{Teach-point $\times$}
\vspace*{0mm}\hspace*{5mm}\closeinteragestart{} 0--5&      -0.054 &      -0.056 &   &   &   &    \\
	& \vspace*{-2mm}{\footnotesize (0.58) } &\vspace*{-2mm}{\footnotesize (0.62) } &   &   &   &    \\
\vspace*{0mm}\hspace*{5mm}\closeinteragestart{} 6--9&      -0.039 &      -0.049 &      -0.018 & -0.19 &   &    \\
	& \vspace*{-2mm}{\footnotesize (0.44) } &\vspace*{-2mm}{\footnotesize (0.47) } &\vspace*{-2mm}{\footnotesize (0.55) } &\vspace*{-2mm}{\footnotesize (0.58) } &   &    \\
\vspace*{0mm}\hspace*{5mm}\closeinteragestart{} 10--13&    0.20 &    0.34 &    0.13 &    0.19 &    0.18 &    0.22    \\
	& \vspace*{-2mm}{\footnotesize (0.38) } &\vspace*{-2mm}{\footnotesize (0.40) } &\vspace*{-2mm}{\footnotesize (0.43) } &\vspace*{-2mm}{\footnotesize (0.45) } &\vspace*{-2mm}{\footnotesize (0.48) } &\vspace*{-2mm}{\footnotesize (0.49) }    \\
\vspace*{0mm}\hspace*{5mm}\closeinteragestart{} 22--29 & -0.57 & -0.56 & -0.69 & -0.56 & -0.74\sym{*}  & -0.69    \\
	& \vspace*{-2mm}{\footnotesize (0.44) } &\vspace*{-2mm}{\footnotesize (0.52) } &\vspace*{-2mm}{\footnotesize (0.45) } &\vspace*{-2mm}{\footnotesize (0.53) } &\vspace*{-2mm}{\footnotesize (0.42) } &\vspace*{-2mm}{\footnotesize (0.48) }    \\
\midrule
Observations   &     9935 &    8452 &    6408 &    5499 &    5340 &    4592    \\
\midrule
\exclcontrol
\exclcontrolcont
\bottomrule
\footnotegap
\multicolumn{7}{L{\footwidth}}{\footnotesize\justify\footattaininter}\\
\end{tabular}
\end{adjustbox}
\end{table}

In Table \ref{tab:teachone}, consistent with our earlier gender results, we find more negative effects for females than males. In column one, closure decreased the educational attainment for females in non-teaching-points villages who were below age 6, between age 6 and 9, and between age 10 and 13 in the year of closure by 0.38 (s.e. 0.24), 0.50 (s.e. 0.24), and 0.65 (s.e. 0.25) years by 2011, respectively. These closely mirror the estimates from Table \ref{regone}, which is due to the fact that 88.5 percent of the 8869 females individuals are from non-teaching-point villages. The effects for females from teaching-point villages are negative as well, however, the effects are insignificant.\footnote{Given that teaching-points go up to grade 4, a larger proportion of the 10 to 13 year-olds at year of closure in teaching-point villages might not have been attending the within-village teaching-point. Our policy effect estimate, however, is still overall negative for children who were 10 to 13 year of age during the year of closure from teaching-point villages, but insignificant.} The weaker effects could be due to the small sample size, but they are also indicative of possibly weaker negative effects of closure for children who moved to consolidated schools from lower quality schools.

For males, results for non-teaching-point males from Table \ref{tab:teachone} are similar to results from Table \ref{regone} for all males given that 87.4 percent of the 9935 males are from non-teaching-point villages. Policy effects on males are noisy and generally insignificant for both teaching-point and non-teaching-point villages. As an exception, we do find significant negative effects, between -0.5 and -0.4 across columns, for males who experienced school closure between 10 and 13 years of age in non-teaching-point villages. For the same age-at-closure group from Table \ref{regone}, the effects were negative and between -0.20 and -0.35, with similar standard errors as here.

Overall, Table \ref{tab:teachone} shows that our results are robust to disaggregating teaching-point and non-teaching-point villages. The negative attainment effects of closure on girls remains when we allow for separate pre-existing educational attainment cohort trajectories and closure effects estimates for teaching-points schools---which were potentially of lower quality and accounted for a larger fraction of closed schools.
\end{spacing}
\clearpage

\subsection{Boarding Interactions}
\label{sec:boarding}
\begin{spacing}{\aptxspc}
In this section, we explore the heterogeneous effects of school closure interacted with boarding status. \textcite{chen_poor_2014} find students who board in school do significantly worse than those who don't board in school after village schools are closed. Different from \textcite{chen_poor_2014}'s approach---defining a boarding status variable by asking students about their individual boarding status---we obtain school level boarding status from the village-head survey. The village-head survey contains three questions about whether boarding is offered or required by the closest primary school from the village in 2011.\footnote{The first question, ``shi fou xu yao zhu su'', can be interpreted in two ways: if students in the village are required to board in the closest primary school to the village, or if students in the village need boarding in the closest primary school to the village, which is more indicative of demand for boarding. The second question, ``shi fou you zhu su sheng'', asks if there are boarding students in the closest primary school. The third question, ``Shi Fou You Ji Su Xue Sheng De Su She'', asks if there is a dorm for boarding students available in the closest primary school. The latter two variables might be more indicative of the supply for boarding.} The answers to these three questions are highly correlated and show similar interactive results with school closure variables for our attainment analysis. In the following section, we present results using the answer from the third question, whether boarding facility (dorm) is avaiable.

Table \ref{tab:tabboarddorm} shows the result of the interaction effect in Equation \eqref{eq:tp}, with \(\text{TP}_v\) replaced by village level variable \(\text{BRD}_v\), which is 1 if there is boarding dorm available in the primary school closest to village in 2011 for either villages with or without school closure. As before, the baseline group consists of children between 14 and 21 years of age in the year of closure. Under the assumption that \(\text{BRD}_v\) is fixed for closure variables since closure, results from Table \ref{tab:tabboarddorm} show if students who go to schools which provide boarding facilities do better or worse than those who attend schools without boarding facilities after village schools are closed.

\begin{table}[htbp]
\centering\caption{Effect of School Closure on Educational Attainment (Dorm Provision)\label{tab:tabboarddorm}}
\begin{adjustbox}{max width=\reginteract\textwidth}
\begin{tabular}{m{\lablcolwidth} >{\centering\arraybackslash}m{1.5cm} >{\centering\arraybackslash}m{1.5cm} >{\centering\arraybackslash}m{1.5cm} >{\centering\arraybackslash}m{1.5cm} >{\centering\arraybackslash}m{1.5cm} >{\centering\arraybackslash}m{1.5cm}}
\toprule
& \multicolumn{6}{L{\innerheadwidth}}{Outcome: grades completed by year 2011} \\                                 \cmidrule(l{5pt}r{5pt}){2-7}          &   \multicolumn{2}{L{3.0cm}}{\footnotesize } & \multicolumn{2}{L{3.0cm}}{\footnotesize 10 $\le$ 2011 Age $\le$ 34} & \multicolumn{2}{L{3.0cm}}{\footnotesize  15 $\le$ 2011 Age $\le$ 34} \\                                  \cmidrule(l{5pt}r{5pt}){2-3} \cmidrule(l{5pt}r{5pt}){4-5} \cmidrule(l{5pt}r{5pt}){6-7}          &   \multicolumn{1}{C{1.5cm}}{1} & \multicolumn{1}{C{1.5cm}}{2} & \multicolumn{1}{C{1.5cm}}{3} & \multicolumn{1}{C{1.5cm}}{4} & \multicolumn{1}{C{1.5cm}}{5} & \multicolumn{1}{C{1.5cm}}{6} \\
\midrule
\formatpanelheadsupregs{7}{L{15cm}}{\baselinegroup}
\formatpanelheadregs{7}{L{\subheadwidth}}{Panel A: \newtablepanfemale}
\formatpanelheadsubregs{7}{L{\subheadwidth}}{No dorm (2011) $\times$}
\vspace*{0mm}\hspace*{5mm}\closeinteragestart{} 0--5 & -0.12 & -0.31 &   &   &   &    \\
	& \vspace*{-2mm}{\footnotesize (0.37) } &\vspace*{-2mm}{\footnotesize (0.40) } &   &   &   &    \\
\vspace*{0mm}\hspace*{5mm}\closeinteragestart{} 6--9 & -0.51 & -0.65\sym{*}  & -0.70\sym{*}  & -0.93\sym{**} &   &    \\
	& \vspace*{-2mm}{\footnotesize (0.32) } &\vspace*{-2mm}{\footnotesize (0.36) } &\vspace*{-2mm}{\footnotesize (0.41) } &\vspace*{-2mm}{\footnotesize (0.45) } &   &    \\
\vspace*{0mm}\hspace*{5mm}\closeinteragestart{} 10--13 & -0.34 & -0.32 & -0.50 & -0.49 & -0.58 & -0.62    \\
	& \vspace*{-2mm}{\footnotesize (0.38) } &\vspace*{-2mm}{\footnotesize (0.44) } &\vspace*{-2mm}{\footnotesize (0.39) } &\vspace*{-2mm}{\footnotesize (0.45) } &\vspace*{-2mm}{\footnotesize (0.51) } &\vspace*{-2mm}{\footnotesize (0.54) }    \\
\vspace*{0mm}\hspace*{5mm}\closeinteragestart{} 22--29&    0.19 &      -0.046 & 0.072 & -0.32 &      -0.024 & -0.42    \\
	& \vspace*{-2mm}{\footnotesize (0.40) } &\vspace*{-2mm}{\footnotesize (0.38) } &\vspace*{-2mm}{\footnotesize (0.45) } &\vspace*{-2mm}{\footnotesize (0.41) } &\vspace*{-2mm}{\footnotesize (0.45) } &\vspace*{-2mm}{\footnotesize (0.43) }    \\
\formatpanelheadsubregs{7}{L{\subheadwidth}}{Has dorm (2011) $\times$}
\vspace*{0mm}\hspace*{5mm}\closeinteragestart{} 0--5 & -0.59\sym{*}  & -0.74\sym{**} &   &   &   &    \\
	& \vspace*{-2mm}{\footnotesize (0.31) } &\vspace*{-2mm}{\footnotesize (0.35) } &   &   &   &    \\
\vspace*{0mm}\hspace*{5mm}\closeinteragestart{} 6--9 & -0.55\sym{*}  & -0.59 & -0.55 & -0.55 &   &    \\
	& \vspace*{-2mm}{\footnotesize (0.33) } &\vspace*{-2mm}{\footnotesize (0.37) } &\vspace*{-2mm}{\footnotesize (0.37) } &\vspace*{-2mm}{\footnotesize (0.41) } &   &    \\
\vspace*{0mm}\hspace*{5mm}\closeinteragestart{} 10--13 & -0.82\sym{***} & -0.81\sym{**} & -0.74\sym{**} & -0.72\sym{**} & -0.66\sym{*}  & -0.60    \\
	& \vspace*{-2mm}{\footnotesize (0.31) } &\vspace*{-2mm}{\footnotesize (0.35) } &\vspace*{-2mm}{\footnotesize (0.33) } &\vspace*{-2mm}{\footnotesize (0.36) } &\vspace*{-2mm}{\footnotesize (0.38) } &\vspace*{-2mm}{\footnotesize (0.42) }    \\
\vspace*{0mm}\hspace*{5mm}\closeinteragestart{} 22--29 & 0.055 &    0.37 & -0.12 &    0.21 & -0.12 &    0.16    \\
	& \vspace*{-2mm}{\footnotesize (0.34) } &\vspace*{-2mm}{\footnotesize (0.36) } &\vspace*{-2mm}{\footnotesize (0.37) } &\vspace*{-2mm}{\footnotesize (0.38) } &\vspace*{-2mm}{\footnotesize (0.38) } &\vspace*{-2mm}{\footnotesize (0.39) }    \\
\midrule
Observations   &     8020 &    7073 &    5127 &    4532 &    4227 &    3740    \\
\midrule
\formatpanelheadregs{7}{L{\subheadwidth}}{Panel B: \newtablepanmale}
\formatpanelheadsubregs{7}{L{\subheadwidth}}{No dorm (2011) $\times$}
\vspace*{0mm}\hspace*{5mm}\closeinteragestart{} 0--5&      -0.033 &    0.15 &   &   &   &    \\
	& \vspace*{-2mm}{\footnotesize (0.34) } &\vspace*{-2mm}{\footnotesize (0.35) } &   &   &   &    \\
\vspace*{0mm}\hspace*{5mm}\closeinteragestart{} 6--9&    0.31 &    0.37 &    0.28 &    0.24 &   &    \\
	& \vspace*{-2mm}{\footnotesize (0.32) } &\vspace*{-2mm}{\footnotesize (0.34) } &\vspace*{-2mm}{\footnotesize (0.37) } &\vspace*{-2mm}{\footnotesize (0.38) } &   &    \\
\vspace*{0mm}\hspace*{5mm}\closeinteragestart{} 10--13 & -0.12 &      -0.098 & -0.27 & -0.31 & -0.37 & -0.42    \\
	& \vspace*{-2mm}{\footnotesize (0.29) } &\vspace*{-2mm}{\footnotesize (0.32) } &\vspace*{-2mm}{\footnotesize (0.31) } &\vspace*{-2mm}{\footnotesize (0.34) } &\vspace*{-2mm}{\footnotesize (0.38) } &\vspace*{-2mm}{\footnotesize (0.39) }    \\
\vspace*{0mm}\hspace*{5mm}\closeinteragestart{} 22--29&    0.52 &    0.63 &    0.51 &    0.82 &    0.45 &    0.74    \\
	& \vspace*{-2mm}{\footnotesize (0.41) } &\vspace*{-2mm}{\footnotesize (0.45) } &\vspace*{-2mm}{\footnotesize (0.45) } &\vspace*{-2mm}{\footnotesize (0.52) } &\vspace*{-2mm}{\footnotesize (0.47) } &\vspace*{-2mm}{\footnotesize (0.55) }    \\
\formatpanelheadsubregs{7}{L{\subheadwidth}}{Has dorm (2011) $\times$}
\vspace*{0mm}\hspace*{5mm}\closeinteragestart{} 0--5 & -0.16 &      -0.067 &   &   &   &    \\
	& \vspace*{-2mm}{\footnotesize (0.29) } &\vspace*{-2mm}{\footnotesize (0.33) } &   &   &   &    \\
\vspace*{0mm}\hspace*{5mm}\closeinteragestart{} 6--9 & -0.25 & -0.21 & -0.24 & -0.22 &   &    \\
	& \vspace*{-2mm}{\footnotesize (0.29) } &\vspace*{-2mm}{\footnotesize (0.31) } &\vspace*{-2mm}{\footnotesize (0.34) } &\vspace*{-2mm}{\footnotesize (0.37) } &   &    \\
\vspace*{0mm}\hspace*{5mm}\closeinteragestart{} 10--13 & -0.52\sym{*}  & -0.44 & -0.47\sym{*}  & -0.38 & -0.51 & -0.46    \\
	& \vspace*{-2mm}{\footnotesize (0.27) } &\vspace*{-2mm}{\footnotesize (0.29) } &\vspace*{-2mm}{\footnotesize (0.28) } &\vspace*{-2mm}{\footnotesize (0.30) } &\vspace*{-2mm}{\footnotesize (0.31) } &\vspace*{-2mm}{\footnotesize (0.32) }    \\
\vspace*{0mm}\hspace*{5mm}\closeinteragestart{} 22--29 & 0.075 &    0.16 & 0.073 &    0.19 &      -0.037 & 0.081    \\
	& \vspace*{-2mm}{\footnotesize (0.29) } &\vspace*{-2mm}{\footnotesize (0.32) } &\vspace*{-2mm}{\footnotesize (0.32) } &\vspace*{-2mm}{\footnotesize (0.35) } &\vspace*{-2mm}{\footnotesize (0.32) } &\vspace*{-2mm}{\footnotesize (0.36) }    \\
\midrule
Observations   &     9054 &    8049 &    5830 &    5223 &    4855 &    4354    \\
\midrule
\exclcontrol
\exclcontrolcont
\bottomrule
\footnotegap
\multicolumn{7}{L{\footwidth}}{\footnotesize\justify\footattaininter}\\
\end{tabular}
\end{adjustbox}
\end{table}

Consistent with our earlier gender results, we find more negative and significant closure effects for females than males in Table \ref{tab:tabboarddorm}. In particular, the effects of closure on female students in villages that have boarding dorms available in the closest school in 2011 are more negative and significant compared to those that have not. From column one, in villages that experienced school closure and had available dorms in 2011, girls whose ages were under age 6, 6-9 and 10-13 at the year of closure have attained less school by 0.59 (s.e. 0.31), 0.55 (s.e. 0.33), and 0.82 (s.e. 0.31) years separately by 2011. The effects of closure were negative but insignificant for females in villages that experienced closure but did not have available dorms in the closest school in 2011. Similar as before, closure effects are noisy for males, and we do not see overall statistically significant effects. The exception is that we see weakly significant negative effects of closure on boys who were 10-13 at year of closure from villages with available dorms in closest school in 2011.

Overall, we see greater negative closure effects when closure is interacted with boarding availability. This finding mirrors the results from \textcite{chen_poor_2014}, where the boarding variable is based on child reported boarding status, which could reflect the combination of household choices, schools' boarding availabilities and requirements. In our context here, there are three possible explanations for the negative effects of the dorm availability interaction variables. First, our information on boarding provision in the closest school is only available in 2011. It is possible that a school did not initially offer boarding at the year of closure, but started offering boarding between the year of closure and 2011 in response to the impact of closure on enrollment and attainment. In other words, there may exist a reverse causality between boarding availability and worse schooling outcomes. To explore this interesting point more fully, we would need to have the history of boarding provisions and the history of school enrollments. Given that we only have enrollment and boarding provision information in 2011, we are not able to explore this point further.

Second, given question wording in our data, it is difficult to disentangle whether a school offers boarding, or whether a school requires boarding. Mandatory boarding might be well-intentioned, but could induce additional pecuniary and non-pecuniary costs on households, especially if boarding is not viewed as a safe option for girls. Some of these potential negative effects for boarding are discussed in \autocite{chen_poor_2014}.

Third, the village school boarding status variable is strongly correlated with the distance to school and school facility quality variables.\footnote{For example, among schools without dorm availability, 56.4 percent are in the 0km distance group and 10.6 percent are in the greater than 3km distance group. Among schools with dorm availability, 26.6 percent are in the 0km distance group and 45.8 percent are in the greater than 3km distance groups.} Boarding status is thus a potential proxy for the joint effects of boarding availability, distance and school facility quality.

Testing our enrollment results from Table \ref{regfive} in Section \ref{sec:mechanism}, we find that boarding status is generally negatively related to enrollment. The effect of boarding is significant when distance to school and school facility quality are not included in the regression, but lose some statistical significance when these variables are included. The inclusion of the boarding status variable, however, does not lead to changes in the overall magnitude and significance of the coefficients on school distance and quality in the enrollment regressions.\footnote{The inclusion of boarding status leads to a small reduction in sample size because boarding status is unknown for some schools. Results available upon request from the authors.}

Boarding provision is a critical dimension of the school closure policy. The availability of boarding options might be beneficial, in principle, if boarding offerings are low cost and attractive. Unfortunately, our dataset does not allow for a full analysis of the causal effect of boarding. Our boarding interaction results should not be interpreted as causal, but rather as correlational and reflecting the possible endogeneity of boarding offerings.

\end{spacing}
\clearpage 

\end{document}